\documentclass[fleqn,usenatbib]{mnras}

\usepackage{newtxtext,newtxmath}
\usepackage[T1]{fontenc}
\usepackage{ae,aecompl}
\usepackage{subfig}
\usepackage{xparse}
\usepackage{letltxmacro}

\LetLtxMacro\oldcitep\citep
\RenewDocumentCommand{\citep}{O{} O{} m}{\oldcitep{#3}}

\usepackage{graphicx}	% Including figure files
\usepackage{amsmath}	% Advanced maths commands
\usepackage{amssymb}	% Extra maths symbols
\usepackage{amstext,verbatim}

\usepackage{mathtools}
\usepackage{xcolor}
\usepackage{float,epsfig,multirow,rotating}
\DeclareMathOperator*{\argmin}{arg\,min}
\DeclareMathOperator*{\argmax}{arg\,max}
\definecolor{Gr1}{HTML}{1B9E77}
\definecolor{Gr2}{HTML}{D95F02}
\definecolor{Gr3}{HTML}{7570B3}
\definecolor{Gr4}{HTML}{E7298A}
\definecolor{Gr5}{HTML}{66A61E}

\newcommand{\norm}[1]{\Vert#1\Vert}

\newcommand{\bftheta}{\boldsymbol{\theta}}

\newcommand{\bfGamma}{\boldsymbol{\Gamma}}

\newcommand{\bfSigma}{\boldsymbol{\Sigma}}

\newcommand{\bfmu}{\boldsymbol{\mu}}
\newcommand{\bfnu}{\boldsymbol{\nu}}

\newcommand{\bfx}{\boldsymbol{x}}
\newcommand{\bfX}{\boldsymbol{X}}

\newcommand{\bmX}{\boldsymbol{\mathcal X}}

\newcommand{\mD}{\mathcal D}
\newcommand{\mB}{\mathcal B}
\newcommand{\mM}{\mathcal M}

\newcommand{\mG}{\mathcal G}

\newcommand{\bfL}{\boldsymbol{L}}

\newcommand{\bit}{\begin{itemize}}
\newcommand{\eit}{\end{itemize}}
\newcommand{\ben}{\begin{enumerate}}
\newcommand{\een}{\end{enumerate}}
\newcommand{\beqn}{\begin{equation}}
\newcommand{\eeqn}{\end{equation}}
\newcommand{\bea}{\begin{eqnarray*}}
\newcommand{\eea}{\end{eqnarray*}}
\newcommand{\bpf}{\begin{proof}}
\newcommand{\epf}{\end{proof}\ms}
\newcommand{\ms}{\medskip}

\DeclareMathAlphabet{\mathsfit}{\encodingdefault}{\sfdefault}{m}{sl}
\SetMathAlphabet{\mathsfit}{bold}{\encodingdefault}{\sfdefault}{bx}{sl}

\title[Five Kinds of Gamma Ray Bursts]{Gaussian-Mixture-Model-based Cluster Analysis Finds Five Kinds of Gamma Ray Bursts in the BATSE Catalog}

\author[Chattopadhyay and Maitra]{
  Souradeep Chattopadhyay,$^{1}$
  and Ranjan Maitra,$^{2}$\thanks{E-mail: maitra@iastate.edu (RM)}
  \\
  $^{1}$Department of Statistics, University of Calcutta, 35, Ballygunge Circular Road,  Kolkata 700019, West Bengal, India\\
$^{2}$ Department of Statistics, Iowa State University, 2438, Osborn
  Drive, Ames, Iowa 50011-1090, USA
}
\date{Accepted XXX. Received YYY; in original form December 29, 2016}

\pubyear{2016}

\begin{document}
\label{firstpage}
\pagerange{\pageref{firstpage}--\pageref{lastpage}}
\maketitle

\begin{abstract}
  Clustering methods are an important tool to enumerate and describe the
different coherent kinds of Gamma Ray Bursts (GRBs). But
  their performance can be affected by a number of factors such 
  as the choice of clustering algorithm and inherent associated
  assumptions, the inclusion of  variables in clustering,  nature of
  initialization methods used or the iterative algorithm or  the
  criterion used to judge the optimal number of groups supported by
  the data. We analyzed GRBs from the BATSE 4Br catalog using
  $k$-means and Gaussian Mixture Models-based clustering methods and
  found that after accounting for all the above factors, all six
  variables -- different subsets of which have been used in the
  literature -- and that are, namely, the flux duration variables  ($T_{50}$,
  $T_{90}$), the peak flux ($P_{256}$) measured in 256-millisecond 
  bins, the total fluence ($F_t$) and the spectral hardness ratios
  ($H_{32}$ and $H_{321}$) contain information on clustering. Further,
  our analysis found evidence of five different 
  kinds of GRBs and that these groups have different kinds of
  dispersions in   terms of shape, size and orientation. In terms of
  duration, fluence and spectrum, the five types of GRBs were
  characterized as intermediate/faint/intermediate, long/intermediate/soft,
intermediate/intermediate/intermediate, short/faint/hard and
long/bright/intermediate. 

\end{abstract}


%%%%%%%%%%%%%%%%%%%%%%%%%%%%%%%%%%%%%%%%%%%%%%%%%%

%%%%%%%%%%%%%%%%% BODY OF PAPER %%%%%%%%%%%%%%%%%%

\section{Introduction}
\label{sec:introduction}
Gamma Ray Bursts (GRBs) are the brightest known electromagnetic events 
known to occur in space and have been studied extensively ever since
their discovery in the late sixties. While the cosmological origin
of GRBs is well-established, questions on their source and
nature remain
unresolved~\citep{chattopadhyayetal07,piran05}. Indeed, as explained
by \citet{shahmoradiandnemiroff15}, researchers \citep[{\em
  e.g.,}][]{mazetsetal81,norrisetal84,dezalayetal92} 
have hypothesized that GRBs really belong to a heterogeneous group with
several sub-populations, but the exact number and descriptive
properties of these groups is an area of active research and
investigation.
Most analyses have traditionally focused on univariate statistical and
descriptive methods for classification,
% GRBs have traditionally been classified, using univariate
% statistical methods,  into two groups comprising of bursts of short
% $(< 2\mbox{s})$ and long $(>2\mbox s)$
% duration~\citep{dezalayetal92,kouveliotouetal93}.  
% Over the years several methodologies has been adopted to obtain a
% proper classification ofGRBs.
with particular focus on the duration of GRBs (as
measured by $\log_{10}T_{90}$ or the time  
within which 90\% of the GRB flux has arrived ).
For example, \citet{kouveliotouetal93} analyzed the
$\log_{10}T_{90}$ distribution of 222 GRBs of the Burst and Transient Source
Experiment (BATSE) 1B Catalog  and found it to have a bimodal
distribution. This led to the establishment of the well-known 
classification of GRBs into the two classes, one  of short bursts (of
durations less than 2s) and the other class of long bursts (bursts
with durations greater than 2s). \citet{pendletonetal97} applied spectral analysis technique to $882$ BATSE GRBs and provided evidence about the existence of bursts populations of two types, the HE (High Energy) bursts and the NHE (no-High Energy) bursts. The 
progenitors of long GRBs have mainly been associated with the collapse
of massive stars \citep{paczynski98,woosleyandbloom06} while those of
short GRBs are thought to be NS-NS, that is the merger of two neutron stars, or
NS-BH, that is the merger of a neutron star with a black
hole~\citep{nakar07}. \citet{horvathetal98} made both two- and
three-Gaussian fits to the $\log_{10}T_{90}$ variable of the 797 GRBs in the
BATSE 3B Catalog and indicated the presence of a third Gaussian
component at a 99.98\% level of significance thus providing evidence
of a third class\citep[see also][]{horvath02}. Similar findings were
also reported on the distribution of
$T_{90}$ with the \textit{BeppoSAX} \citep{horvath09}, \textit{Swift}/BAT 
\citep{horvathetal08} and \textit{Fermi}/GBM \citep{tarnopolski15}
datasets, with the observation holding regardless of whether
${\chi}^2$-fitting
\citep{horvathetal98,tarnopolski15} or maximum
likelihood 
\citep{horvath02,horvathetal08,horvath09,horvathandtoth16} was used in analysis.
\citet{zitounietal15} analyzed 248
\textit{Swift}/BAT GRBs with known redshifts and confirmed a
preference of statistical tests for three groups instead of two. More
recently however, \citet{zhangetal16} and \citet{kulkarnianddesai17}
studied the duration distributions of the  BATSE,
\textit{Swift} and \textit{Fermi} GRB datasets and concluded that only
the \textit{Swift} dataset potentially supports a three-Gaussians
model, while two-Gaussians models 
are strongly  supported by the BATSE and \textit{Fermi}
datasets. 
Three kinds of GRBs were also  found in the {\it Swift} GRB datasets
by \citet{deugartepostigoetal11} and \citet{horvathandtoth16}.

\citet{mukherjeeetal98} explain that many studies in astronomy have
typically only used univariate and bivariate statistical analyses,
potentially providing an incomplete understanding of the
relationships between the different variables in the GRB
datasets. Building on the review of multivariate statistical methods
provided by \citet{feigelsonandbabu98} for the benefit of more
thorough analysis of datasets in astronomy,
\citet{mukherjeeetal98} used both
non-parametric hierarchical clustering and a more formal
Model-based Clustering (MBC) approach (with Gaussian mixtures and using six and
three parameters) on 797 BATSE 3B  
catalog GRBs and found evidence
in favour of three groups. \citet{chattopadhyayetal07} carried out
clustering using the $k$-means algorithm and MBC with Dirichlet
Process mixture modeling on the larger BATSE 4B catalog (as per a reviewer of this article) of 1594 GRBs using the six 
variables used by \citet{mukherjeeetal98} and supported the presence
of a third group. Similar findings were reported by
\citet{veresetal10} and \citet{horvathetal10}, but their analysis used
the smaller {\it Swift/BAT} dataset and just two variables 
({\em i.e.} $\log_{10} T_{90}$ and the log-hardness ratio $\log_{10}
H_{32}$ where $H_{32} = F_2/F_3$, with 
$F_2$ and $F_3$ being the time-integrated fluences in the 50-100 keV and
100-300 keV spectral channels, respectively). \citet{horvathetal10} also argued
lack of significant evidence for a fourth cluster by means of a
likelihood ratio $\chi^2$-test on twice the difference in
loglikelihood for three and four clusters. (Note, however, that the
use of the $\chi^2$ test on twice the difference in loglikelihoods
between two models assumes that the larger model is nested within the
null model, an assumption that does not generally hold for MBC or other
non-hierarchical clustering algorithms). Our own experiments using
$k$-means and the Jump statistic~\citep{sugarandjames03} with the 
BATSE GRB dataset and variables used by \citet{chattopadhyayetal07}
did not replicate their results. This led us to review and perform a
detailed investigation and cluster analysis of the GRBs in the 
BATSE 4Br catalog.

Cluster analysis~\citep{kettenring06,xuandwunsch09,everitt11} is widely used in
many disciplines to group observations into homogeneous classes or
{\em clusters}. Clustering is an unsupervised learning 
approach wherein classification rules are obtained in the absence of a
response variable. As such, it is a difficult problem in
general~\citep{maitra01}. Many clustering algorithms exist but they
can all be broadly grouped into the hierarchical and the
non-hierarchical kinds. The first case comprises both
agglomerative and divisive clustering algorithms where groups of
observations are  formed  in a tree-like hierarchy
with the property that observations that are together at one level are
also together higher up the tree. These algorithms typically have
criteria set to measure the discrepancy between two entities and also
to specify how these distances change upon merging between any two
sub-groups. As pointed out by \citet{chattopadhyayetal07}, the assumption of a
hierarchy is restrictive and such a  scheme is
methodologically unable to recover or repair from a partitioning
happening higher up in the tree. 

Non-hierarchical partitional algorithms, on the other hand, lack the
regimented structure of their hierarchical counterparts and usually
rely on optimizing an objective function, which in the case of MBC, is 
the observed loglikelihood function given the observations. For a specified
number of groups, the optimization problem is typically multimodal and
solved by iterative greedy algorithms, therefore careful
initialization is important. Various  approaches~\citep[see { \em
    e.g.}][]{akaike73,akaike74,schwarz78,rousseeuw87,sugarandjames03,maitraetal12}
exist to determine the number of homogeneous groups supported by the
data.  

Even within non-hierarchical clustering, the choice of algorithm ({\em
  e.g.} $k$-means or MBC) is important and hinges on the types and
reasonableness of assumptions that underlie the different kinds of
groups. For instance, the $k$-means algorithm assumes homogeneous
spherically-dispersed groups of roughly similar sizes while MBC allows
for greater flexibility in the shapes, orientations and volumes as well
as different sizes in the distributions of each group. Another aspect
which assumes tremendous significance in the context of all the different
studies published using different numbers and sets of variables in the astronomy
literature~\citep[{\em e.g.}]{mukherjeeetal98,veresetal10,horvathetal06,horvathetal10,deugartepostigoetal11,horvathandtoth16,zhangetal16}, is
that of the specific parameters (variables in statistics jargon) that
should be used in clustering. Actually, incorporating  redundant
information by including variables that do not add to clustering
information can potentially impact and even 
give rise to 
spurious cluster assignments~\citep[refer to][for both illustrations and
  potential solutions in different
  contexts.]{rafteryanddean06,maugisetal09,wittenandtibshirani10},
Thus, selection of the most relevant
variables having discriminating information is very  important in the
context of clustering.

\begin{comment}
Implications of proper methods 
for variable selection is of utmost importance since many of the
variable selection methods available suffer from some drawbacks in
some scenario. This can cause the method to return erroneous
results. Apart from this clustering algorithms heavily depends on the
nature of initialization and thus improper initialization can have
considerable effect on the final results. The K-means algorithm
assumes the clusters to be homogeneous and spherical deviation, from
which can cause considerable errors in the final results. All these
considerations call for a thorough analysis to ascertain the
feasibility of K-means in this scenario. 
\end{comment}

This paper is organized in three further
sections. Section~\ref{sec:background} provides an overview of partitional
clustering algorithms and discusses issues  arising from improper or
inadequate initialization, methods for 
choosing the number of groups and for finding the most relevant variables
for inclusion for clustering. GRBs in the BATSE 4Br catalog are
clustered and analyzed using these methods in
Section~\ref{sec:GRB}. The paper concludes with 
some discussion. Additionally, an online supplement provides the 
interested reader with commented R~\citep{R} 
and (where appropriate, {\tt html} code) for performing  our cluster analysis. 

\section{Overview of Clustering Methods}
\label{sec:background}
We first briefly but comprehensively discuss the many
issues bedeviling cluster analysis, especially in the context of GRBs,
and strategies to combat them. As mentioned in
Section~\ref{sec:introduction}, there is a large amount of literature on 
clustering, so here we focus on issues in methods that are commonly used in
astronomy and that are easily implemented using the 
open-source statistical software R~\citep{R} and its packages. We
restrict attention only to the non-hierarchical $k$-means and MBC
algorithms given hierarchical clustering's
inflexibility in allowing for wider classes of models
\citep{chattopadhyayetal07}. 
\subsection{The $k$-means clustering algorithm}
\label{sub:kemans}
Given $n$ $p$-dimensional observations
$\bfx_1,\bfx_2,\ldots,\bfx_n$, the $k$-means algorithm~\citep{macqueen67}
groups the observations 
into a pre-determined ($K$) number of groups
$\mG_1,\mG_2,\ldots,\mG_K$ by minimizing the objective function
\begin{equation}
  W_K = \sum_{k=1}^K\sum_{i=1}^n\zeta_{ik}^{(K)}\norm{\bfx_i - \bmu_k}^2,
  \label{kmeans}
\end{equation}
where $\norm{\bfx} = \sqrt{\bfx^{T}\bfx}$ and, for each $i=1,2,\ldots
,n$ and
$k=1,2,\ldots,K$, we have that $\zeta_{ik}^{(K)}$ is 1 if $\bfx_i$ belongs
to the $k$th group 
$\mG_k$ and is 0 otherwise. In this problem, $\zeta_{ik}^{(K)}$s and
$\bmu_k$s are all statistical parameters over which $W_K$ is
optimized. Optimizing \eqref{kmeans} is an NP-hard
problem~\citep{gareyandjohnson79} with an iterative solution provided
by the $k$-means algorithm~\citep{lloyd82,forgy65} having the
following steps: 
\begin{enumerate}
\item {\em Initialization.} Select $K$ initial seeds for $\bmu^\circ_1,\bmu^\circ_2,\ldots,\bmu^\circ_K$. 
\item \label{assignment} {\em Assignment.} Assign each observation
  to the $\bfmu_k^\circ$ closest to it. That is, assign 
  $\bfx_i$ to $\mG_k$ where $k =
  \argmin_l\norm{\bfx_i-\bfmu^\circ_l}$. Set $\zeta_{ik}^{(K)} = 1$ for
  when $\bfx_i\in\mG_k$ and 0 otherwise.
\item \label{updates} {\em Parameter updates}. Update $\bfmu^\circ_k$
  to the respective group means. That is, update 
  $\bfmu^\circ_k =
  \sum_{i=1}^n\zeta_{ik}^{(K)}\bfx_i/ 
  \sum_{i=1}^n\zeta_{ik}^{(K)}$.
\item {\em Iteration and Convergence.} Repeat Steps~\ref{assignment}
  and \ref{updates} until no further rearrangement is possible. 
\end{enumerate}
The above $k$-means  algorithm is easily described and  commonly used,
but the statistics community and software use a more efficient 
variant  of the  algorithm 
due to \citet{hartiganandwong79} which keeps track of observations
having no possibility of changing immediately, and takes them out of
contention from the current update calculations. The R function {\tt
  kmeans()} implements this variant as its default. 

An important aspect of the $k$-means algorithm is that it treats
contributions from each variable uniformly: thus variables with smaller
magnitudes would tend to be swamped out by the variables having higher
magnitudes in the calculation of Equation~\eqref{kmeans}. Therefore,
unless all variables are known to be on the same scale, it is 
customary for the variables to be scaled individually before
optimization.

The $k$-means algorithm needs specification of $K$ to
proceed. There are several approaches  to determining an optimal $K$,
but a quick method that we have also found to work well with $k$-means
is the jump statistic 
\citep{sugarandjames03} that was also used by
\citet{chattopadhyayetal07} in their $k$-means clustering of the BATSE
4B GRBs. Given a clustering solution with $K$ groups, the jump
statistic involves computing an overall minimum 
distance measure ($d_K$) which is estimated by the distortion $d_K'$ 
that, in the $k$-means scenario, can be taken to be the optimized value of
\eqref{kmeans}, that is $d_K'\equiv W_K$. \citet{sugarandjames03}
contend that the {\em distortion curve} obtained by plotting $d_K'$
against $K$ will monotonically decrease with
increasing $K$ until $K$ is greater than the true number of groups,
after which the curve will level off with a smaller slope. The jump
statistic defined by \citet{sugarandjames03} is defined as $J_{K}=
d_K'^{-Y} - d'^{-Y}_{K-1}$. For $K=1$, $J_K = d_K'^{-Y}$.
The value of $K$ which gives the largest jump statistic yields the
$K$ most supported by the data.
The exact choice of $Y$ is left to the user, with no clear guideline
but the  choice of $Y=p/2$ was used in their experiments, and also adopted by
\citet{chattopadhyayetal07} in their GRB analysis. A
significance-based bootstrap approach that estimates the $p$-value of a more
complicated (higher $K$) solution than needed for describing the
data was suggested by \citet{maitraetal12}. This method provides
$p$-values for solutions with all possible pairs of $K$ and $K'>K$
being tested and has the added advantage of even assessing the case
of no clustering. 

  \subsection{Model-based clustering}
  \label{sub:mbc}
  One drawback of $k$-means clustering is that it is an optimization
  algorithm with no grounding in variability or the mechanism that
  generated the data~\citep[see however, \citealt{celeuxandgovaert92}
    for writing $k$-means as a 
    Classification-EM algorithm in the context of a GMM and][for
    framing $k$-means in a 
    nonparametric distributional
    setting]{maitraandramler09,maitraetal12}. Model-based  
  clustering~\citep{fraleyandraftery02,melnykovandmaitra10} provides a
  principled approach to the problem of clustering by postulating
  that, for a given total number of components $K$, the observations
  $\bfx_1,\bfx_2,\ldots,\bfx_n$ are realizations from the mixture
  model~\citep{mclachlanandpeel00} with density
  \begin{equation}
    f(x;\nu) = \sum_{k=1}^{\textit{K}}\pi_k f_k(\bfx;\bfnu_k)
    \label{mixmodel}
  \end{equation}
  where $f_k(\bfx;\bfnu_k)$ represents the density of the $k$th group
  parameterized by $\bfnu_k$ 
  and $\pi_k$ represents the mixing proportion of the $k$th group, that is,
  $\pi_k = Pr[\bfx_i \in \mG_k]$ for $k=1,2,\ldots,K$ and $\sum_{k=1}^K\pi_k
  = 1$. The most commonly-used mixture model is the Gaussian mixture
  model (GMM), where each $f_k(\bfx;\bfnu_k)$ is taken to be the
  multivariate Gaussian 
  density $\phi(\bfx;\bmu_k, \bfSigma_k)$ with mean $\bmu_k$ and dispersion
  matrix  $\bfSigma_k$. Estimation is via the Expectation-Maximization (EM)
  algorithm \cite{dempsteretal77,mclachlanandkrishnan08} which has the following steps:
  \begin{enumerate}
  \item {\em Initialization}. Obtain starting values
    $\{(\bfSigma_k^\circ,\bfmu_k^\circ,\pi_k^\circ);k=1,2,\ldots,K\}$. 
  \item {\em E-step updates}. For $k=1,2,\ldots,K$ and
    $i=1,2,\ldots,n$, calculate the posterior probability that the $i$th 
    observation arises from the $k$th group:
    \begin{equation}
      \label{eq.post}
      \pi^\circ_{ik} =
      \frac{\pi^\circ_k\phi(\bfx_i;\bfmu_k,\bfSigma_k)}{\sum_{l=1}^K\pi^\circ_l
        \phi(\bfx_i;\bfmu_l,\bfSigma_l)}.
    \end{equation}.
  \item {\em M-step updates}. For $k=1,2,\ldots,K$, obtain updates:
    \begin{eqnarray}
      \pi^\circ_k & = \frac{\sum_{i=1}^n\pi^\circ_{ik}}{\sum_{k=1}^K\sum_{i=1}^n\pi^\circ_{ik}},\\
      \bfmu^\circ_k &=
      \frac{\sum_{i=1}^n\pi^\circ_{ik}\bfx_i}{\sum_{i=1}^n\pi^\circ_{ik}}, \mbox{
        and }\\
      \bfSigma^\circ_k & =
      \frac{\sum_{i=1}^n\pi^\circ_{ik}(\bfx_i-\bfmu^\circ_k)(\bfx_i-\bfmu^\circ_k)^T}{\sum_{i=1}^n\pi^\circ_{ik}}.
    \end{eqnarray}
  \item Alternate
    between the E- and M-steps until numerical convergence.
  \end{enumerate}
  Faster versions of the EM algorithm that reduce redundant
  computations exist: 
  indeed, the R package {\tt   EMCluster} utilizes the Alternative
  Partial Expectation Conditional Maximization (APECM)
  algorithm~\citep{chenandmaitra11,chenetal13} that provides a
  substantial speedup. Note also that the EM algorithm itself only
  provides estimates for the GMM, with clustering obtained only in a
  post-processing step by assigning $\bfx_i$ to the class for which the
  converged E-step posterior probability is the highest. Thus, upon
  convergence, $\bfx_i$ is
  assigned to the class $k$ where $k =
  \argmax_{l}\pi^\circ_{il}$. 

  MBC also assumes a given number of
  components. There are several approaches to selecting 
  $K$~\citep{melnykovandmaitra10}, but the most popular is
  to choose the $K$ having the highest Bayes' Information Criterion
  (BIC)~\citep{schwarz78} which is 
  calculated as the maximized log likelihood function (obtained by the
  converged EM) penalized by subtracting $(m\log n)/2$, where $m$ is
  the number of unconstrained parameters in the $K$-component mixture
  model. BIC  is easily calculated and has appealing
  consistency properties~\citep{keribin00}. Further, it can be 
  cast~\citep{kassandraftery95} as a quick and convenient
  approximation to
  the {\em Bayes Factor}~\citep{neathandcavanaugh12}   which is a
  popular approach to estimating the relative posterior
  odds between  competing models. The {\em Bayes Factor} is an
  important model-selection metric so we discuss it in some
  detail here.

  Suppose that we have two competitors 
  $\mathcal{M}_{1}$ and $\mathcal{M}_{2}$ (under consideration, out of
  $M$ possible models: $\mM_\ell$ for $\ell =1,2,\ldots M$).
  In the current context, let  $\mM_1$  be the model
$\sum_{k=1}^K\pi_k\phi(\bfx;\bfmu_k,\bfSigma_k)$  and  let $\mM_2$ be the model
$\sum_{k=1}^{K'}\eta_k\phi(\bfx;\bftheta_k,\bfGamma_k)$ (with $K\neq K'$, though
other formulations, {\em e.g.} $K=K'$ but $\bfSigma_k \equiv
\sigma^2I$ for $k = 1,2,\ldots,K$ while $\bfGamma_k$ are unstructured
for $k=1,2,\ldots,K'$, and generalizations are possible).
Let $\pi(\mathcal{M}_{1})$ and $\pi(\mathcal{M}_{2})$ be the prior 
chance  of occurrence of models $\mathcal{M}_{1}$ and
$\mathcal{M}_{2}$  respectively. Also, 
  $\pi(\mathcal{M}_{1}|\mD)$ and $\pi(\mathcal{M}_{2}|\mD)$ are the posterior
  probabilities of $\mathcal{M}_{1}$ and $\mathcal{M}_{2}$
  given  $\mD$. For $m=1,2,\ldots,M$, the posterior probability of 
  $\mathcal{M}_{m}$ given $\mD$ is 
  \begin{equation}
    \pi(\mathcal{M}_{m}|\mD)=\frac{\pi(\mD|\mathcal{M}_{m})\pi(\mathcal{M}_{m})}{\sum_{l=1}^{M} \pi(\mD|\mathcal{M}_{l})~\pi(\mathcal{M}_{l})}
  \end{equation}
 where $\pi(\mD|\mathcal{M}_{i})$ is the likelihood of the dataset
 $\mD$ given the model  $\mathcal{M}_{i}$. For the two models
 $\mathcal{M}_{1}$ and $\mathcal{M}_{2}$, the ratio
 ${\pi(\mathcal{M}_{1})}/{\pi(\mathcal{M}_{2})}$ is known as the
 prior odds in favor of $\mathcal{M}_{1}$ while
 ${\pi(\mathcal{M}_{1}|\mD)}/{\pi(\mathcal{M}_{2}|\mD)}$ is
 the posterior odds in favor of $\mathcal{M}_{1}$. The \textit{Bayes
   Factor} for models $\mathcal{M}_{1}$ and $\mathcal{M}_{2}$ is
 defined as
 \begin{equation}
   \mB_{12} = \frac{\pi(\mathcal{M}_{1 }|\mD)
     \pi(\mM_2)}{\pi(\mathcal{M}_{2}|\mD) \pi(\mM_1)},
   \label{BF}
 \end{equation}
that is, the  ratio of the posterior odds in favor of
$\mathcal{M}_{1}$ to the prior odds in favor $\mathcal{M}_{1}$.
Intuitively, it is easy to see that if $\mB_{12} > 1$ 
for the two models $\mathcal{M}_{1}$ and $\mathcal{M}_{2}$, then we
prefer $\mathcal{M}_{1}$ over $\mathcal{M}_{2}$.  
  \begin{comment}
    Twice the logarithm of the Bayes Factor  (Computed as the ratio
    of integrated likelihoods for the two models) is approximately equal
    to the difference between the BIC values for the two models that are
    being compared. Clustering is achieved by assigning each observations
    to one of the groups using some pre-specified rule. A popular way is
    through Bayes Rule which allocates observations to the clusters on the
    basis of their posterior probabilities.For a complete review on model
    based clustering see \citet{melnykovandmaitra10}.
  \end{comment}
  The ratio $\mB_{12}$ can, at times, be hard to compute but, under the
  assumption of noninformative and flat priors, the BIC can be
  easily   used to approximate  this ratio because $2\log {\mB}_{12}$
  approximately equals the difference between BIC   values of the two
  models $\mM_1$ and $\mM_2$ being compared. The R package
  {\tt  
    mclust}~\citep{fraleyetal12}  uses BIC to decide
  between different $K$-component models having different dispersion
  assumptions. We conclude here by noting that
  \citet{kassandraftery95} also provide some guidance on the
  difference between BICs in choosing a more complex model:
  specifically, they recommend the more complicated model positively, strongly
  and very strongly accordingly as the improvement in BIC is between 2 and 6, 6
  and 10 and beyond 10, respectively. Differences in BIC that are less
  than 2 are  worthy of no more than a bare
  mention~\citep{kassandraftery95}. We use these criteria to 
  determine $K$ with Gaussian-Mixture-Models-based Clustering 
  (GMMBC) of GRBs.  
  
  \subsection{Issues with $k$-means and MBC algorithms}
  \subsubsection{Initialization}
  Both $k$-means and MBC are iterative methods that
  find local optima in the vicinity of their initialization. As such,
  the choice of initial values to start these algorithms has great
  impact on its performance. We refer to \citet{maitra09} for examples
  on the pitfalls 
  of poor initialization and also for references to possible
  remedies. For both algorithms, one common fix is to start the
  algorithm at several random initial values and run to convergence from
  each starting point. Then the solution with the lowest
  optimal value of equation~\eqref{kmeans} in the case of $k$-means and
  highest optimized loglikelihood value for MBC is taken to
  be the optimal solution. %For $kselect $K$ random observations from
  %the sample and then start $k$-means with
  %these as starting values and in the case of
  %MBC,  from mean  several valid points  
  A commonly-used initializer for each of these $k$-means runs sets
  $\bfmu_k^\circ$ to be a random sub-sample of $K$ values from
  $\bfx_1,\bfx_2,\ldots,\bfx_n$. For  GMMBC,
  such initialization method can  also be employed, but only to obtain
  $\bfmu_k^\circ$s, and then Euclidean distances can be used to make
  assignments from where initializing estimates of $\pi^\circ_k$s and
  $\bfSigma^\circ_k$s are obtained. A more sophisticated approach to
  initializing GMMBC uses the {\em emEM} 
  algorithm \citep{biernackietal03} which starts the EM 
  algorithm from several random 
  starts, runs each short ({\em em}) step to lax convergence, and then
  runs the optimal among them to stricter numerical convergence in a
  long {\em EM} step. 
  (See also \citet{maitra09} for a  variant of {\em emEM} called {\em
    Rnd-EM} which 
    runs the short {\em em} steps for only one iteration.)
  An alternative deterministic approach,
  implemented in {\tt mclust} uses model-based hierarchical clustering
  to initialize the GMMBC. 
  \subsubsection{Inherent structural assumptions on the groups}
  The $k$-means optimization
  function~\eqref{kmeans} does not explicitly
  make use of a specific model, the algorithm itself prefers homogeneous
  spherically-dispersed groups, and can be viewed as a special case of a
  Classification-EM algorithm using a GMM with equal mixing proportions
  and homogeneous spherical
  dispersions~\citep{celeuxandgovaert92}. This can have an effect on 
  clustering performance even when the true $K$ is known. In order to
  demonstrate the effect of  deviation from homogeneity and spherical
  assumptions we simulate a 2-D data set with three groups, all of
  which have spherical dispersions, but with the first group having
  much larger dispersion than the other two. We partition this dataset into
  three groups using both $k$-means and GMMBC given $K=3$ and display the
  results in
  \begin{figure*}
    \mbox{ 
      %\subfloat[]{\label{fig:a}\includegraphics[width=0.33\textwidth]{t1_original_done.pdf}}
      \subfloat[][\label{fig:1a}]{\includegraphics[width=0.33\textwidth]{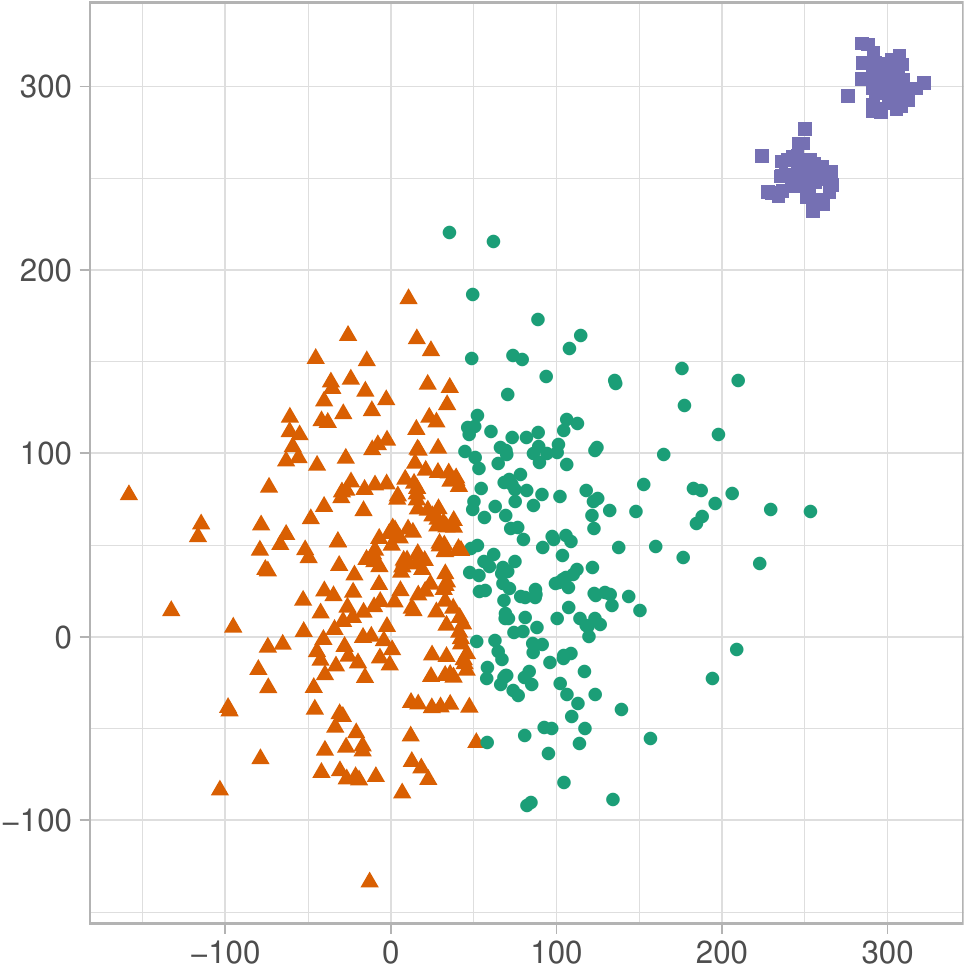}}
      \subfloat[][\label{fig:1b}]{\includegraphics[width=0.33\textwidth]{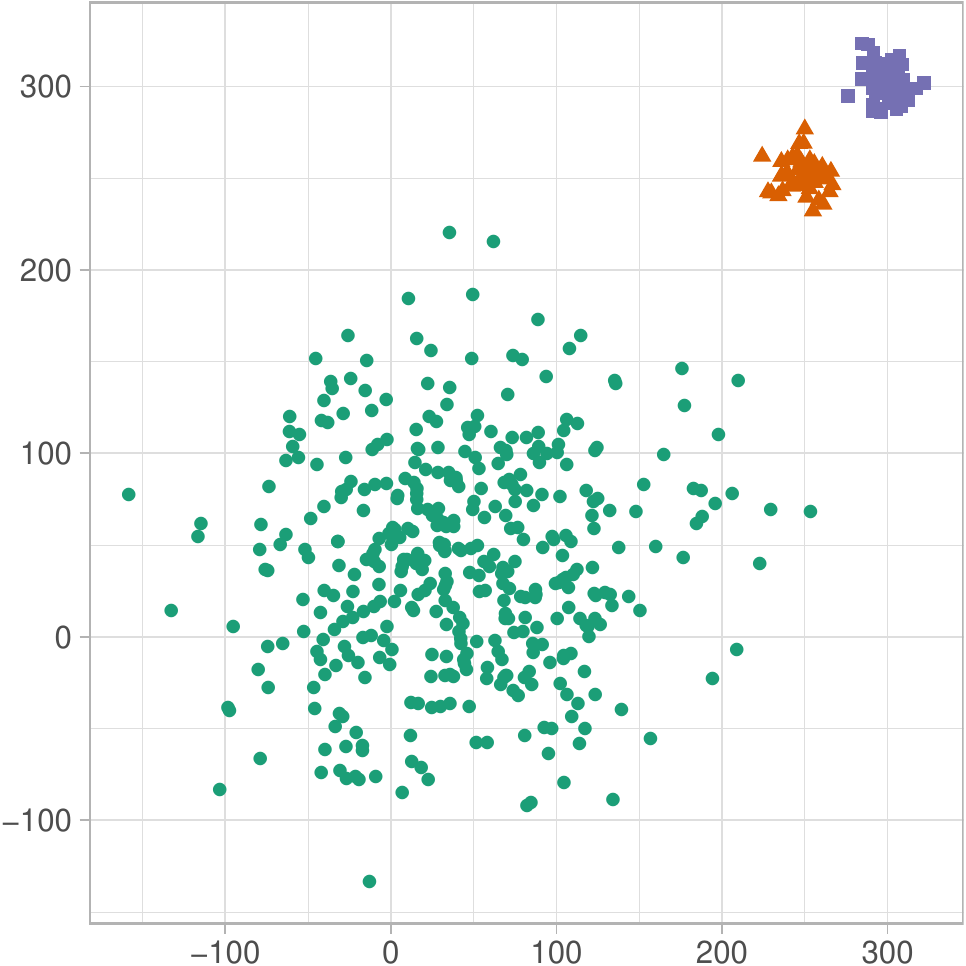}}
      \subfloat[][\label{fig:1c}]{\includegraphics[width=0.33\textwidth]{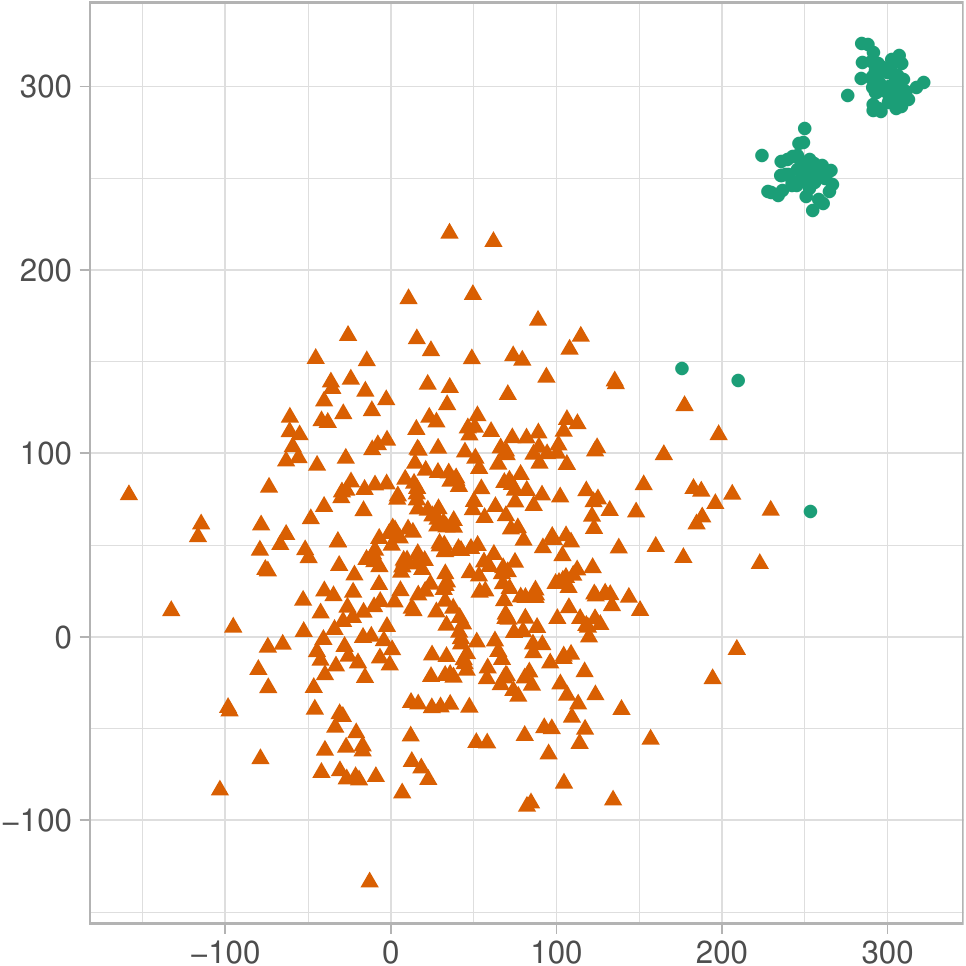}}
    }
    \caption{Clustering of the simulated dataset using (a) $k$-Means
      with $K=3$ and (b)  Model Based Clustering with $K=3$ and (c)
      $k$-means clustering with $K=2$, which is also the optimal number of
      groups, as per the jump statistic.}
    \label{fig:kmeans.MBC}
  \end{figure*}
  Figure \ref{fig:kmeans.MBC}. The results clearly indicate that there
  is a distinction in the optimized solutions obtained using
  (well-initialized) $k$-means   and GMMBC. In the first case~(Figure
  \ref{fig:1a}), the $k$-means  
  solution splits the larger cluster into two while combining the two
  smaller true clusters into one simply because of its predilection for
  homogeneous spherical groups. GMMBC has the ability to
  model mixing proportions and dispersions more generally, providing 
  the   correct solution here~(Figure~\ref{fig:1b}). To further illustrate
  pitfalls arising from potential model misspecification, we consider 
  using well-initialized $k$-means solutions for
  $K=1,2,\ldots,10$. The jump statistic chooses $K=2$ as the optimal
  solution, which Figure~\ref{fig:1c} shows more or less puts observations in
  the larger cluster into the first group and 
  observations in the two   smaller groups into the second. Three
  observations from the larger 
  group that are closer to the combined center of the two (true) smaller groups
  are also misclassified. The choice of $K=2$ as the
  optimal grouping makes sense under the   assumption of spherical
  homogeneous clusters governing $k$-means   because the larger group 
  is substantially well-separated from the two smaller ones which at
  some resolution can be grouped together as one cluster with a
  similar  spherical dispersion structure as the larger one. 
  This example illustrates  the 
  importance of the assumptions underlying the algorithms
  used in clustering and the challenges that may consequently arise in
  their interpretation.  
  \subsubsection{Variable Selection in Clustering}
  \label{r:d}
  An important issue in clustering is in deciding the
  variables that are relevant for the 
  purpose. Several authors~\citep[{\em
      e. g.}][]{horvath02,horvathetal08,zitounietal15,zhangetal16} 
  have used only $\log_{10} T_{90}$ while others~\citep[{\em e.g.}
  ][]{veresetal10,horvathetal04,horvathetal06,horvathetal10} have used  $\log_{10} 
  T_{90}$ and $\log_{10}H_{32}$. \citet{mukherjeeetal98} and
  \citet{chattopadhyayetal07} have used between three and six
  variables in their   investigations. Redundancy in variables
  included for clustering can considerably degrade
  performance~\citep[see][who provided a variable selection approach
    for GMMBC]{rafteryanddean06}.
  Specifically, \citet{rafteryanddean06} formulated  variable
  selection  in  terms of model selection, and proposed an effective
  way to  remove irrelevant variables from a dataset. Their
  methodology partitions a   set of variables ${\bmX}$ into three
  subsets ${\bmX}^{(1)}$, 
 ${\bmX}^{(2)}$ and ${\bmX}^{(3)}$ where ${\bmX}^{(1)}$
 consists of the set of variables already selected for clustering,
 ${\bmX}^{(2)}$ consists of the variable(s) under consideration
 for inclusion or exclusion from the set of clustering  variables and
 ${\bmX}^{(3)}$ denotes the other remaining 
 variables. The decision to include or exclude ${\bmX}^{(2)}$
 from the set of clustering variables is taken based on the following
 two  models on the entire dataset ${\bmX}$: 
  \begin{equation*} 
  \begin{split}
  \mathcal{M}_1:~P({\bmX}|z_1) 
     			&=~P({\bmX}^{(1)}, {\bmX}^{(2)}, {\bmX}^{(3)}|
z_1)\\ 
&=~P({\bmX}^{(3)}|{\bmX}^{(2)},
{\bmX}^{(1)})P({\bmX}^{(2)}|{\bmX}^{(1)})P({\bmX}^{(1)}|z_1)\\
\mathcal{M}_2 :~P({\bmX}|z_1)
     			&=~P({\bmX}^{(1)}, {\bmX}^{(2)}, {\bmX}^{(3)}|
z_1)\\
&=~P({\bmX}^{(3)}|{\bmX}^{(2)}, {\bmX}^{(1)})P({\bmX}^{(2)},{\bmX}^{(1)}|z_1)
\end{split}
\end{equation*}
where $z_1$ is the set of unobserved cluster memberships.  Model
$\mathcal{M}_1$ implies that
${\bmX}^{(2)}$ gives no additional
information about clustering while model $\mathcal{M}_2$ implies
that ${\bmX}^{(2)}$ provides additional information about
clustering beyond that provided by the already-included
${\bmX}^{(1)}$. $\mM_1$ and $\mM_2$ are compared using
the {\it Bayes  Factor}~\eqref{BF} with
\begin{comment}
\begin{equation}
\mB_{12}=~\frac{P({\bmX}|\mathcal{N}_1)}{P({\bmX}|\mathcal{N}_2)}
\end{equation}
\end{comment}
$B_{12} > 1$  providing evidence that the set  ${\bmX}^{(2)}$ is
redundant. As before, BIC provides a quick approximation. This results
in the greedy search algorithm of \citet{rafteryanddean06}
which chooses or deselects variables as per improvement in BIC. At
each stage the best  
  combination of the number of 
  groups and clustering model is chosen. A brief description of the
  forward- and backward-stepwise selection   algorithm is as follows: 
  \begin{enumerate}
  \item
    \label{step1}
    Select the first clustering variable as the one that provides the
    maximum evidence of univariate clustering.   
  \item
    \label{step2}
    The second clustering variable is selected such that it gives
    the highest  evidence of bivariate clustering after including the first
    variable that was selected.  
  \item
    \label{step3}
    The next variable is proposed such that it shows the maximum
    evidence of clustering with the previously chosen
    variables included. The variable is accepted as a
    clustering variable if the evidence favours this outcome over not
    including it as a     clustering variable. 
  \item
    \label{step4}
    The variable (from the current set of variables) for which the
    evidence of  clustering including all the selected
    variables versus clustering including all the
    variables except the proposed variable is weakest is proposed to
    be removed from the current set of    variables. This 
    variable is removed if the evidence for clustering with its
    inclusion is     weaker than the evidence to the contrary.
  \item
    \label{step5}
    Steps \ref{step3} and \ref{step4} are iterated until two
    consecutive steps are rejected, at which point the procedure
    terminates. 
  \end{enumerate}
  This algorithm is implemented by the R package {\tt
    clustvarsel}~\citep{clustvarselR} and we use it here. 
  \subsection{Validity of Obtained Groupings}
  A difficult aspect of clustering, exacerbated for  multivariate
  datasets, is cluster validation. Here, we  
  discuss a few graphical and numerical ways for assessing our groupings. 
  \subsubsection{The silhouette width}
  \citet{rousseeuw87} developed the silhouette width as a popular
  but computationally intensive way for judging distance-based
  clustering results. The basic objective  is
  to compare the similarity of an object to its own group 
  members with that to observations in other groups. The
  silhouette value $\zeta_i$ for the $i$th observation lies in
  $[-1,1]$ with a high value   
  indicating a close match of the observation with its own group and a
  poor match with others, thus satisfying the 
  primary clustering goal of finding distinct homogeneous
  groups. The $\zeta_i$s are calculated for each observation
  with the distribution of these indices providing a measure of
  cluster validity. Operationally, $\zeta_i$  is calculated for the
  $i$th observation $\bfx_i$ assigned to 
  $\mG_k$ (say) via the following steps: 
  \begin{enumerate}
  \item Calculate the average distance to all other observations in
    $\mG_k$ (its own group). Call this average distance
    $\varpi_i$. Then, for $\bfx_i\in\mG_k$, we have   $\varpi_i =
    \sum_{j\in\mG_k}d(\bfx_i,\bfx_j)/(n_k-1)$ where $n_k$ is the number of
    observations in $\mG_k$.
  \item Also, for each group $\mG_l,l\neq k$, 
    obtain the average distance, $\vartheta^{(l)}_i$ of $\bfx_i$ to all $\bfx_j\in\mG_l$. That is, calculate $\vartheta^{(l)}_i =
    \sum_{j\in\mG_l}d(\bfx_i,\bfx_j)/n_l$. Let $\vartheta_i =
    \min_{l\neq k}\vartheta_i^{(l)}$  be the minimum of these
    average distances of $\bfx_i$ to the other groups. 
  \item The silhouette index for the $i$th observation is then
    \begin{equation*}
      \varsigma_i =
      \frac{\vartheta_i - \varpi_i}{\max (\varpi_i, \vartheta_i)}. 
    \end{equation*}
  \end{enumerate}
  Clearly, $-1 \leq\varsigma_i\leq1$ $\forall$ $i$.
  In the above, $d(\bfx_i,\bfx_j)$ is the distance between $\bfx_i$ and $\bfx_j$,
  for which the Euclidean or any standard metric is used.
  However,  while it is straightforward to use the
  Euclidean distance for calculating $\varsigma_i$s for   $k$-means
  clusterings, the distance to be used with results from
  GMMBC and other clustering methods is not always clear.

  \subsubsection{Graphical Displays of Supervised Principal Components}
  Meaningful graphical  displays are challenging
  propositions for datasets having more than two dimensions, so
  projections onto 2- or 3-dimensions have to be made in a way that
  the main features are presented. For grouped data, the main features
  to be presented   are the between-groups separability  and
  within-groups homogeneity.  The goal then is to find the projection
  that best shows the separation of the groups in the data. The most popular
  statistical approach for dimensionality reduction is Principal
  Components Analysis (PCA) independently developed
  by~\citet{pearson1901} and \citet{hotelling33a,hotelling33b}
  which 
  projects (possibly) correlated variables into (a possibly lower number
  of) uncorrelated variables called \textit{Principal
    Components}.  PCA is an unsupervised learning tool sensitive to
  outliers, but more importantly in the context of grouped data, does
  not use this information and as such is   not very useful in the
  context of finding  
  projections that provide this sense of separation between 
  groups. Weighted PCA~\citep{korenandcarmel04}
  is an alternative to PCA that can handle outliers depending on the choice  
  of weights. The main objective is to find a $q$-dimensional 
  projection ($q < p$, where $p$ is the dimensionality of $\bfx_i$s) that
  maximizes  
  \begin{eqnarray}
    \sum_{i<j}\upsilon_{ij}\delta^2_{ij}(q)
    \label{sq.dist:SPCA}
  \end{eqnarray}
  where $\delta^2_{ij}(q)$ denotes the Euclidean distance between
  two observations $\bfx_i$ and $\bfx_j$  in the $q$-dimensional
  projection space and ${(\upsilon_{ij})}^{n}_{i,j=1}$ denotes the
  symmetric pairwise non-negative weights (or
  dissimilarities). By convention, $\upsilon_{ii} \equiv 0$ $\forall$ $i$.
 \citet{korenandcarmel04} propose to robustify PCA by choosing
  \begin{eqnarray}
    \upsilon_{ij} = \frac{1}{\delta_{ij}}
    \label{spca:weight}
  \end{eqnarray}
  where $\delta_{ij}\equiv\delta_{ij}(p)$ denotes the Euclidean
  distance between  $x_i$ and $x_j$ in the original space.
  This choice of weights for weighted PCA yields  \textit{Normalized
    PCA} and can result in 
  well-balanced projections~\citep{korenandcarmel04}. 
  It still remains to describe how the projection into the
  $q$-dimensional space   should be   carried out. Recall that the
  objective is to obtain the   $q$-dimensional projection 
  that maximizes \eqref{sq.dist:SPCA}. This is done by defining  a
  $n\times n$ \textit{Laplacian} matrix 
  \begin{eqnarray}
    L^{(\upsilon)}_{ij} = \begin{cases}
      \sum_{j=1}^{n}\upsilon_{ij}  & i=j\\
      -\upsilon_{ij}& i\neq j
    \end{cases}
    \label{spca:laplacian}
  \end{eqnarray}
  and then obtaining the $q$-dimensional projections in terms of the
  eigenvectors corresponding to the $q$ largest eigenvalues of the
  matrix $\bfX^T\bfL^{(\upsilon)}\bfX$, where $\bfX^T = (\bfx_1\vdots
  \bfx_2\vdots\ldots\vdots \bfx_n)$ is the  
  $p\times n$ matrix containing the data.

  Our description hitherto has not accounted for available label
  information in the data. This label information can be  used to
  inform the weights and obtain  the 
  discriminating projection in 
  $q$-space that will yield the projected distances $\delta_{ij}(q)$
  that will separate out the groups, as far as possible, in the projected
  space. \citet{korenandcarmel04} suggest tweaking the
  $\upsilon_{ij}$s for the $(\bfx_i,\bfx_j)$ pairs where both observations 
  have the same class labels. Thus, they modify
  \begin{eqnarray}
    \upsilon^{(\ell)}_{ij} (\tau) = \begin{cases}
      \tau\upsilon_{ij}& \text{if \textit{i} and \textit{j} have the same label}\\
      \upsilon_{ij}& \text{otherwise}
    \end{cases}
    \label{spca:lab}
  \end{eqnarray}
  where $0 \leq \tau \leq 1$.
  A typical choice of $\tau$ is 0, though other values are possible.
  Proceeding with this specification of $\upsilon^{(\ell)}_{ij}$ leads
  us to modify $\bfL^{(\upsilon)}$ in Equation~\eqref{spca:laplacian} to 
  $\bfL^{(\upsilon^{(\ell)})}$ and to get the projections in the
  same manner as for  normalized PCA. Projections of the data thus
  obtained provide us with supervised principal components and the
  methodology is called Supervised PCA (SPCA).
  %We have rescaled the data before obtaining the components by pre
  %multiplying observations from each class with square root of the
  %inverse of variance covariance matrix of each class.\\
 Note that calculation of  $\delta_{ij}$s   in the original space has
 been proposed using Euclidean distance. This distance is again
 natural to use  
  $k$-means-clustering results, but the distance to be used for
  results from GMMBC is not always clear.
  \subsubsection{Measuring distinctiveness of groups via the overlap}
  \label{sec:overlap}
  An overlap measure typically indicates the extent
  to which clusters obtained through a method are
  distinct from another and thus can be used to judge the goodness of
  the clustering. In the context of GMMs, \citet{maitraandmelnykov10}
  \begin{comment}
  For example \citet{dasgupta99} a defined a overlap measure termed
  as \textit{c}-separation as: two p-variate Gaussian densities
  $N_{p}(\bfmu_{i},\bfSigma_{i})$ and 
  $N_{p}(\bfmu_{j},\bfSigma_{j})$ are \textit{c}-separated if
  $||\bfmu_{i}-\bfmu_{j}||\geqslant\textit{c}\sqrt{p\ max\ (d_{max}(\bfSigma_{i}),d_{max}(\bfSigma_{j}))}$,
  where $d_{max}(\bfSigma)$ is the largest eigenvalue of $\bfSigma$. The
  author mentioned there that there is significant to moderate to scant
  overlap between at-least two clusters for c = 0.5, 1.0 and 2.0.\\ A
  better measure has been defined by \citet{maitraandmelnykov10}
  \end{comment}
  defined overlap between two Gaussian clusters as the sum of their 
  misclassification probabilities. For the general GMM, these measures
  are somewhat involved~\citep[see Theorem 1    of][]{maitraandmelnykov10} but for the special case of the
  $k$-means solutions the pairwise overlap between the $k$th and the
  $l$th cluster can be defined as $\omega_{kl} =
  = 2\Phi(-\norm{\bfmu_k-\bfmu_l}/2\sigma)$ where $\Phi(\cdot)$ is the
  $p$-variate Gaussian density, $\bfmu_k$ and $\bfmu_l$ are the $k$th and the
  $l$th cluster means and $\sigma$ is the common (homogeneous) standard
  deviation for each group, estimated unbiasedly as $W_K/((n-K)p)$
  with $W_K$ as the optimized value of~\eqref{kmeans}.
  In either case, for a dataset partitioned into $K$   groups, we  can
  obtain a $K \times K$ matrix $\Omega$ of pairwise overlap
  measures. Summarizing this matrix is not easy, so 
  \citet{melnykovandmaitra11}~(see manual) developed the generalized
  overlap $\ddot\omega$ by borrowing a summary measure from
  \citet{maitra10}. Specifically, they proposed $\ddot\omega =
  (\lambda_{(1)}-1)/{(K-1)}$, where $\lambda_{(1)}$ is the largest
  eigenvalue of $\Omega$. Smaller values of $\ddot\omega$ are expected
  to   indicate the   most distinctive groupings. The R package {\tt
    MixSim}~\citep{melnykovetal13} calculates the pairwise and 
  generalized overlap measures through the {\tt overlap()} and {\tt
    overlapGOM()} functions, respectively. Referring back to the
  example in Figure~\ref{fig:kmeans.MBC} we calculate $\ddot\omega$ to
  be 0.126 for the clustering of Figure \ref{fig:1a}, $2.932\times 10^{-5}$
  for the grouping of 
  Figure \ref{fig:1b} and 0.005 for the partitioning in Figure
  \ref{fig:1c},  respectively.
  \begin{comment}
  The lower value of $\ddot\omega$ for  Figure \ref{fig:1c}
  relative to that in Figure \ref{fig:1b} makes sense when viewed in
  the context of the distinctiveness of clusters.
  The two spherical
  clusters of  Figure \ref{fig:1c} are far more separated in general
  than the correct three-groups solution of  Figure \ref{fig:1b}. This
  example illustrates
  \end{comment}
  Despite the good agreement here with the correct clustering
  solution, we note that $\ddot\omega$ is only a worthwhile diagnostic
  in the assessment of the obtained clustering and not necessarily a 
  mechanism to determine the best clustering solution for which
  BIC here provides the correct answer. 

\section{Cluster Analysis of GRBs}
\label{sec:GRB}
The BATSE catalog provides temporal and spectral information for many
GRBs. Of interest to us are the parameters:
\begin{description}
\item[$T_{50}$:]{The time by which 50\% of the flux arrive.}
\item[$T_{90}$:]{The time by which 90\% of the flux arrive.}
\item[$P_{64}$, $P_{256}$, $P_{1024}$:]{The peak fluxes measured in
  bins of 64, 256 and 1024 milliseconds, respectively.} 
\item[$F_{1}$, $F_{2}$, $F_{3}$, $F_{4}$:]{The four time-integrated
  fluences in the  20-50, 50-100, 100-300, and $>$ 300 keV spectral channels,
  respectively.} 
\end{description}
\citet{mukherjeeetal98} identified three more composite variables
used by researchers for studying GRBs. These are:
\begin{description}
  \item [$F_t$=$F_1+F_2+F_3+F_4$:]{The total fluence of a
GRB.}
  \item [$H_{32}= F_3/F_2$:]{Measure of spectral hardness using the
    ratio of $F_2$ and $F_3$.}
  \item [$H_{321}= F_3/(F_1 + F_2)$:]{Measure of spectral hardness
    based on the ratio of channel fluences $F_1,F_2,F_3$.}
\end{description}
The current BATSE catalog, that is, the BATSE 4Br
catalog~\citep{paciesasetal99} contains bursts from the BATSE 3B
catalog studied by 
\citet{mukherjeeetal98} along with 515 additional bursts between 20
September 1994 and 29 August 1996. The BATSE 4Br catalog also contains
revised locations for 208 bursts from the BATSE 4B catalog analyzed by
\citet{chattopadhyayetal07}. The BATSE 4Br Catalog (and also the BATSE
4B and older catalogs) has several zero entries for the four 
integrated time fluences $F_1$, $F_2$, $F_3$, $F_4$. There is also one
zero entry for the peak fluxes. 
\begin{table}
  \caption{Number ($n_j$) of observations with zeroes in each of
    the BATSE 4Br catalog parameters (denoted by $X_j$).}  
  \label{tab:data}
    \begin{tabular}{cccccccccc}
\hline\hline
$X_j$&$T_{50}$&$T_{90}$&$P_{64}$&$P_{256}$&$P_{1024}$&$F_1$&$F_2$&$F_3$&$F_4$\\
\hline
$n_j$&0&0&1&1&1&29&12&6&339\\\hline\hline
\end{tabular}
\end{table}
Table~\ref{tab:data} provides the number of zero observations for each
field in the BATSE 4Br catalog. How these zero entries should be
included in the analysis can determine the quality of our results,
especially if these zeroes are indicators for anomalous or missing
values rather than numerical values. This has particular impact in the
context of variables derived using $F_4$, for which there are as many
as 339 zero values. \citet{mukherjeeetal98} and \citet{chattopadhyayetal07} have
performed their analyses on the BATSE 3B and 4B catalogs,
respectively, after dropping these zero entries. \citet{horvathetal06}
analyzed the BATSE 
4Br catalog, but they restricted their attention only to
$\log_{10}T_{90}$ and $\log_{10}H_{32}$ for which 1956 GRBs have
non-zero observations in both the $F_2$ and $F_3$ parameters that
go into calculating $H_{32}$.

It would be instructive to find out how the zeroes in the parameters
occur. To obtain more information in this regard, we contacted the
BATSE GRB team with our questions. A team member, Charles
A. Meegan has,  in 
personal communication, explained that ``the zero values ultimately
derive from the fact that a model background, determined from data
before and after the burst, must be subtracted from the signal during
the burst.  Occasionally, fluctuations in the background lead to a
negative value for the burst fluence.  This is most prevalent in the
lowest and highest channels of the weaker bursts, where the intensity
is generally low..  Since a negative fluence is unphysical, these
values were set to zero in the catalog.'' He went on to add that in
these cases, the quoted error bar is to be interpreted as a $1-\sigma$
upper limit. Further details on  how the background  calculations and
subtractions are done is provided in  
\citet{pendletonetal94}'s analysis of the first BATSE catalog.
Thus, the recorded zeroes in the BATSE catalog for these parameters are not numerical values, but rather records of uncertain values. Table~\ref{tab:data} provides further support of this assertion because most zeros are in the fluences $F_1$ and $F_4$. Therefore, the only appropriate approach to treat these GRBs with zero parameters as essentially having missing observations in those parameters. (We have also verified that the 1594 GRBs used in \citet{chattopadhyayetal07} are among the 1599 BATSE 4Br GRBs without any zero parameters -- the balance five GRBs have trigger numbers 107, 2450, 6368, 6404 and 6645.) Therefore, in the absence of clustering methods for partially observed data, we exclude GRBs missing any parameter from our cluster analysis. 

The missing observations in the four integrated time fluences  mean  that computation of the composite variables $F_t$, $H_{32}$ and  $H_{321}$ is not possible for all the 1973 GRBs in the BATSE 4Br  catalog. (The one case with zeroes for the peak flux parameters also  has zero readings for the integrated time fluences and is therefore part of the GRBs missing $F_1$, $F_2$, $F_3$ or $F_4$.) . Thus after
 excluding the incomplete GRBs the 4Br catalog has 1599 GRBs
 containing complete information on 
the six variables $T_{50}$, $T_{90}$, $P_{256}$, $F_t$, $H_{32}$ and
$H_{321}$. 
\begin{figure}
\includegraphics[width=\columnwidth]{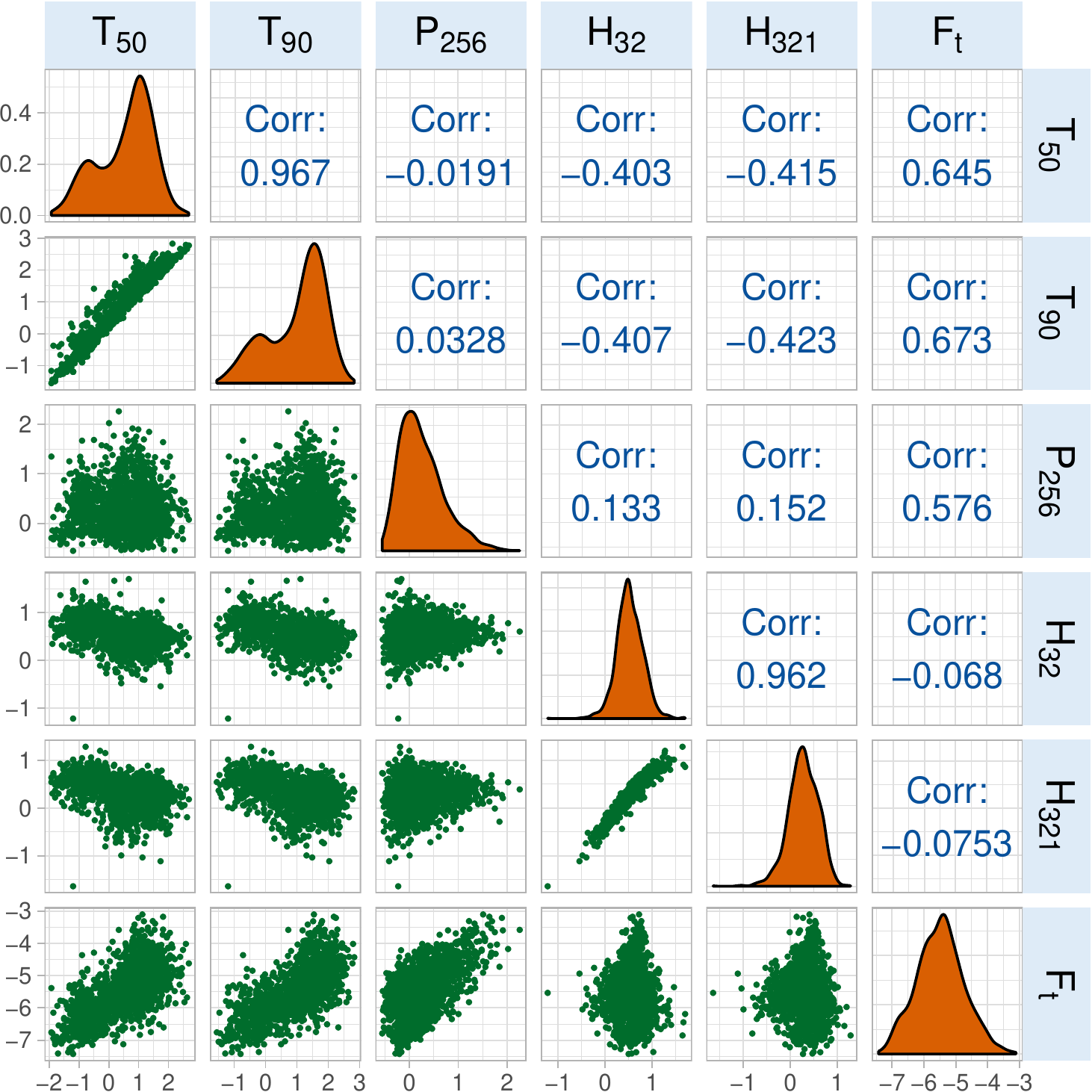}
\caption{A matrix of scatterplots (the lower triangular portion),
  density plots ( the diagonal) and correlation coefficients(the upper
  triangular portion) of the base-10 logarithms of the six parameters
  $T_{50}$, $T_{90}$, 
  $P_{256}$, $H_{32}$, $H_{321}$, $F_t$ using 1599 GRBs of the BATSE
  4Br catalog.}
\label{fig:Bivariate}
\end{figure}
\citet{mukherjeeetal98} used these six
variables, namely $\log_{10}T_{50}$, $\log_{10}T_{90}$,
$\log_{10}P_{256}$, 
$\log_{10}F_{t}$, $\log_{10}H_{32}$ and $\log_{10}H_{321}$ for
hierarchical clustering and three ($\log_{10} T_{90}$, $\log_{10}
F_t$ and $\log_{10} H_{321}$) variables for GMM-based analysis on 797
GRBs (having  observations on all these parameters) from the
BATSE 3B catalog. \citet{chattopadhyayetal07} used all six of these
variables in their cluster analysis of the GRBs from the BATSE 4B catalog. We also  
revisit cluster analysis of the GRBs using the BATSE 4Br catalog using
these six variables. Analysis of these six variables means that we can
only consider for analysis the 1599 GRBs for which readings on all the
parameters are available. However, this leaves out 374 GRBs. Of these,
only 44 GRBs have $F_t$ missing. Thus,
there are 1929 GRBs with information on five parameters 
$T_{50}$, $T_{90}$, $P_{256}$, $H_{32}$ and $H_{321}$. We analyze
these GRBs separately also in order to get an indication on whether the
clustering properties of this larger set of GRBs are similar to that
of the 1599 GRBs with data on all six parameters.

We first briefly discuss the univariate and bivariate
relationships between the six parameters in the BATSE 4Br catalog. 
Figure \ref{fig:Bivariate} displays the bivariate scatterplots along with
the correlation coefficients and univariate density plots of the six
parameters. The two duration
variables $\log_{10}T_{50}$ and $\log_{10}T_{90}$
and the  two hardness ratios
$\log_{10}H_{32}$ and $\log_{10}H_{321}$
show very high positive association between themselves. High positive
associations are also seen between the duration variables and the total 
fluence $\log_{10}F_t$. The peak flux $\log_{10}P_{256}$
and $\log_{10}T_{90}$ exhibits a weak positive association among themselves. The association of $\log_{10}P_{256}$ with
$\log_{10}T_{50}$ is weakly negative while the association between either
duration variable and each hardness ratios is also moderate. Note also
that the logarithmic transformations on each of the parameters has
reduced skewness appreciably, as seen in the univariate density plots
displayed along the diagonal of Figure~\ref{fig:Bivariate}.

Beyond the associations, the scatterplots of Figure~\ref{fig:Bivariate}
show the limitations posed by bivariate and univariate summaries, as
was also discussed by 
\citet{mukherjeeetal98}. Both $\log_{10}T_{50}$ and $\log_{10}T_{90}$
are bimodal in their univariate densities, but none of the bivariate
figures show much grouping. Thus, any grouping in the GRBs, if they
exist, are in dimensions higher than two and can not be recovered by
considering only univariate or bivariate summaries. A reviewer has
pointed out that \citet{bagolyetal98} have used the first two
principal components to perform clustering in a bivariate framework:
note, however, that \citet{chang83} has demonstrated theoretically as
well as by simulations and in applications to real data, that the
principal components with the largest eigenvalues do not necessarily
contain the most information about the cluster structure, and that
taking a subset of principal components can lead to a major loss of
information about the groups in the data. Similar concerns have also
been expressed by other authors~\citep[see, {\em
  e.g.}][]{greenandkrieger95,yeungandruzzo01,kettenring06}. We now perform
cluster analysis on the 1599 GRBs with observations using all six
parameters  $\log_{10}T_{50}$, $\log_{10}T_{90}$, $\log_{10}P_{256}$, 
$\log_{10}F_{t}$, $\log_{10}H_{32}$ and $\log_{10}H_{321}$.

\subsection{Clustering GRBs Using all Six Parameters}
We first perform $k$-means clustering of the 1599
GRBs in the BATSE 4Br and then move on to GMMBC.
\subsubsection{$k$-means clustering}
\label{GRB:kmeans}

We revisited \citet{chattopadhyayetal07}'s analysis (done on 1594
complete observations of the BATSE 4B catalog) by performing
$k$-means clustering with $K=1,2,\ldots,20$
groups on the 1599 BATSE 4Br observations. Similar to the approach of
\citet{mukherjeeetal98}, 
\citet{chattopadhyayetal07} and other authors, and from the density
plots of Figure~\ref{fig:Bivariate}, we analyzed all parameters  in
the logarithmic scale. Further,  because these parameters  
measure different quantities, they were also individually scaled to
have the same standard deviation. To allay the effects of
initialization, for each $K$ we initialized our $k$-means algorithms
using both deterministic and stochastic methods. We first initialized
$k$-means with the results 
obtained upon performing hierarchical clustering using the
\citet{ward63} criterion and then cutting the resulting tree at $K$
groups. The algorithm was then run to convergence from these
hierarchically-obtained means. An
alternative approach ran $k$-means to convergence from each of $10Knp$
random starts, as per \citet{macqueen67} with each start simply being
$K$ (unique) randomly-chosen GRBs. The best of all these
converged $k$-means solutions -- where best means the solution providing the
smallest value of $W_K$ as per Equation~\eqref{kmeans} - is
taken to be the $k$-means solution for that $K$. The objective behind
so many initializations is to reduce the chance that our $k$-means solutions
for any $K$ have not arrived at a global optimum. We also followed
\citet{chattopadhyayetal07} in using the Jump
statistic~\citep{sugarandjames03} to decide on the optimal number of
groups. 

\begin{figure}
\mbox{ 
\subfloat[]{\label{fig:dstrn}\includegraphics[width=0.5\columnwidth]{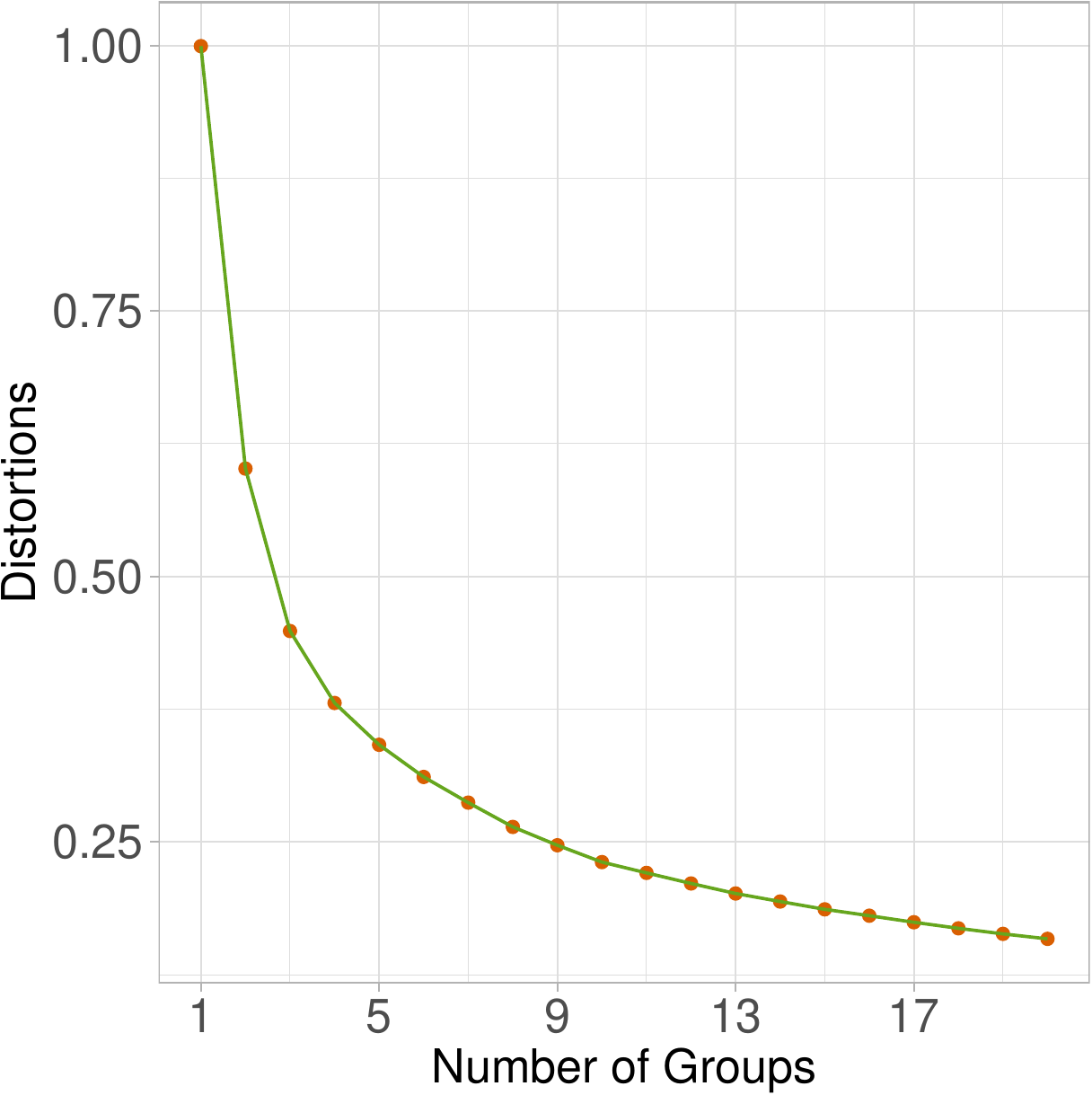}}
\subfloat[]{\label{fig:jump}\includegraphics[width=0.5\columnwidth]{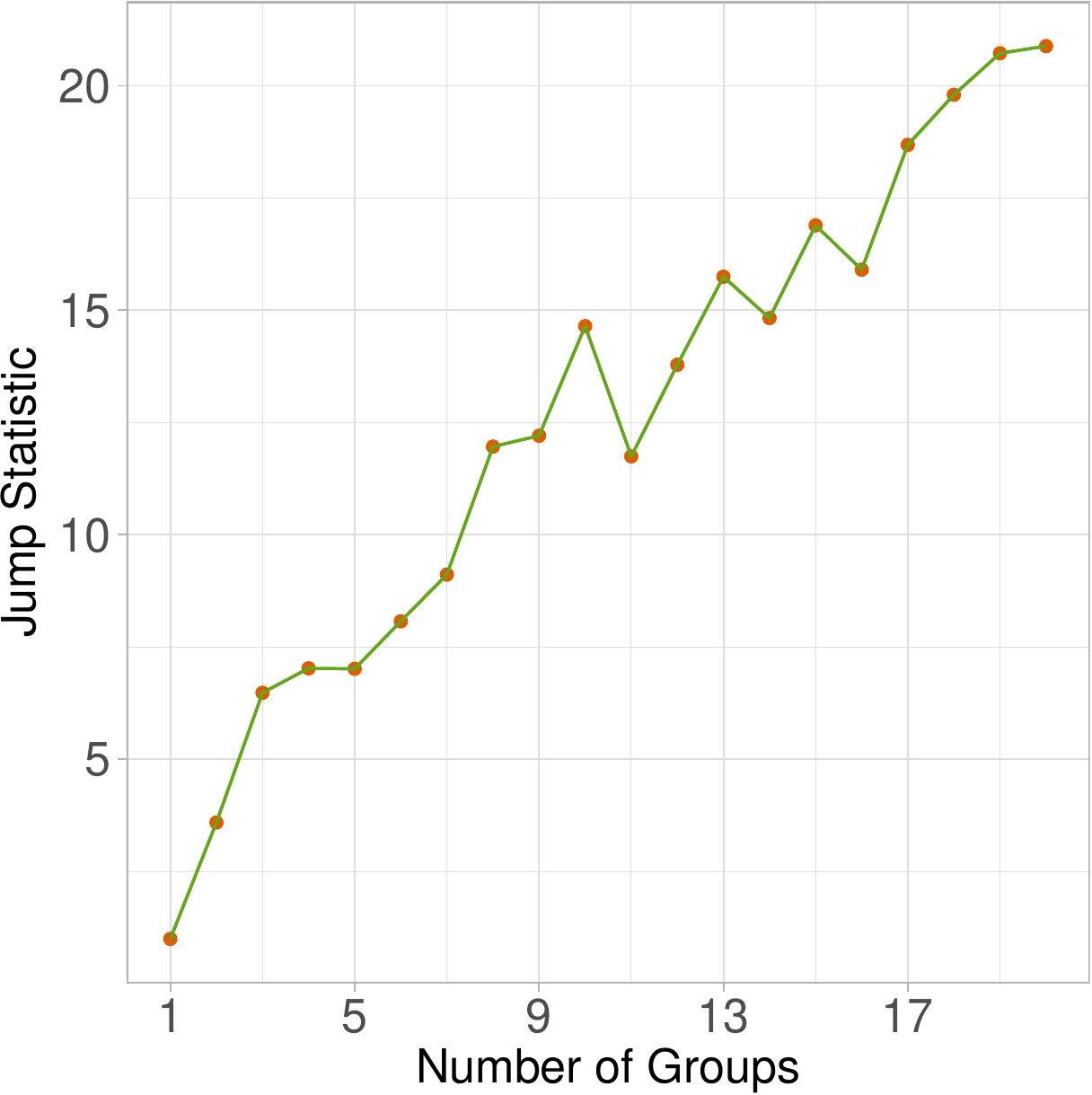}}}
\caption{(a) The distortion and  (b) jump curves for $k$-means
  clustering solutions of the GRBs.}
\end{figure}

Figures~\ref{fig:dstrn} and \ref{fig:jump} respectively display the  distortion
and the jump curves of the $k$-means solutions for
$K=1,2,\ldots.20$. There is not much leveling off of the kind found
in  \citet{chattopadhyayetal07} for either the distortion curve or the
jump statistic. Indeed, the distortion curve keeps on trending down
with decreasing $K$, while the jump statistic generally trends up with  
increasing $K$. Our results are somewhat contrary to those of
\citet{chattopadhyayetal07}. Looking back at their results, it appears
from Figure 1 of their paper that $K=4$ had the highest value of the
jump statistic, and the distortion curve did not level off even in
that paper. We also see the near-imperceptible dip in the jump curve
for $K=5$ but that is very quickly reversed for $K=6$ and beyond. We also tried
the methods of \citet{maitraetal12} to assess the 
question as to whether a larger $K'$-means clustering solution fits the data
significantly better than a smaller $K$-means one ($K'>K$) and were
able to reject the null hypothesis of no significant improvement on
fitting the larger model over the smaller model for all
$(K,K')$-pairs,  with $1\leq K < K'\leq20$. All the tests reported
negligible $p$-values which means that significantly better fits are
provided by larger $K$. This points to the possibility that
actual groups in the GRBs may have general-shaped and unequal
dispersions (spreads). Stipulating a homogeneous spherical
structure on them in this situation (as $k$-means inherently does)
leads to significantly better fits with solutions 
that model each general dispersion structure with sets
of homogeneous spherical dispersions, each centered in different regions of
the ellipsoidal-structured groups.

\begin{figure}
  \centering
  \mbox{
    \subfloat[]{\label{fig:k-violin}\includegraphics[height=0.33\textwidth,width=0.5\textwidth]{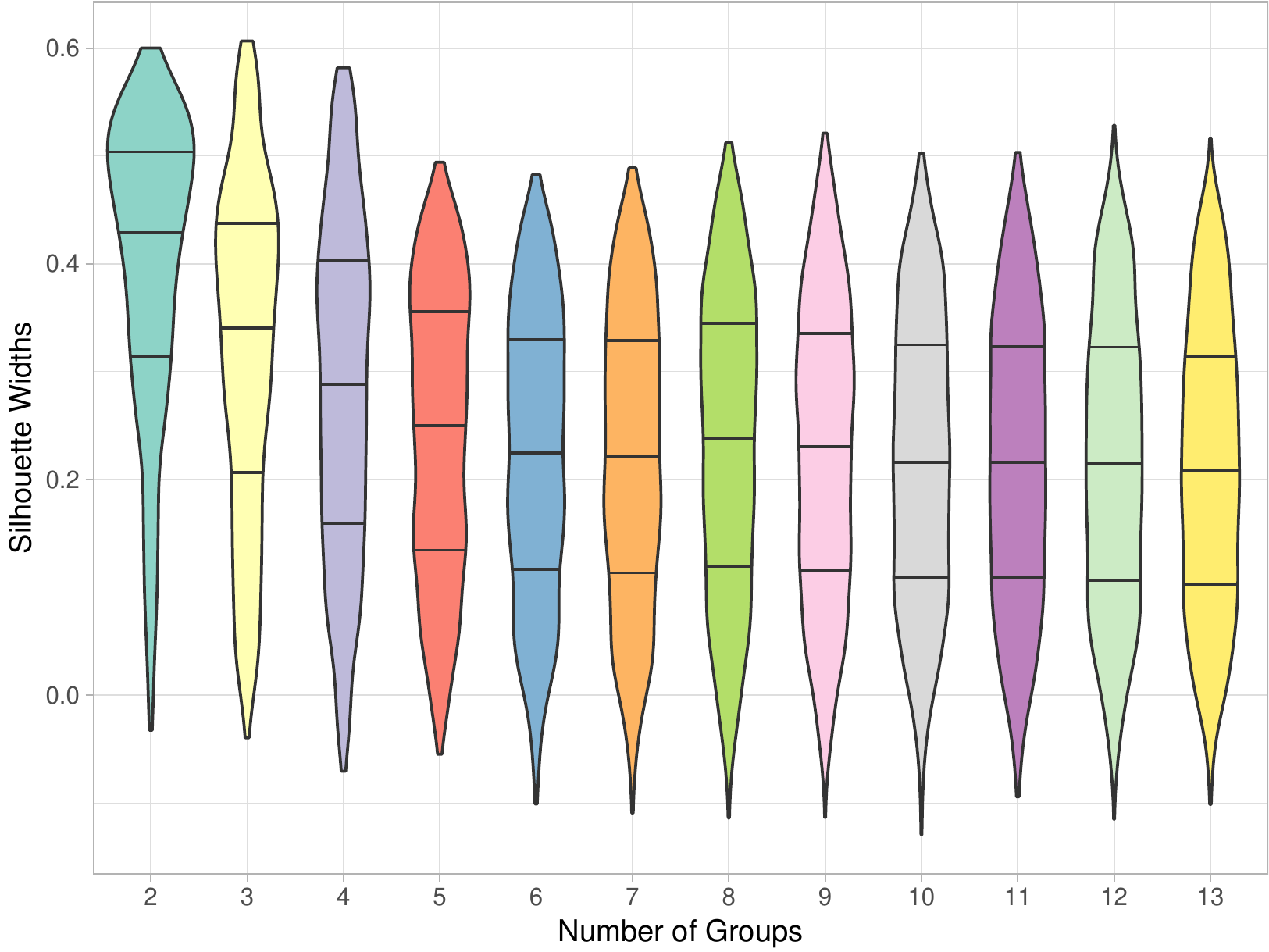}}
    }
  \mbox{
    \subfloat[]{\label{fig:k-2sil}\includegraphics[width=0.25\textwidth]{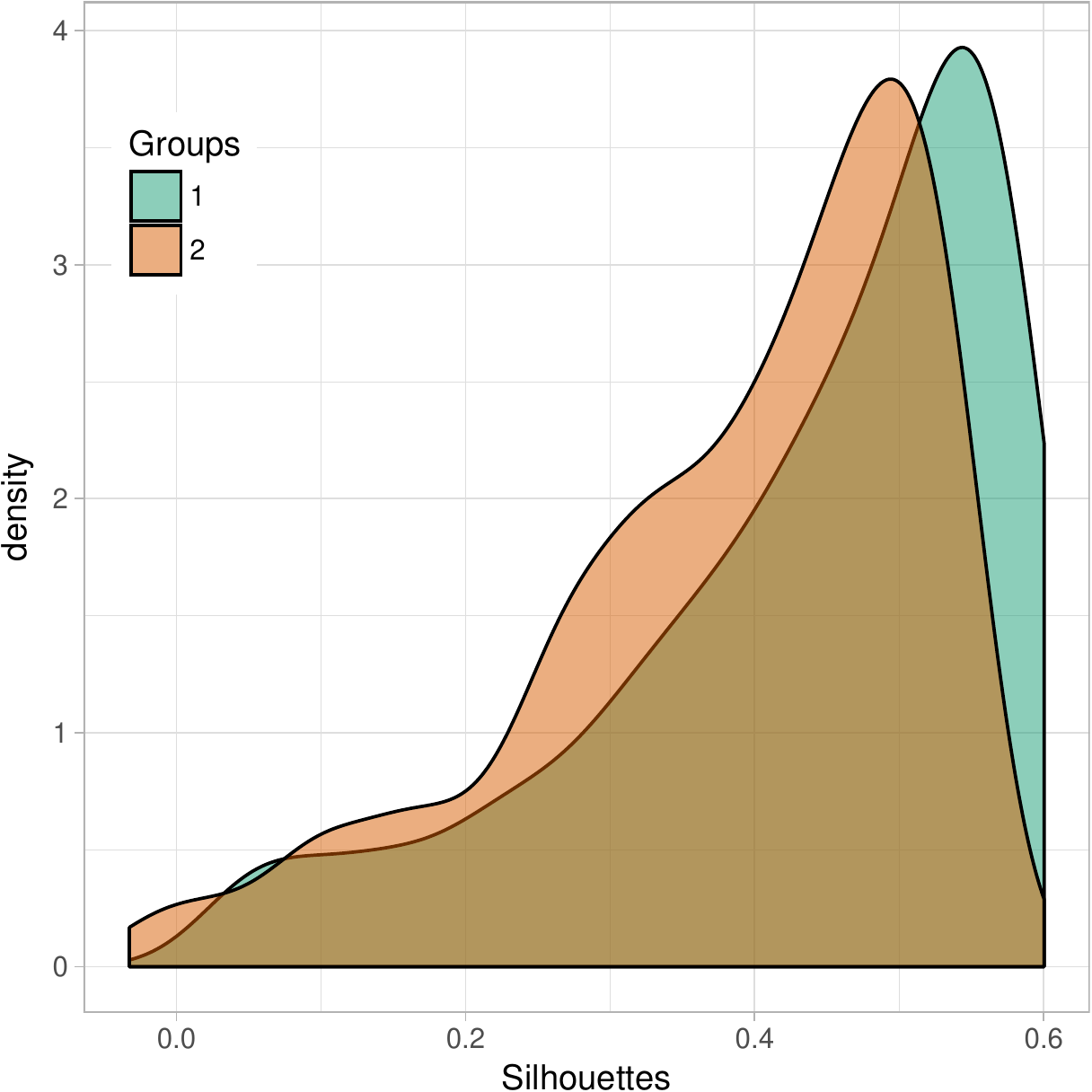}}
    \subfloat[]{\label{fig:k-2means}\includegraphics[width=0.25\textwidth]{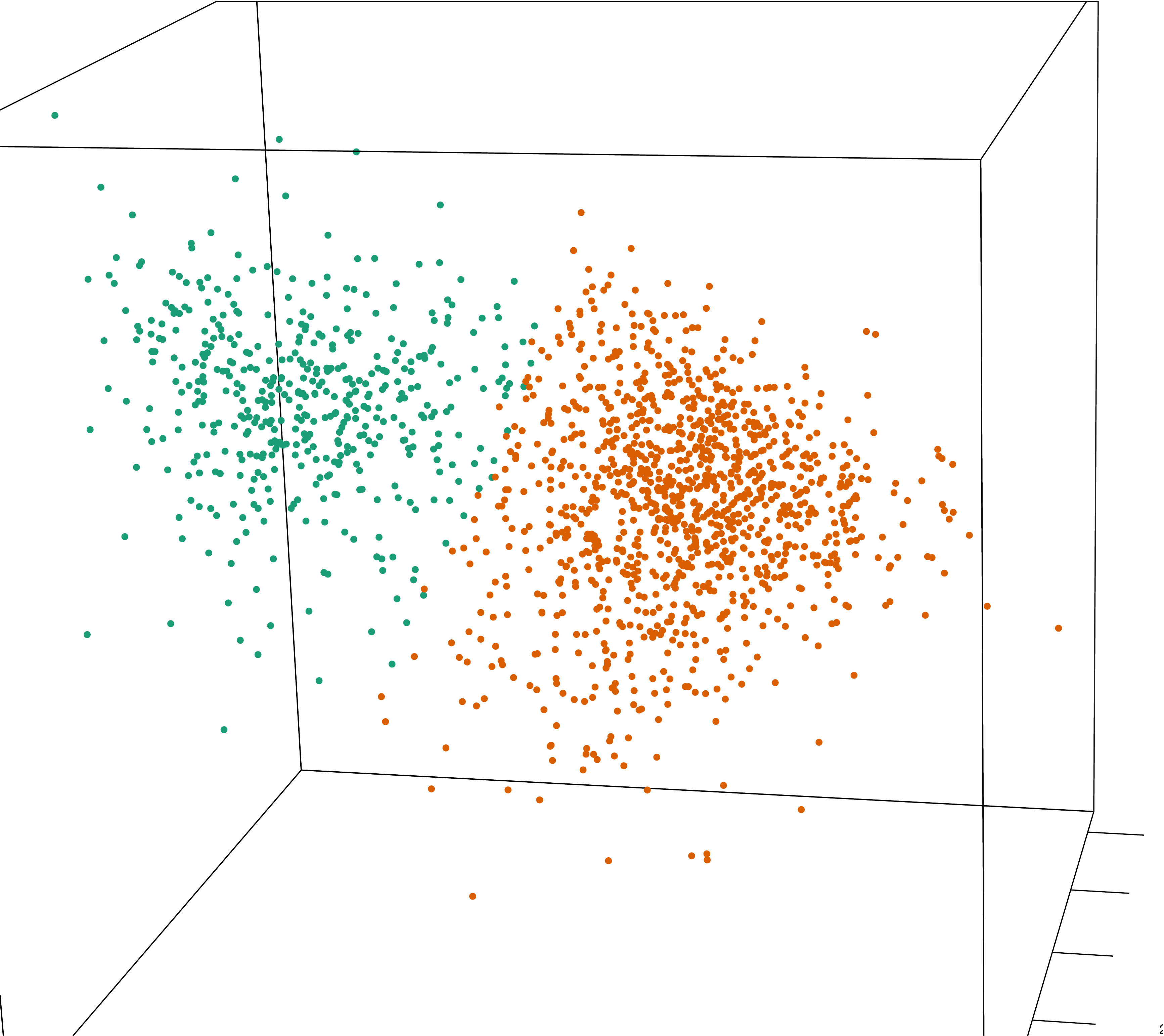}}}
  \mbox{
    \subfloat[]{\label{fig:k-3sil}\includegraphics[width=0.25\textwidth]{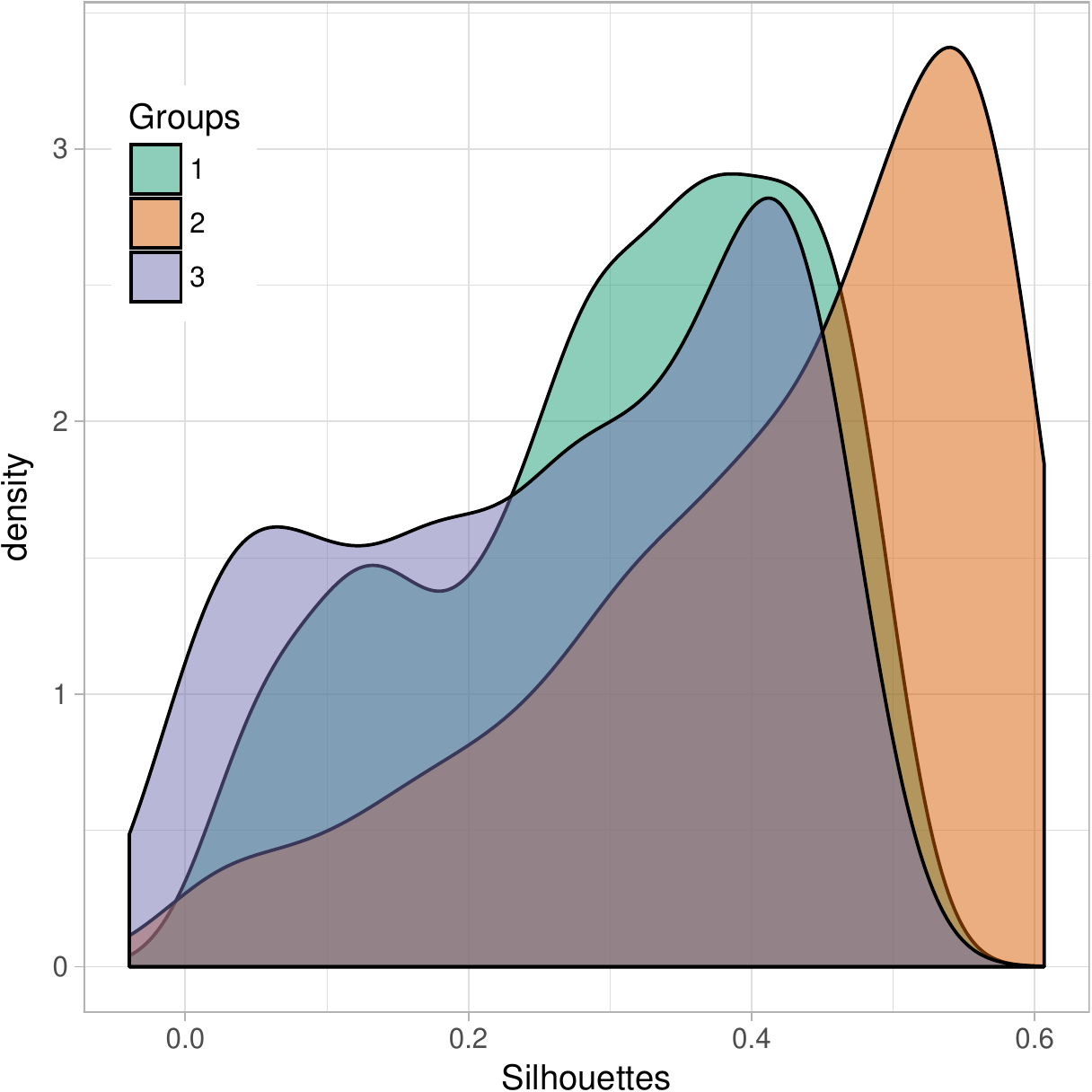}}
    \subfloat[]{\label{fig:k-3means}\includegraphics[width=0.25\textwidth]{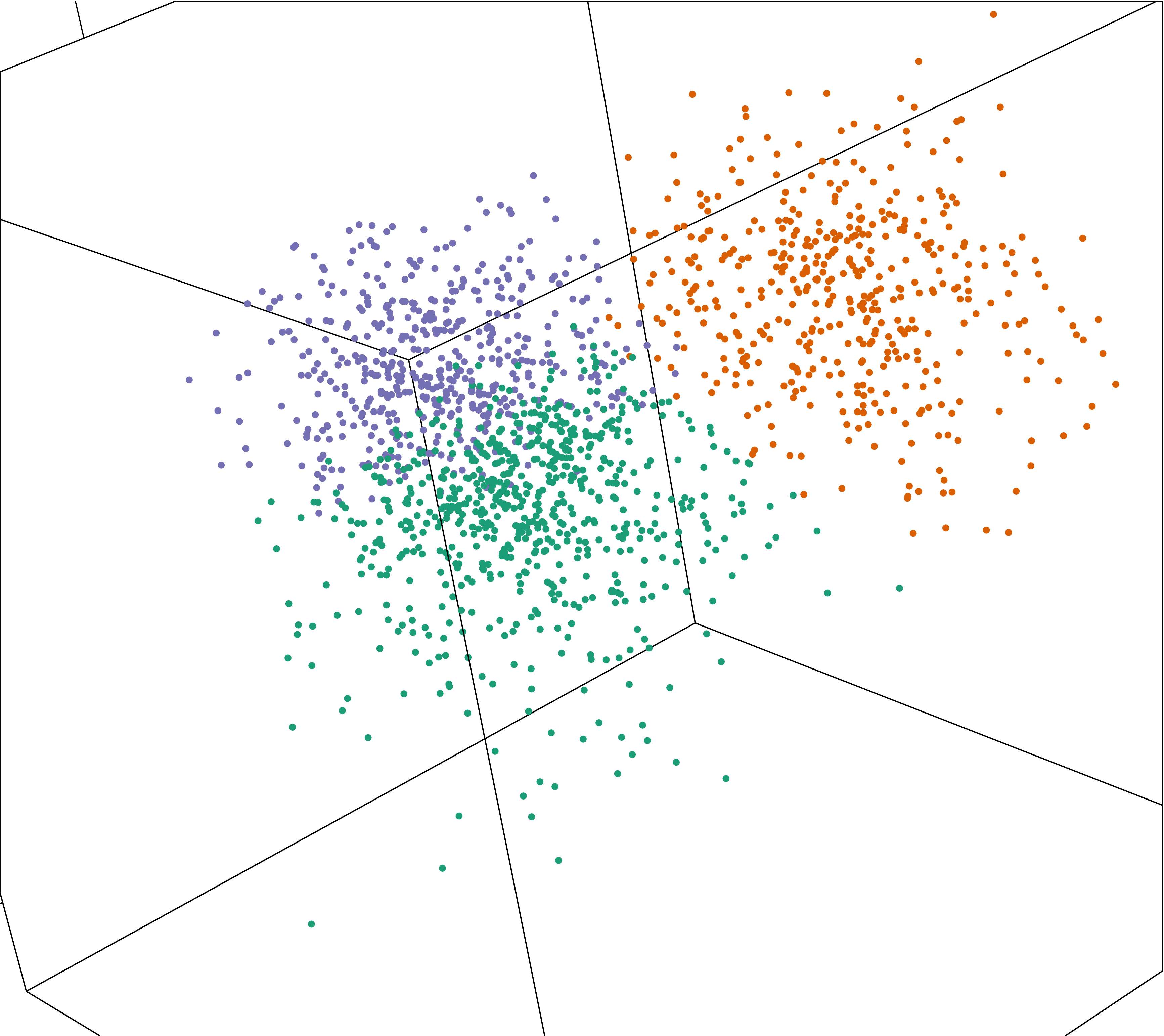}}
  }
\caption{(a) Distributions of the silhouette widths obtained for the
  $k$-means clustering of the GRB, for $2,3,\ldots,13$ groups. (b)
  Group densities of the silhouette widths and (c) three-dimensional
  projection plot of the SPCAs for the 2-means solutions.
  (d) Group densities of the silhouette widths and (e) three-dimensional
  projection plot of the SPCAs for the 3-means solutions.}
\end{figure}
We also evaluated cluster validity via the silhouette widths for each
observation and at each $K$. We display in Figure~\ref{fig:k-violin} the
individual silhouette widths through a kernel density plot (violin plot).
Like a boxplot, a violin plot~\citep{hintzeandnelson98} 
represents distributions of data through medians and the quartiles and
extrema,  but additionally displays the 
density of the data (here, the  silhouette widths ) at different  
values. For clarity, we only display the silhouette indices for
$K=2,3,\ldots,13$ but the values are fairly similar for higher
$K$. There appears to be greater validity for $K=2$ than for other
$K$, but there is considerable overlap between the  
distributions of the silhouette indices for $K=2$ and the ones for the
other $K$.

We now further investigate the groups formed by the 2- and 3-means
solutions. Figure~\ref{fig:k-2sil} displays the densities of the
silhouette widths of the GRBs assigned to each of the groups in the
2-means solution. The distribution shows that most of the silhouette
widths for both groups have moderate to moderately high values and so
have high cluster 
validity. However, there is a sizable minority of GRBs that do not
have high silhouette widths, and some even have negative values. This
finding is supported by Figure~\ref{fig:k-2means} which is a
three-dimensional display of the three best SPCA projections. The
display in Figure~\ref{fig:k-2means} presents what in our view is the
perspective showing the best separation between the groups (see the
supplementary materials for {\tt HTML} code providing the
interested reader the ability to try out other perspectives). The two 
groups are similarly-shaped and sized with somewhat distinct cores,
but there are also many observations from one group that could easily
have arisen from the other. Indeed, it would be quite difficult in the figure
to demarcate  the two groups in the absence of
color. Figure~\ref{fig:k-3sil} and Figure~\ref{fig:k-3means} show
corresponding distributions of the group silhouette widths and the
three-dimensional SPCA projections ({\tt HTML} code in the supplement)
for the 3-means clustering of the GRBs. Once again, there are small 
values for many of the silhouette widths, indicating some discomfort
with the clustering. Note also that the most separated group as per
Figure~\ref{fig:k-3means} has substantially moderately high silhouette
widths, while the other two groups are similar, mostly having moderate
values.  This illustration indicates the pitfalls with using $k$-means
clustering on the GRBs. As the number of groups increases, $k$-means prefers
breaking up the observations into smaller and smaller equi-sized
spherically dispersed groups, but as viewed by the small silhouette
widths in the Figure~\ref{fig:k-violin}, even this is not completely
adequate. Numerically, the generalized overlap for $K=2$ is 0.028
while that for $K=3$ is 0.044. Therefore, our investigations indicate
not much support for  \citet{chattopadhyayetal07}'s finding of a 
preferred $3$-means  solution for grouping GRBs. Indeed, given our
results, we do not find much appeal for  $k$-means-type clustering
solutions for 
understanding the heterogeneities in GRBs. We therefore proceed with
GMMBC for further analysis of GRBs. 
\subsubsection{Gaussian Mixture Models-based Clustering}
\label{GRB:GMM}
\citet{mukherjeeetal98} performed GMMBC on the BATSE 3B
catalog data using 
three of the six variables (log$T_{50}$, log$T_{90}$, log$P_{256}$, log$F_{t}$,
log$H_{32}$ and log$H_{321}$). %They removed  $T_{50}$ and $H_{32}$ on
                               %account of high redundancy and kept
                               %only three variables $T_{90}$, $F_{t}$
                               %and $H_{321}$ for the purpose of
%GMMBC. This was done through
Using visual inspection \citep[See Figure 1 of][]{mukherjeeetal98},
the authors identified highly redundant variables and performed GMMBC
using only  $T_{90}$, $F_{t}$ and $H_{321}$. Other  authors
\citep{horvath02,horvathetal08,zitounietal15,zhangetal16} 
  have used only $\log_{10} T_{90}$ while others~\citep[{\em e.g.}
  ][]{veresetal10,horvathetal04,horvathetal06,horvathetal10} have used  $\log_{10}  
  T_{90}$ and $\log_{10}H_{32}$ in their respective cluster analyses.
  \citet{chattopadhyayetal07} used all six variables in their
  GMMBC. Recent advances in this 
  field motivated us to reexamine the issue of redundancy 
  among these six variables in the context of clustering and using the
  formal  GMMBC-based variable 
  selection methods discussed in Section \ref{r:d}. We used   
  {\tt clustvarsel}  \citep{clustvarselR} to perform GMMBC
variable selection  using  $\log_{10}T_{50}$, $\log_{10}T_{90}$, 
$\log_{10}P_{256}$, $\log_{10}F_{t}$, $\log_{10}H_{32}$ and
$\log_{10}H_{321}$. The results of the forward- and-backward-selection
variable selection algorithm are presented in 
Table~\ref{tab:1}
\begin{table}
\caption{Results of the forward- and backward-variable selection step
  for determining redundancy of  $\log_{10}T_{90}$, $\log_{10}T_{50}$,
  $\log_{10}H_{321}$, $\log_{10}H_{32}$, $\log_{10}P_{256}$,
  $\log_{10}F_{t}$ in GMMBC.}
\centering
\begin{tabular}{rlrcr}
\hline\hline
Step & Variable  & Step Type & BIC Difference & Decision\\
\hline
1 & $\log_{10}T_{90}$ & Add & 452.95 & Accepted \\
2 & $\log_{10}T_{50}$ & Add & 395.74 & Accepted \\
3 & $\log_{10}H_{321}$ & Add & 176.59 & Accepted \\
4 & $\log_{10}H_{321}$ & Remove & 176.95 & Rejected \\
5 & $\log_{10}H_{32}$ & Add & 443.06 & Accepted \\
6 & $\log_{10}T_{50}$ & Remove & 273.56 & Rejected \\
7 & $\log_{10}P_{256}$ & Add & 260.28 & Accepted \\
8 & $\log_{10}T_{50}$ & Remove & 235.61 & Rejected \\
9 & $\log_{10}F_{t}$ & Add & 185.52 & Accepted \\
10 & $\log_{10}T_{50}$ & Remove & 194.60 & Rejected \\
%11 & $\log_{10}T_{50}$ & Remove & 194.599 & Rejected \\ [1ex]
\hline
\end{tabular}
%Notes: According to the model based done through clustvarsel package in R the selected subset of variables are $\log_{10}T_{90}$, $\log_{10}T_{50}$, $\log_{10}H_{321}$, $\log_{10}H_{32}$, $\log_{10}P_{256}$, $\log_{10}F_{t}$.
\label{tab:1}
\end{table}
and indicate scant support for the theory of redundancy among the six
variables for clustering: we therefore proceed, using all of them
(in the logarithmic scale) in our GMMBC. 

\begin{figure}
\includegraphics[width=0.45\textwidth]{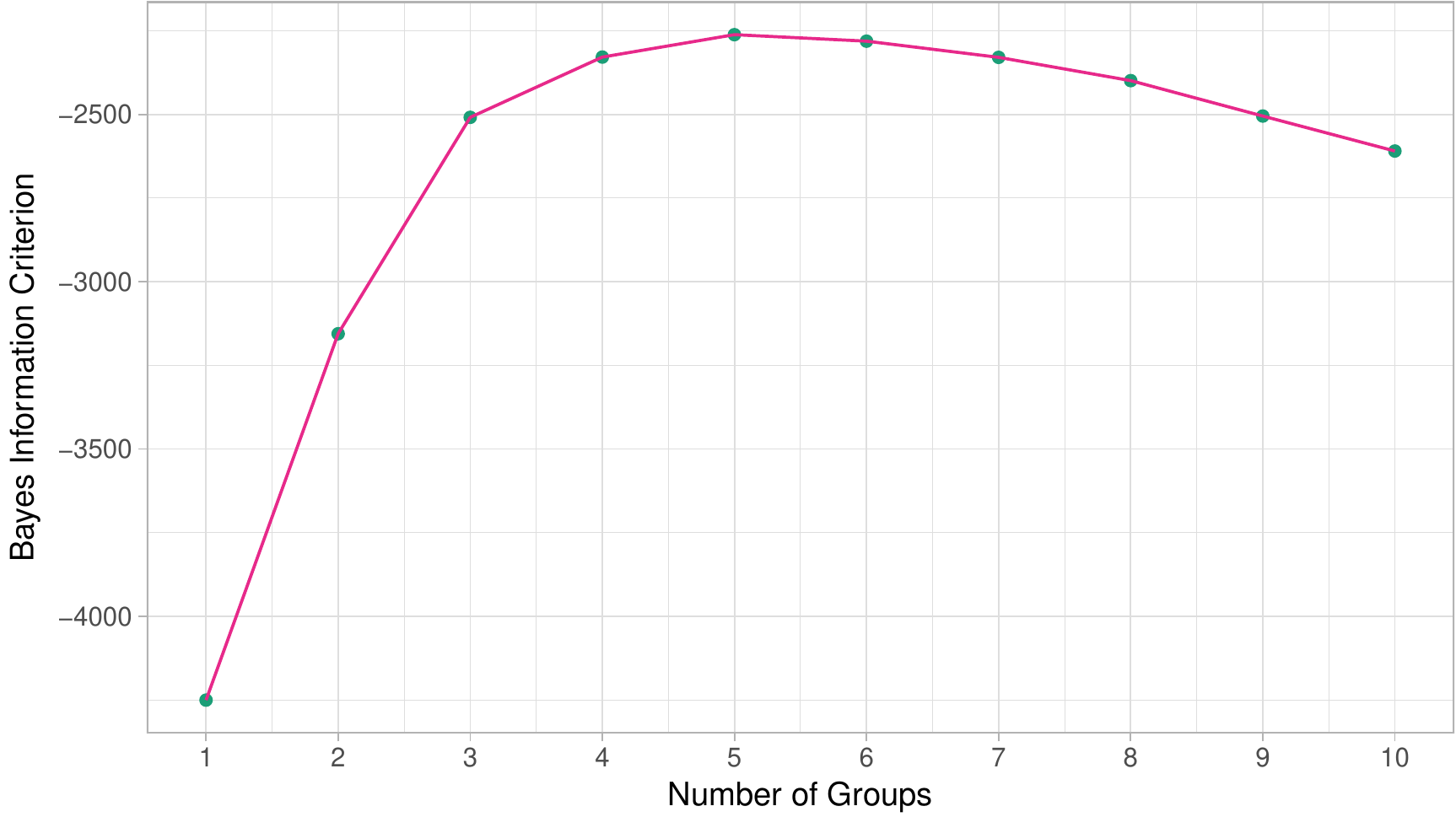}
\caption{Plot of BIC with  $K$ upon performing GMMBC of the 1599 GRBs
  in the BATSE 4Br catalog.} 
\label{fig:BIC}
\end{figure}

\citet{chattopadhyayetal07} performed GMMBC but with a Dirichlet
Process prior to 
decide on the number of components. However, there is not much
software matching the modeling flexibility of {\tt mclust}~\citep{fraleyetal12}
or the enhanced initialization and fast computational approaches of
{\tt
  EMCluster}~\citep{Chen2015EMClusterpackage,Chen2015EMClustervignette}. We
therefore  use these packages for GMMBC of the GRBs for each
$K=1,2,\ldots,9$ and chose for each $K$ the solution with the highest
loglikelihood. The BIC was also calculated for each $K$: these are
displayed in Figure~\ref{fig:BIC}. 
\begin{comment}
We are motivated to carry out
model based clustering using all of the six variables to ascertain how
partitioning occurs when all of the six variables are used. To perform
model based clustering using these six variables we have used the
EMCluster package in R. EMCluster provides several effective
initialization methods for model based clustering of finite mixture
Gaussian distribution with unstructured dispersion in both of
unsupervised and semi-supervised clustering.(See
\citet{Chen2015EMClusterpackage} and \citet{Chen2015EMClustervignette}
for a complete documentation on EMCluster). To select the best model
we have used BIC. Figure 3 shows the BIC for different values of
$K$.\\
\end{comment}
From that figure, it is clear that amongst all GMMs, a mixture of five Gaussian
densities provides the best fit to the GRB dataset. Thus, there is
evidence of five kinds of GRBs in the BATSE catalog. This finding is
at variance with studies published in the literature which have
identified at most three distinct  groups. \citet{chattopadhyayetal07}
used all six variables and the BATSE 4B GRBs and found three
groups, but most of the other classifications used only the duration
variables. (We note that our results here used the BATSE
4Br catalog, but we also separately performed the analysis on the
older BATSE 4B dataset with 1594 complete observations and obtained
a similar five-groups solution. This provides greater confidence in
our findings.) 

\paragraph{Validity of the GMMBC Solution:} 
\label{validity}
The distance-based silhouette widths or the SPCA displays can not be
calculated for results obtained using GMMBC with general dispersion matrices. So
we only discuss the overlap measures between the different groups
and the generalized overlap measures for $K$-components-fitted GMMBC
solutions ($K=2,3,4,5$) as reported in
\begin{figure}
  \centering
  \mbox{
    \hspace{-0.045\textwidth}
    \subfloat[]{\label{tab:overlap}
      \begin{tabular}{cc}
        \hline\hline
        $K$&$\ddot\omega$\\ \hline
        2 & 0.17 \\%0.169\\
        3 &0.12 \\ %0.122 \\
        4 & 0.11\\ %0.107\\
        5&0.10\\ %0.097\\
%        6&0.09\\%0.0944674
%        7&0.08\\%0.082855
%        8&0.08\\%0.08496672
%        9&0.07\\%0.07133114
        \hline\hline
      \end{tabular}
    }
    \centering
    \subfloat[]{\label{fig:overlap}
      \raisebox{-.5\height}
               {\includegraphics[width=0.37\textwidth]{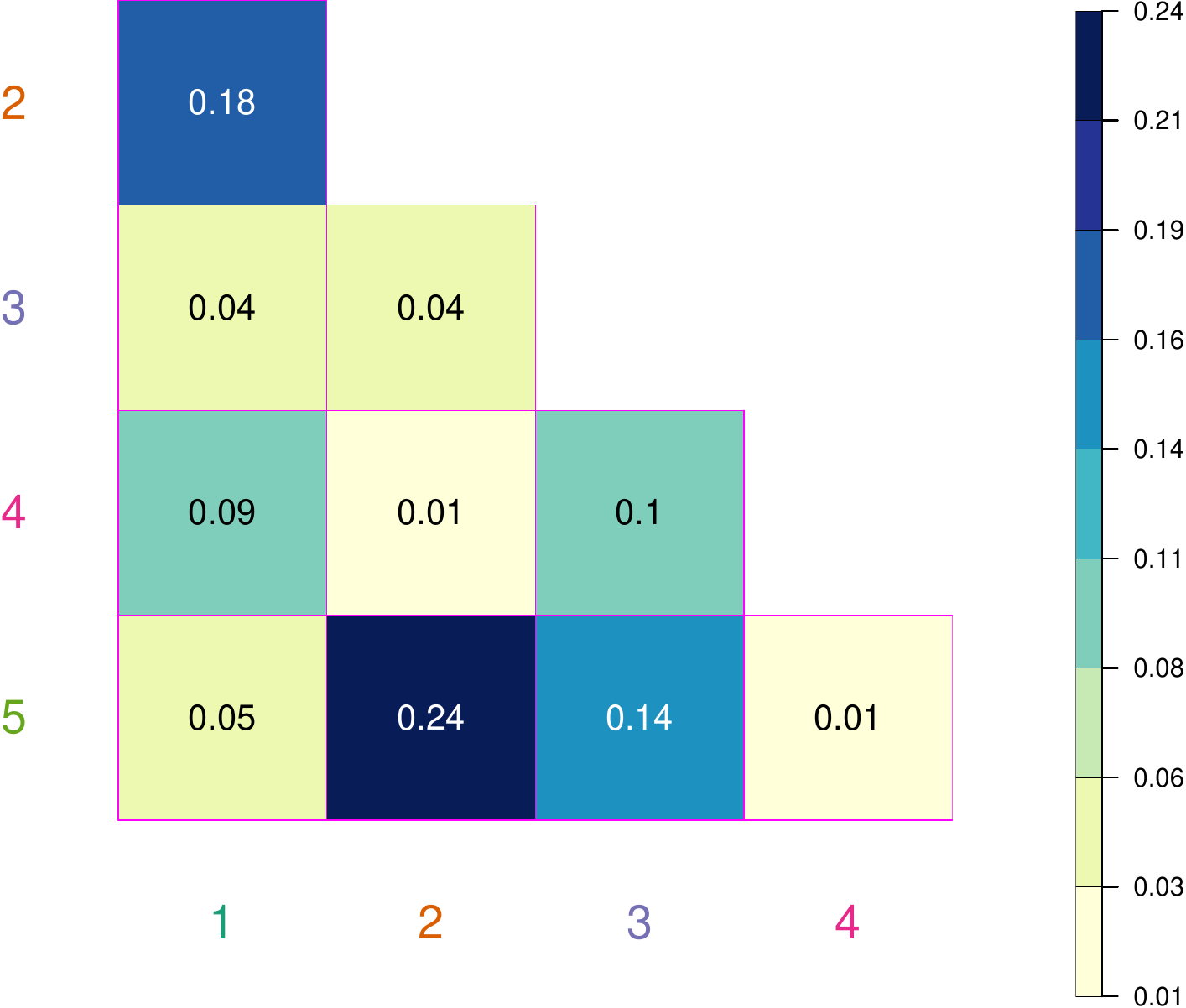}}    }
    %  }
    %  \mbox{
    \centering
  }
  \caption{(a) Generalized overlap ($\ddot\omega$) measures for the
    2-, 3-, 4- and   5-components GMMBC solutions of the 1599  BATSE 4Br GRBs.
    (b) Pairwise overlap measures between the $k$ and the 
    $l$th groups in the 5-component GMMBC solution as 
    indicated by the margins. Color in the margins correspond
    to the group indicator that is the same for all displays and tabulations 
    involving the five-component GMMBC fits to the dataset with six parameters.}
\end{figure}
the table in
Figure~\ref{tab:overlap} where the $K=5$ model marginally presents the most
separated components over $K=4$, with both solutions providing more
distinct components than the GMMs with  $K=2$ or 3 components. We note
that the generalized (and also our pairwise) overlap measures
are based on the population mixture model components with parameters
estimated from the data. Thus, the values are calculated under the
assumption of the GMM that provides the best fit  to the data.  

The BIC and the generalized overlap measures provide additional
indication on the presence of five distinct kinds of GRBs. We now
comprehensively evaluate the five classes of GRBs that we have
obtained from our analysis.

\paragraph{Analysis of  Results}
\label{GRB:6analysis}
Figure~\ref{fig:overlap} displays the pair-wise overlap measures
between the five groups in the 
five-component GMMBC fit to the data. The pairwise overlap measures
indicate that the fourth group is the most distinct from all the
others while the fifth group has substantial overlap with the second
and third groups. The second and third groups are
fairly distinct from each other however,  so one may consider
describing these groups themselves as a mixture of
three-components~\citep{baudryetal10}, but 
descriptions and characterizations of such merged groups are harder
and less interpretable. 
Figure~\ref{fig:overlap} also indicates distinctiveness between the
first and the third  groups, and (to a lesser extent) between the
first and the fifth 
groups.

\begin{figure}
   \includegraphics[width=0.5\textwidth]{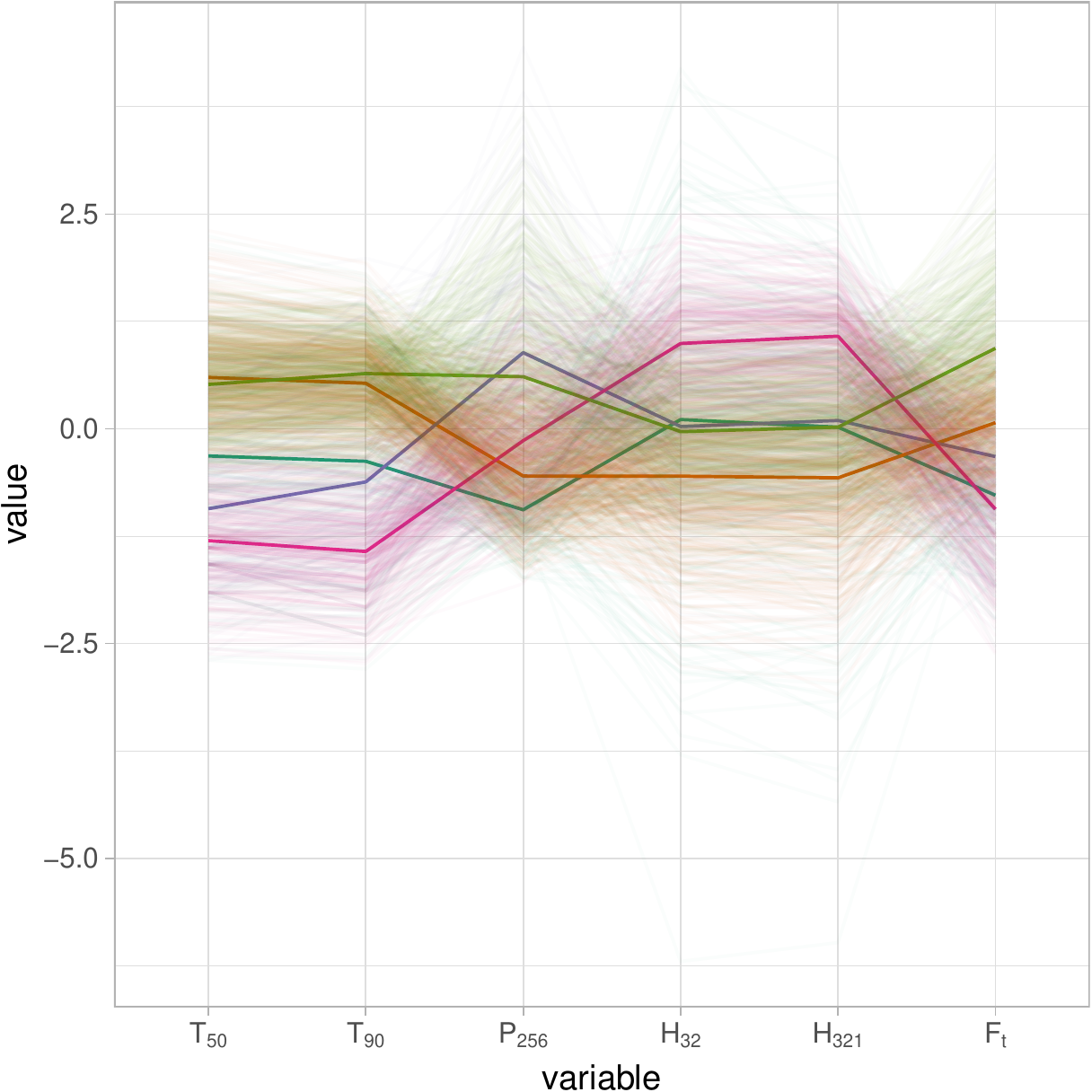}
   \caption{Parallel coordinate plot of the 1599 BATSE 4Br GRBs colored as per
     their group indicators. The  solid lines represent the group medians for each of  the six variables displayed. Variables are in the logarithmic scale.}
\label{pcp.6var.plot}
\end{figure}

Table~\ref{tab:nks} tabulates the number of observations in each of
the five groups. (The color for each group indicator in the table corresponds to
the identities for each group and, for easy reference, hold for  all
displays and tabulations that refer to the 5-component GMMBC 
clusterings.) Clearly, the second and fifth groups have the most
GRBs while the third group has the fewest GRBs. 
Table~\ref{tab:means} provides the five group means from the GMMBC of
GRBs. We also display {\em parallel coordinate plots} of
all the observations with lines colored as per
their classifications~(Figure \ref{pcp.6var.plot}). A parallel
coordinate plot~\citep{inselberg85,wegman90} is an effective way to
visualize data containing multiple dimensions, where lines link the
observation value for each coordinate. The 
coordinates themselves are displayed vertically on the same scale and are
equi-spaced. These coordinate axes are called the parallel axes and a
point in the $p$-dimensional space is represented as a 
polyline with vertices on these parallel axes. The position of the
\textit{i}th axis corresponds to the \textit{i}th coordinate of the
point~\citep[see][]{inselberg85,wegman90}. 
We use the parallel coordinates plot in Figure \ref{pcp.6var.plot} to
visually inspect the five groups and draw conclusions. 
   \begin{table}
   \caption{(a) Number of GRBs and (b) Means of the six parameter
     values in each of the five groups identified by GMMBC.} 
      \mbox{\subfloat[Number of observations in each group]{\label{tab:nks}{
            \centering
            \begin{tabular}{rrrrrr} \hline\hline
              Group & {\color{Gr1} 1} & {\color{Gr2} 2} &  {\color{Gr3} 3} & {\color{Gr4} 4} &  {\color{Gr5} 5} \\ \hline
           Number of observations & 174 & 551 & 149 & 292 & 433 \\ \hline
         \end{tabular}
          }
     }
   }
      \mbox{\subfloat[Mean parameter values for each group]{\label{tab:means}{
         \begin{tabular}{rrrrrrr}           \hline\hline
           $k$ &$\log T_{50}$&$\log T_{90}$&$\log P_{256}$&$\log H_{32}$&$\log
           H_{321}$&$\log F_{t}$\\ \hline
           \color{Gr1} 1&0.337&0.703&-0.150&0.536&0.232&-6.074\\
           \color{Gr2} 2&1.142&1.519&0.058&0.372&0.101&-5.405\\
           \color{Gr3} 3&-0.234&0.547&0.697&0.545&0.314&-5.594\\
           \color{Gr4} 4&-0.657&-0.312&0.240&0.795&0.617&-6.159\\
           \color{Gr5} 5&1.032&1.557&0.519&0.511&0.274&-4.838\\
           \hline
         \end{tabular}
       }
     }
   }
   \end{table}
\begin{figure*}
\mbox{ 
\subfloat[Group 1]{\label{fig:cor1}\includegraphics[width=0.333\textwidth]{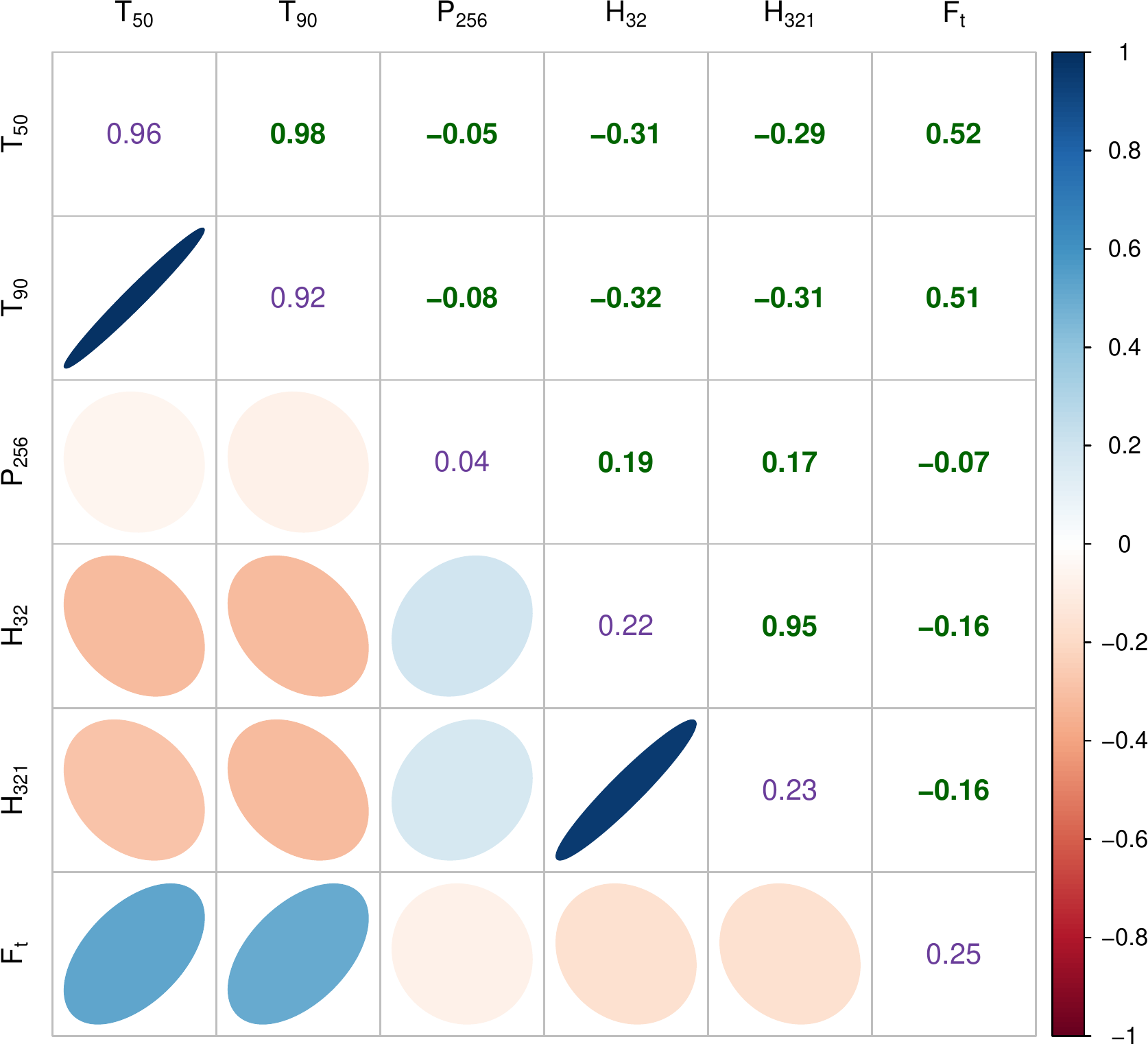}}
\subfloat[Group 2]{\label{fig:cor2}\includegraphics[width=0.333\textwidth]{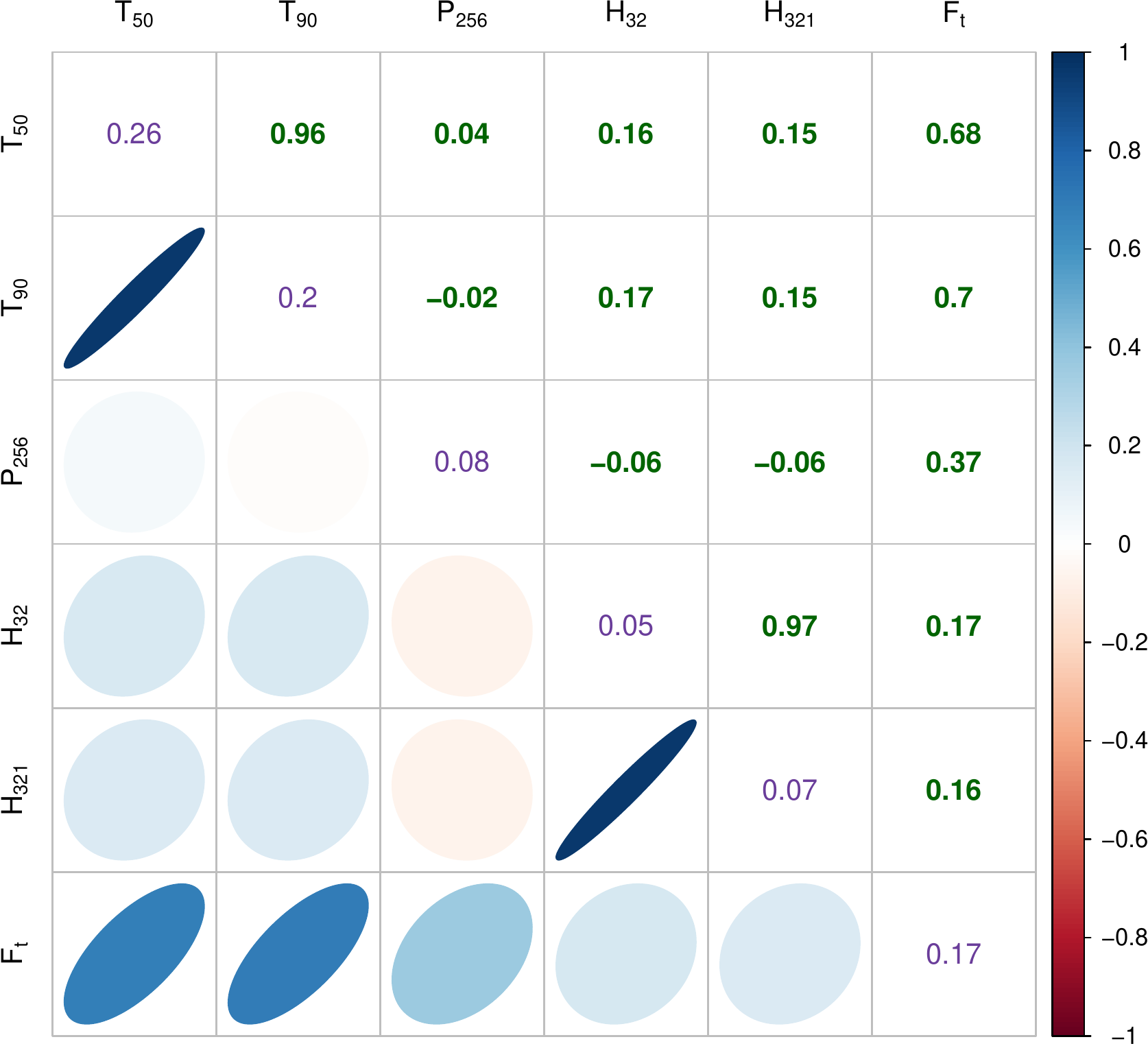}}
\subfloat[Group 3]{\label{fig:cor3}\includegraphics[width=0.333\textwidth]{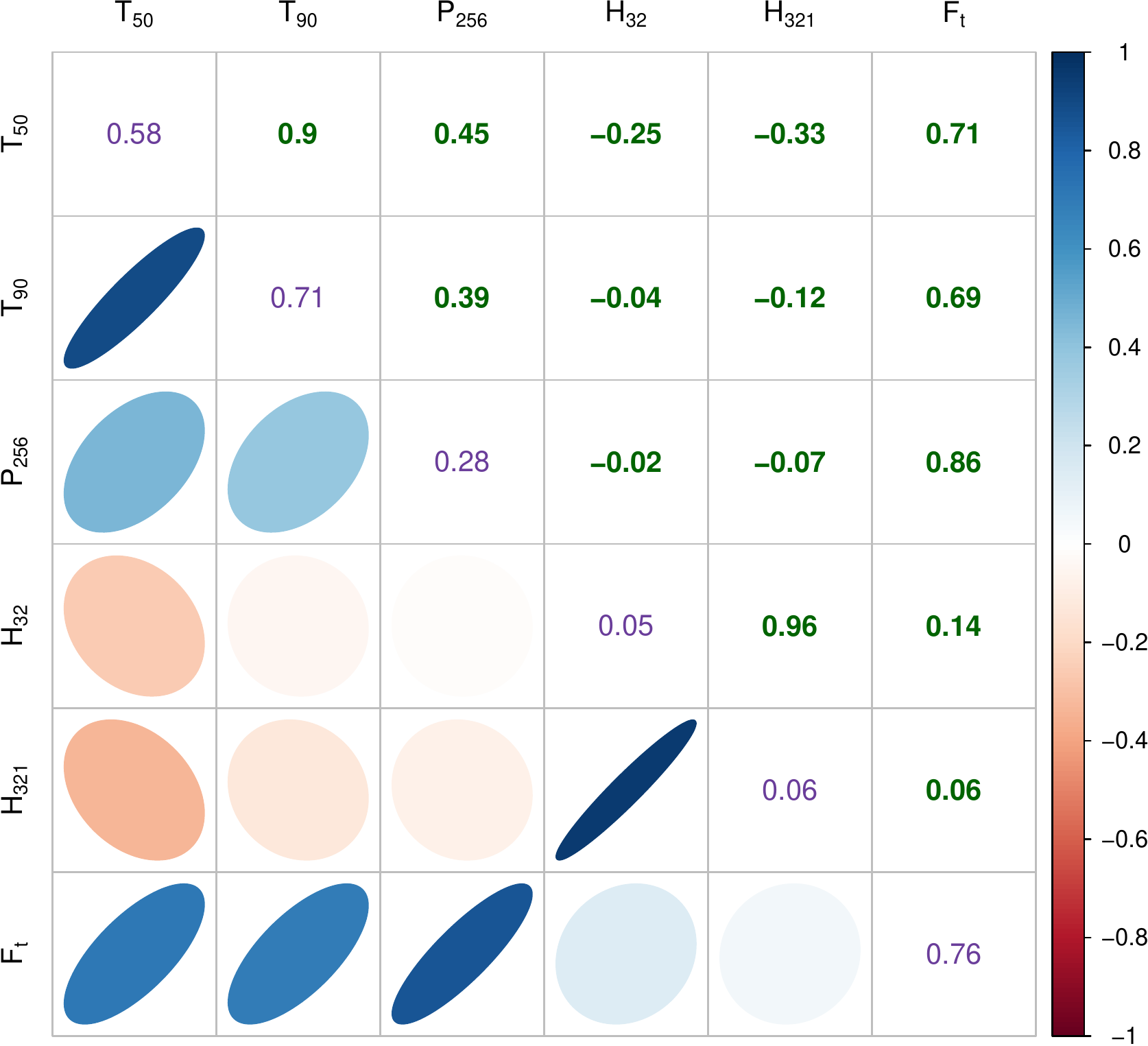}}}
\mbox{
\subfloat[Group 4]{\label{fig:cor4}\includegraphics[width=0.333\textwidth]{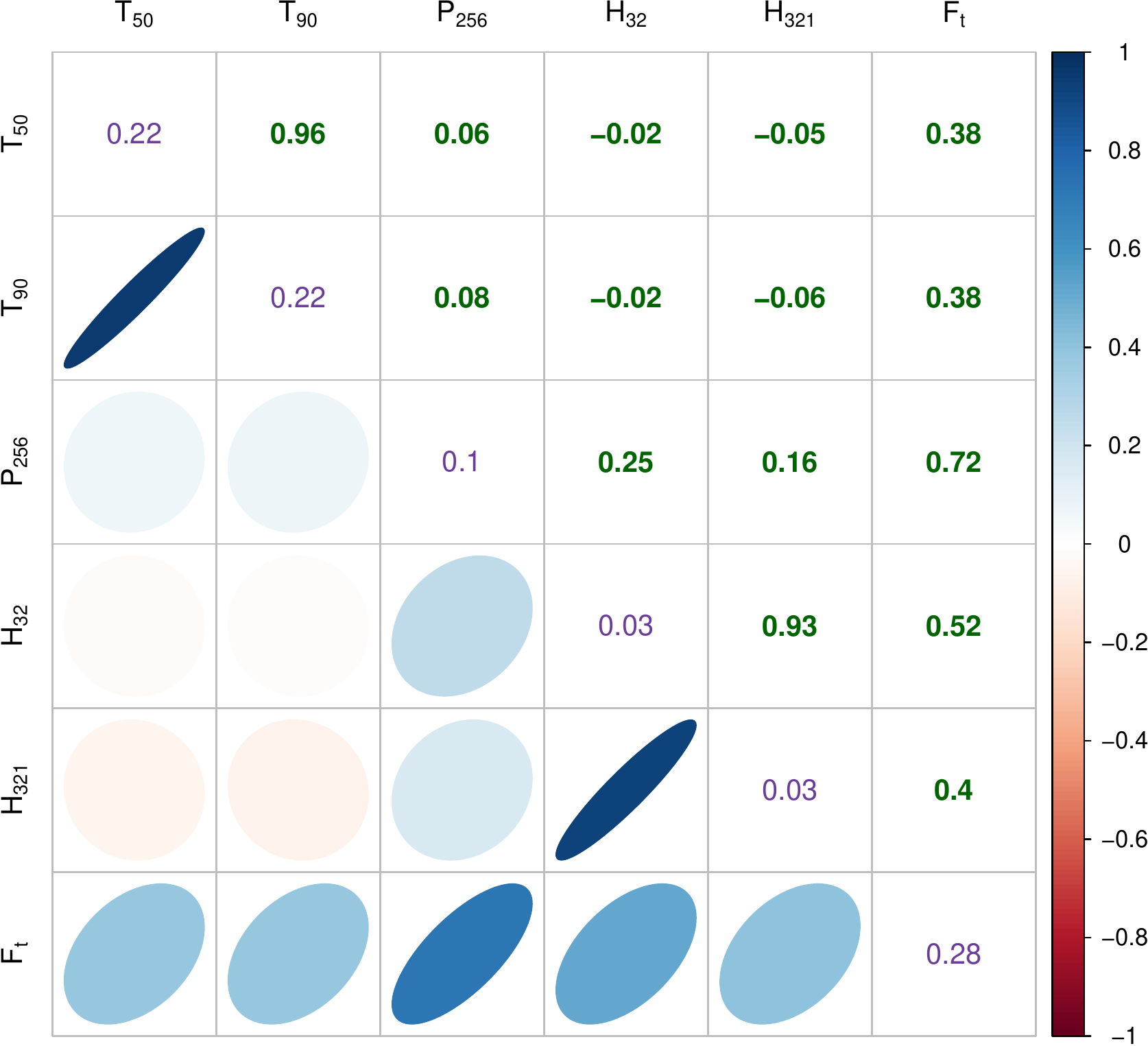}}
\subfloat
    [Group 5]{\label{fig:cor5}\includegraphics[width=0.333\textwidth]{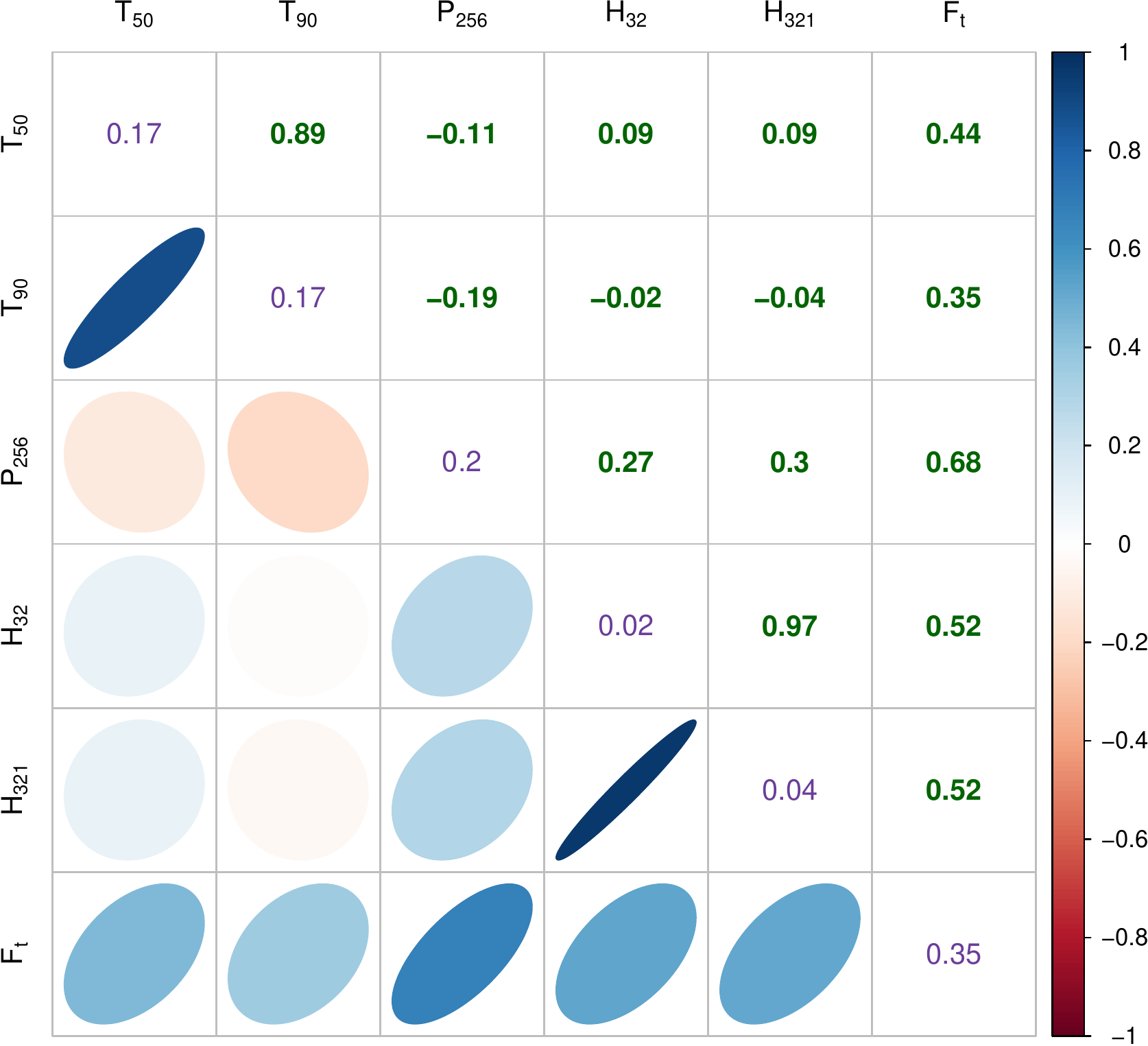}}}
\caption{Variances and displays of the estimated correlations for each
  of the five   groups obtained from the 5-component GMMBC solution of
  the 1599 GRBs. For each group, the off-diagonal elements
  display correlation between the variables while the diagonals
  display the variances. Both correlations and variances are
  calculated for the variables in the base-10 logarithmic scale.}
  %  Here $T_{50}=log_{10}T_{50}$,  $T_{90}=log_{10}T_{90}$,
  %  $P_{256}=log_{10}P_{256}$, $F_{t}=log_{10}F_{t}$},
  %  $H_{32}=log_{10}H_{32}$, $H_{321}=log_{10}H_{321}$.
  \label{corrplot.6var}
\end{figure*}

To assess the properties of the five groups optimally found by GMMBC
and BIC, we study 
the duration variable 
 $\log_{10}T_{90}$ and the fluence variable $\log_{10}F_{t}$. These
 two variables were used by
 \citet{chattopadhyayetal07} and \citet{mukherjeeetal98} to
 understand properties of groups obtained by them.  We  also
 consider interpretation of our results 
 using these two variables to  facilitate easy comparison with the
 findings of  \citet{chattopadhyayetal07} and
 \citet{mukherjeeetal98}. Our fourth group (which is also the most
 distinct as per the pairwise overlap)  consists of bursts of the
 shortest duration (about 0.5s) while 
 burst durations of the first and third groups are around
5s and 3s, respectively. The fluences are also the highest for the second
and fifth groups, both also having the longest durations of
bursts. The fourth group (with shortest-duration bursts) has 
the lowest fluence. As mentioned in Section \ref{sec:introduction},
the popular classification scheme classifies GRBs
 as short bursts ($T_{90} < 2 s$) and long bursts ($T_{90} >
 2s$). Following this framework,  bursts of the fourth group can be
 designated  as short duration 
 bursts.~\citet{chattopadhyayetal07} further classified bursts with
 $T_{90}>2s$ into two groups, the long-duration low-fluence bursts and
 the long-duration high-fluence bursts. Following their rationale, our
 first and third groups can be designated as having long-duration bursts
 with low fluence while our second and fifth groups can be categorized
 as  those with long  duration bursts with fluence higher than that of
 the first and third groups. It is also of interest to compare the
 GMMBC results of  \citet{mukherjeeetal98} on the smaller complete
 BATSE 3B GRBs with 
 our groupings. Indeed, the three groups obtained by
 \citet{mukherjeeetal98} have a good amount of similarity (in terms of
 the group means) with three of our groups. For example, the group mean of
 $\log_{10}T_{90}$ obtained by \citet{mukherjeeetal98} for their three
 groups were around $1.55$, $-0.42$ and $0.71$ which are quite similar
 to the means of our first, fourth and fifth groups. Actually, our
 second and fifth  groups  are quite similar for $\log_{10}T_{90}$
 to the mean of the first group of
 \citet{mukherjeeetal98}. 
 \begin{comment}
 Also, the fourth group obtained by us resembles
 closely (in terms of $\log_{10}T_{90}$) to their second group while.
 \end{comment}
 Our first group also shows
 considerable similarity to the third group (again in terms of
 $\log_{10}T_{90}$ ) obtained by \citet{mukherjeeetal98}. The other 
 variables also exhibit similar resemblance with the results
 of \citet{mukherjeeetal98}.

% Unlike \citet{chattopadhyayetal07}
 \citet{mukherjeeetal98} described group properties using the hardness
 ratios in addition to  duration and total fluence. Thus, they 
 classified their three groups using the three properties
 Duration/Fluence/Spectrum. Using this rule the three groups obtained
 by them were summarized as long/bright/intermediate, short/faint/hard
 and intermediate/intermediate/soft. Adopting this rule, we find that
 our  five groups can be classified in terms of 
 intermediate/faint/intermediate,
 long/intermediate/soft, intermediate/intermediate/intermediate, 
 short/faint/hard and long/bright/intermediate.  

In addition to studying the means for understanding group properties, we
also calculated the correlations between the six variables in each of the 
five classes.
 Figure \ref{corrplot.6var} displays the correlation structures for
 the five GMMBC groups. The diagonals of each display
 provide the estimated group variances of the six variables obtained
 by GMMBC. The $(i, j)$th cell in the upper off-diagonal part 
of the display provides the numerical value of the correlation
coefficient between the $i$th and $j$th variables 
while the lower off-diagonal $(j, i)$th cell  provides a diagrammatic
representation of the same correlation coefficients. The extent of
linear relationship between 
any two variables in each of the clusters can then be easily
represented  using these visual representations.
The relationships between the six variables for each of the five
groups can also be understood by looking at correlation displays of
Figure \ref{corrplot.6var}. For each of the five groups, the two
duration variables $\log_{10}T_{50}$ and $\log_{10}T_{90}$ are very
strongly correlated as are the hardness ratios $\log_{10}H_{32}$ and
$\log_{10}H_{321}$. On the other hand,
$\log_{10}T_{90}$ and $\log_{10}H_{321}$ exhibit moderately negative
association in the first and third groups while they are weakly
(negatively) correlated in the fourth and the fifth
groups and moderately positively correlated in the second group.
The duration $\log_{10}T_{90}$ 
and fluence $\log_{10}F_t$ exhibit moderate positive association in
the fourth and fifth groups and strong positive association in the
other three groups. There is weak negative association between
$\log_{10}P_{256}$ and  $\log_{10}F_t$ in the first group, a
moderately positive association in the second group and strong
positive correlation between these two parameters in the other groups.
In summary, our identified
groups have similar properties in the common variables as the groups
in \citet{mukherjeeetal98}, but we also identify additional structure by
including the additional variables ignored by them but declared
relevant by variable selection.
\begin{comment}
\paragraph{Additional Comments}
\label{comments}
A reviewer suggested  dropping $\log_{10}P_{256}$ from our
analysis. GMMBC of the 1599 BATSE 4Br GRBS using  $\log_{10}T_{90}$, 
$\log_{10}T_{50}$, $\log_{10}H_{321}$, $\log_{10}H_{32}$,
$\log_{10}F_{t}$ also yielded five groups as per BIC. These groups
were, however, less distinct (with higher overlap measures) than those 
obtained using $\log_{10}T_{90}$,
$\log_{10}T_{50}$, $\log_{10}H_{321}$, $\log_{10}H_{32}$,
$\log_{10}F_{t}$ and $\log_{10}P_{256}$, which is understandable given
that Table~\ref{tab:1} shows that that each of the six variables carry  
necessary and relevant information for clustering beyond that
provided by the other five variables.

\end{comment}

\subsection{Clustering the 1929 GRBs using complete information on five parameters}
\label{GRB:5var}
We now perform GMMBC using the five variables on the
1929 GRBs for which complete observations on 
$\log_{10}T_{50}$,
$\log_{10}T_{90}$, $\log_{10}P_{256}$, $\log_{10}H_{32}$ and
$\log_{10}H_{321}$ are available. Our objective here is  to assesswhether the groups found by clustering the 1599 GRBs with six parameters
adequately explain the kinds of GRBs generally available in the BATSE 4Br
catalog. After obtaining the GMMBC for the optimal $K$ as per BIC, we 
identify the groups and study their properties in
relation to the groups identified in Section~\ref{GRB:GMM}.
Our investigative framework mirrors Section~\ref{GRB:GMM}
so we report results in brief.
\subsubsection{Results} 
\begin{figure}
  \includegraphics[width=0.45\textwidth]{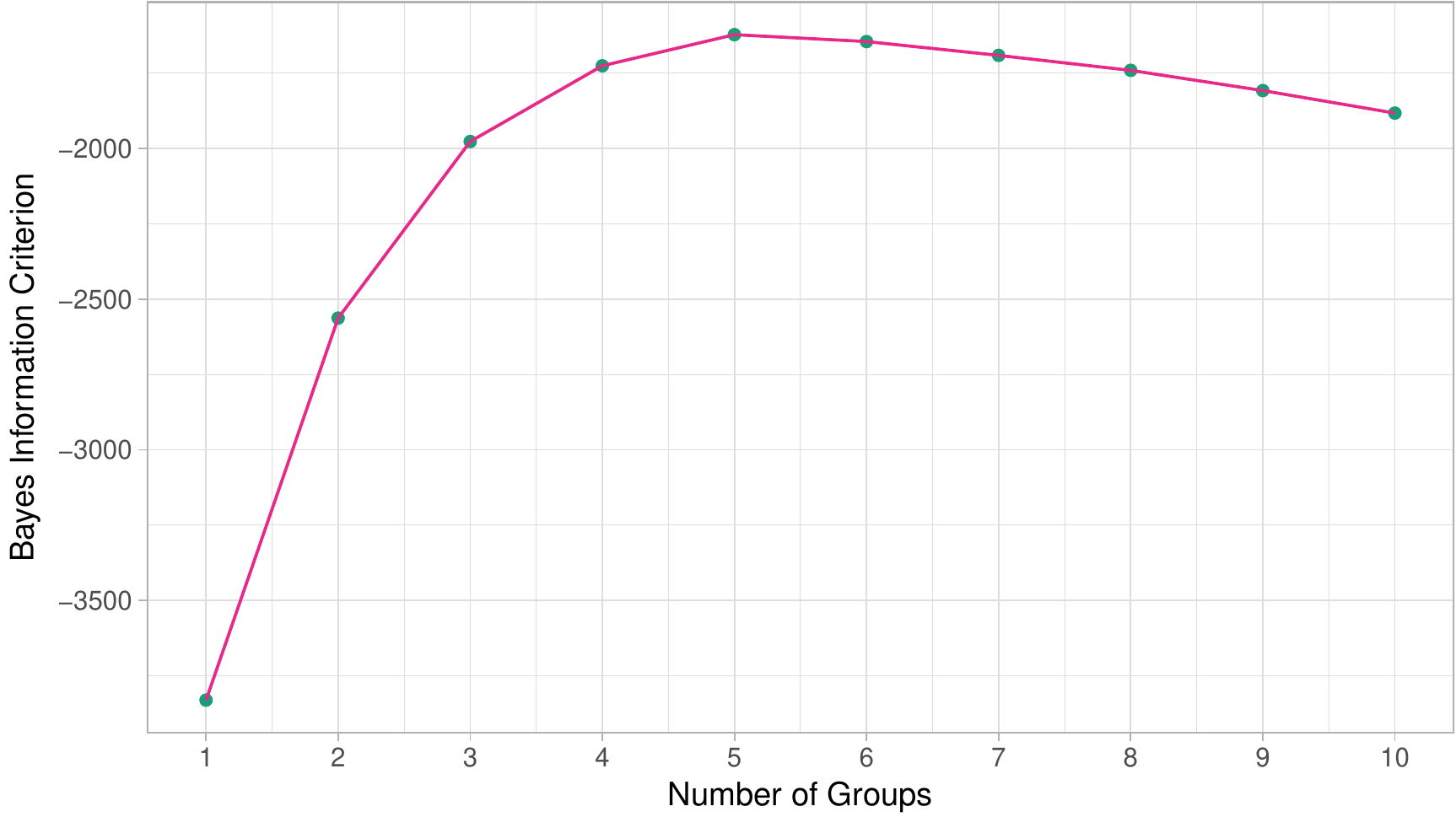}
  \caption{Plot of BIC against  $K$ for GMMBC with
    5 parameters.}
  \label{fig:BIC 5var}
\end{figure}

\begin{figure}
  \centering
  \mbox{
    \hspace{-0.045\textwidth}
    \subfloat[]{\label{tab:overlap5}
      \begin{tabular}{cc}
        \hline\hline
        $K$&$\ddot\omega$\\ \hline
        2 & 0.14 \\%0.143\\
        3 &0.11 \\ %0.110 \\
        4 & 0.11\\ %0.114\\
        5&0.11\\ %0.113\\
        \hline\hline
      \end{tabular}
    }
    \centering
    \subfloat[]{\label{fig:overlap5}
      \raisebox{-.5\height}
               {\includegraphics[width=0.37\textwidth]{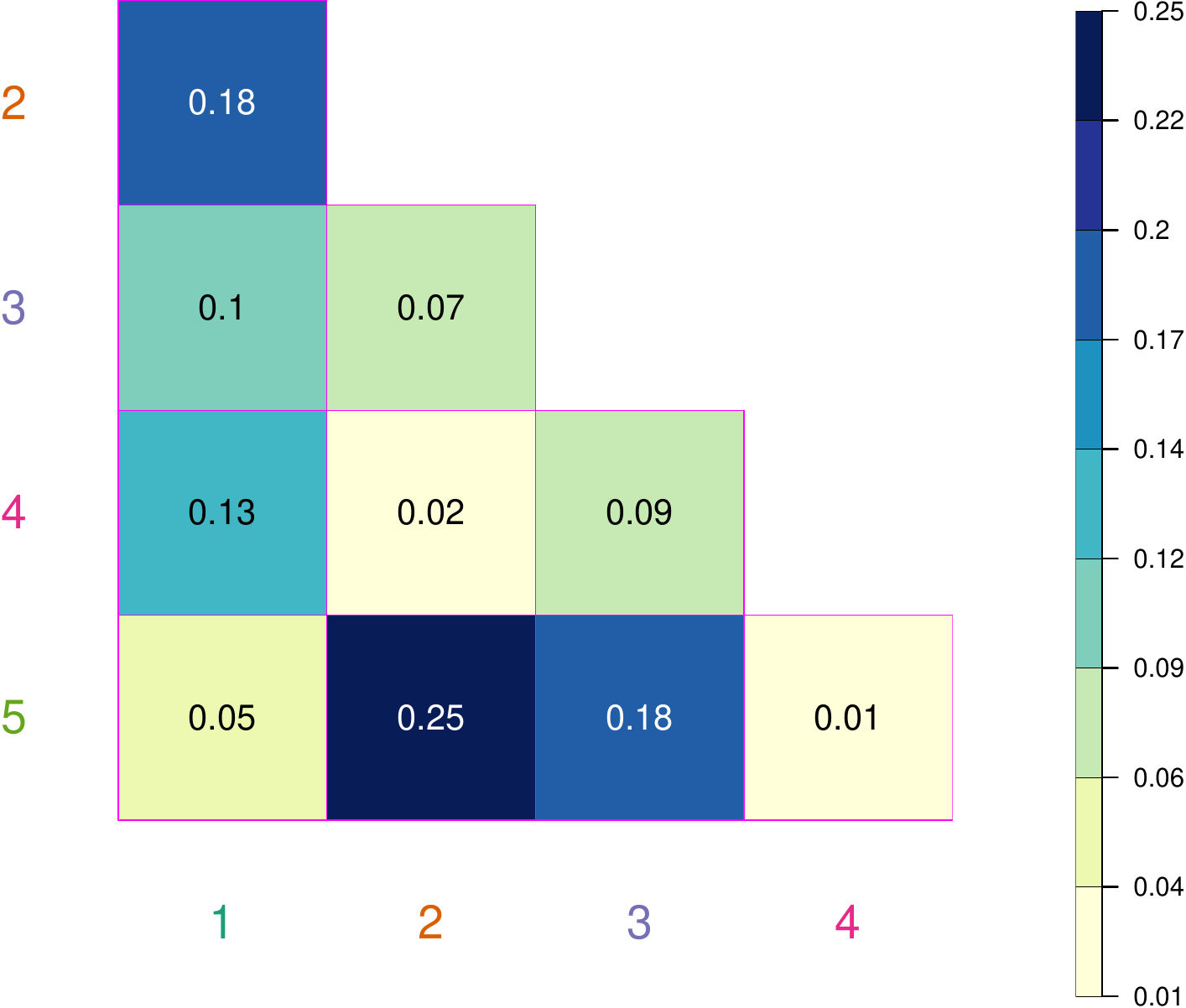}}    }
    %  }
    %  \mbox{
    \centering
  }
  \caption{(a) Generalized overlap ($\ddot\omega$) measures for the
    2-, 3-, 4- and   5-component GMMBC solutions of the 1929  BATSE
    4Br five-parameter GRBs.
    (b) Pairwise overlap measures between the $k$ and the 
    $l$th groups of the 5-groups solution as 
    indicated by the margins. Here margin color corresponds
    to the group indicator that is the same for all displays and tabulations 
    involving the five-component GMMBC fits to the five-parameter dataset.}
\end{figure}

We again first scout for redundancy among these five variables for clustering the 1929 GRBs using the variable selection algorithm 
of Section~\ref{r:d} with the {\tt clustvarsel} package in R. Table~\ref{tab:5varmc} indicates that all
five variables contain relevant clustering information, and none of
them are redundant, a result that  is not
wholly surprising in the light of the results of Table~\ref{tab:1}.
Proceeding as before with GMMBC using the five variables $\log_{10}T_{50}$,
$\log_{10}T_{90}$, $\log_{10}P_{256}$, $\log_{10}H_{32}$ and
$\log_{10}H_{321}$, we find that the BIC is again optimal for $K=5$, as
per Figure~\ref{fig:BIC 5var}. The generalized overlap measures for $K=2,3,4,5$ are reported in the table of
Figure~\ref{tab:overlap5} and show negligible differences between $K=3,4,5$ for each of which
$\ddot\omega$ is around 0.11. Thus, though BIC clearly indicates that $K=5$ is
the clear winner, there is not much difference in terms of separation
of the clusters between either of the $K=3,4,5$ solutions. (We remind
the reader of our comment at the end of Section~\ref{sec:overlap} that
$\ddot\omega$ is a diagnostic for evaluating the distinctiveness and
quality of the obtained clustering while BIC is a metric for choosing
$K$.) Note also that the separation is far higher for $K=5$ than for
$K=3$ in the table of Figure~\ref{tab:overlap}, which indicates that
the additional relevant clustering information provided by
$\log_{10}F_t$ (which can not be used here because of missing
observations) provides more distinct groups.
\begin{table}
  \caption{Results of stepwise variable selection
    algorithm for determining redundancy of  $\log_{10}T_{90}$, $\log_{10}T_{50}$,
    $\log_{10}P_{256}$, $\log_{10}H_{321}$, $\log_{10}H_{32}$,
    for GMMBC of the 1929 five-parameter  GRBs.}
  \centering
  \begin{tabular}{rlrcr}
    \hline\hline
    Step & Variable  & Step Type & BIC Difference & Decision\\
    \hline
    1 & $\log_{10}T_{90}$ & Add & 533.01 & Accepted \\
    2 & $\log_{10}T_{50}$ & Add & 478.64 & Accepted \\
    3 & $\log_{10}P_{256}$ & Add & 247.12 & Accepted \\
    4 & $\log_{10}P_{256}$ & Remove & 247.52 & Rejected \\
    5 & $\log_{10}H_{321}$ & Add & 252.34 & Accepted \\
    6 & $\log_{10}H_{321}$ & Remove & 245.97 & Rejected \\
    7 & $\log_{10}H_{32}$ & Add & 711.53 & Accepted \\
    8 & $\log_{10}T_{50}$ & Remove & 271.56 & Rejected \\ [1ex]
    \hline
  \end{tabular}
  \label{tab:5varmc}
\end{table}
%    $K$&2&3&4&5\\ \hline
%    $\omega$&0.143&0.110&0.114&0.113\\\hline\hline
%  \end{tabular}
\paragraph{Analysis:}
\label{sec:analysis5}
Figure~\ref{fig:overlap5} displays the pairwise overlaps between the
five groups  in the five-component GMM fit to the 1929 GRBs. As in
Figure~\ref{fig:overlap}, the fourth group is the most distinct. Also,
the fifth group shows a substantial amount of 
overlap with the second and third groups. Note that most of the
pairwise overlap measures for 
the five groups in the  six-parameter analysis (Figure~\ref{fig:overlap})
are considerably lower than the pairwise overlaps for the five groups
in the current five-parameter analysis indicating that the groups are
now less distinct upon exclusion of $\log_{10}F_t$, which
as per Table~\ref{tab:1} contains  relevant clustering
information.

GMMBC of the 1599 GRBs using complete
observations on six parameters ($\log_{10}T_{50}$, $\log_{10}T_{90}$,
$\log_{10}P_{256}$, $\log_{10}H_{32}$, $\log_{10}H_{321}$ and
$\log_{10}F_t$) and the 1929 GRBs using complete observations on the
five parameters (with $\log_{10}F_t$ excluded from the above list) both yielded
five groups as per BIC.  It would be of interest to compare the two
groupings.  Table~\ref{tab:6:5} tabulates the 
groups assigned to the 1929 GRBs under each of the two
classifications. Indeed, 330 of these GRBs could not be classified under
the GMMBC of Section \ref{GRB:GMM}  because of missing $F_t$s (and are
assigned NA under the grouping of Section \ref{GRB:GMM}). It is
interesting to note that a clear
majority of the 330  GRBs that could not be clustered in
Section~\ref{GRB:GMM} appear to be of the second kind, with the other
kinds being fairly evenly represented but for the third group which
only has 18 GRBs. 
The high values in the diagonal elements indicate that the overall
grouping structure agrees quite well under both analyses. It is
however, interesting to note that the second and the 
fifth groups have the most mismatches for both cases. (These are also
the two largest classes in both groupings.) A plausible reason for
these mismatches may be the
loss of relevant clustering information by our 
having to exclude $F_t$ in order to perform GMMBC of the 1929 GRBs. 
\begin{table}
  \caption{Number of 1929 GRBs assigned to each grouping using GMMBC
    of 1599 GRBs using complete information on all six parameters
    (Grouping I) and GMMBC of 1929 GRBs using complete information on
    five parameters (Grouping II). GRBs which are missing the sixth
    parameter ($F_t$) can not be assigned under Grouping I and are
    placed in the NA category.}
  \begin{tabular}{cc|cccccc|c}\hline\hline
    && \multicolumn{6}{c}{Grouping I (from GMMBC of 1599 GRBs)} \\
    &&\color{Gr1} 1&\color{Gr2} 2&\color{Gr3} 3&\color{Gr4} 4&\color{Gr5} 5&NA&Total\\ \hline
    \multirow{6}{*}{\begin{sideways}{Grouping II}\end{sideways}}
    &\color{Gr1} 1&78&3&7&1&0&40&129\\
    &\color{Gr2} 2&71&456&1&3&36&200&767\\
    &\color{Gr3} 3&1&11&130&3&17&18&180\\
    &\color{Gr4} 4&18&0&6&285&3&34&346\\
    &\color{Gr5} 5&6&81&5&0&377&38&507\\ 
    &Total&174&551&149&292&433&330\\ \hline
  \end{tabular}
  \label{tab:6:5}
\end{table}

\begin{table}
  \caption{Mean values for the five parameters in each of the five
    groups obtained     by GMMBC of the 1929 GRBs.}
  \begin{tabular}{rrrrrrr}
    \hline\hline
    $k$ &$\log T_{50}$&$\log T_{90}$&$\log P_{256}$&$\log H_{32}$&$\log
    H_{321}$&\\ \hline
    \color{Gr1} 1&-0.109&0.323&-0.122&0.548&0.174\\
    \color{Gr2} 2&1.093&1.470&-0.057&0.365&0.095\\
    \color{Gr3} 3&1.116&1.609&0.461&0.464&0.222\\
    \color{Gr4} 4&-0.063&0.662&0.678&0.459&0.222\\
    \color{Gr5} 5&-0.640&-0.299&0.197&0.769&0.592\\
    \hline
  \end{tabular}
  \label{tab:means5}
\end{table}
\begin{figure}
  \includegraphics[width=0.5\textwidth]{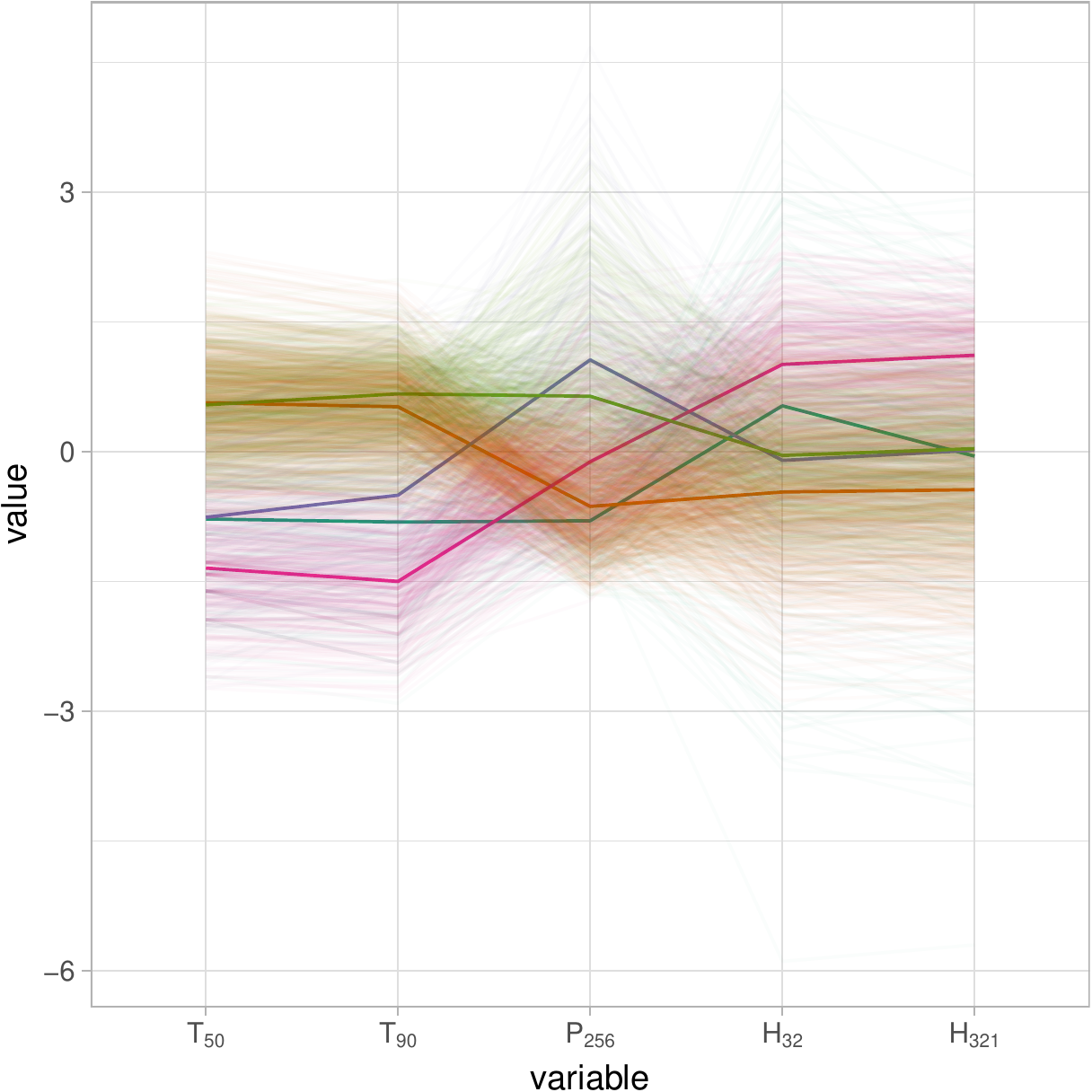}
  \caption{Parallel coordinate plot of 1929 BATSE 4Br GRBs colored as per
    their group indicators. The  solid lines represent the median of
    the five variables shown in 
    the plot. Variables are in the logarithmic scale.}
  \label{pcp.5var.plot}
\end{figure}
The five group means of the current analysis are presented in
Table~\ref{tab:means5}. In Section \ref{GRB:6analysis} we had used the
three parameters duration ($T_{90}$), fluence ($F_t$) and hardness
($H_{321}$) to classify the five groups obtained in that section. Such
classification 
is not possible here due to the omission of $F_t$. If we are to
classify the groups using only the duration variable  $T_{90}$, then
the fifth group will be classified as the group containing short duration
bursts while the first and fourth groups will be classified as the group
containing bursts of intermediate duration. The remaining two groups
(Groups 2 and 3) are those with the long duration
bursts. Comparing the group means of the
current analysis with those obtained in the six-parameter 
analysis (Table~\ref{tab:means}) shows good agreement in the case
of $\log_{10}H_{32}$ but not for many of the other cases. In general,
the second group means are reasonably close for all common parameters. 
This sort of discrepancy for the other groups and parameters is not very
surprising because the exclusion of $\log_{10}F_t$ has resulted in less
distinct clusters. We facilitate a visual inspection of the
five groups obtained in the five-parameters analysis by means of a
parallel coordinate plot  (Figure~\ref{pcp.5var.plot}). Indeed, the
medians for many of the parameters are similar to those for the
six-parameter case (note that the median is a more robust insensitive
measure of the central tendency than the mean) which indicates that
the disagreement in the means is perhaps because of the presence of
noise owing to the reduced distinctiveness in the five-parameter
groupings potentially on account of the exclusion of $\log_{10}F_t$. 
\begin{figure*}
  \mbox{ 
    \subfloat[Group 1]{\label{fig5:cor1}\includegraphics[width=0.33\textwidth]{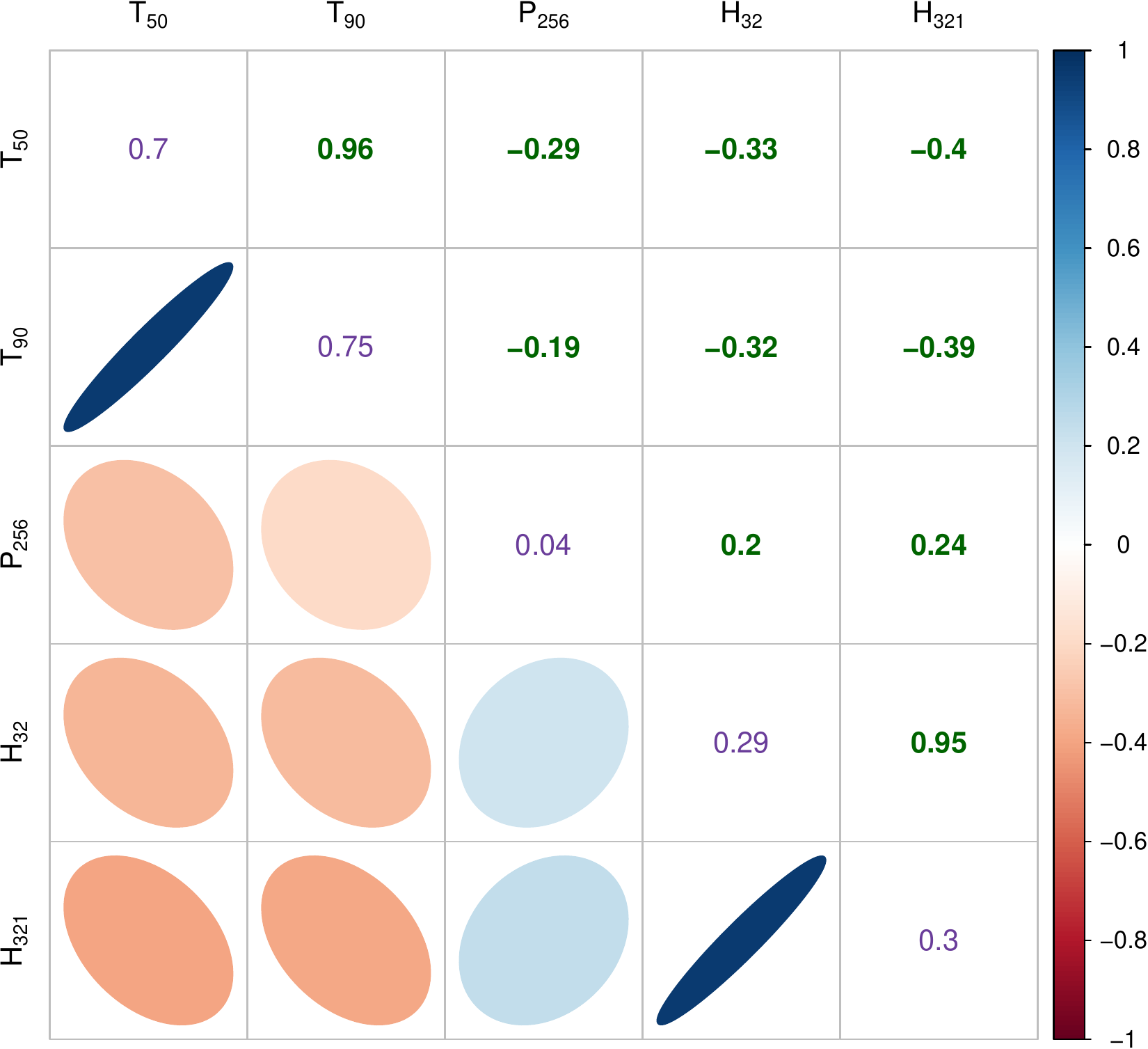}}
    \subfloat[Group 2]{\label{fig5:cor2}\includegraphics[width=0.33\textwidth]{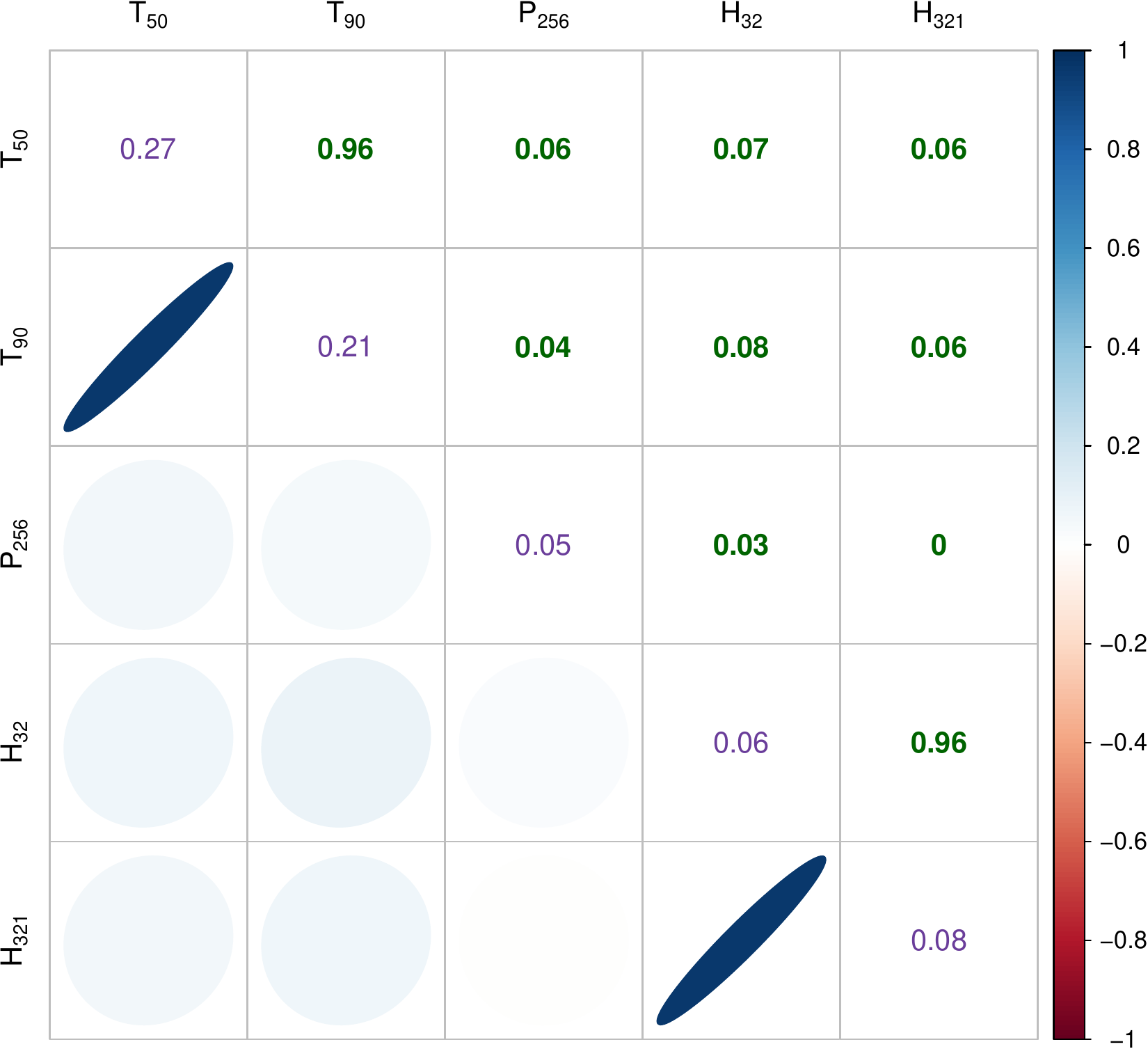}}
    \subfloat[Group 3]{\label{fig5:cor3}\includegraphics[width=0.33\textwidth]{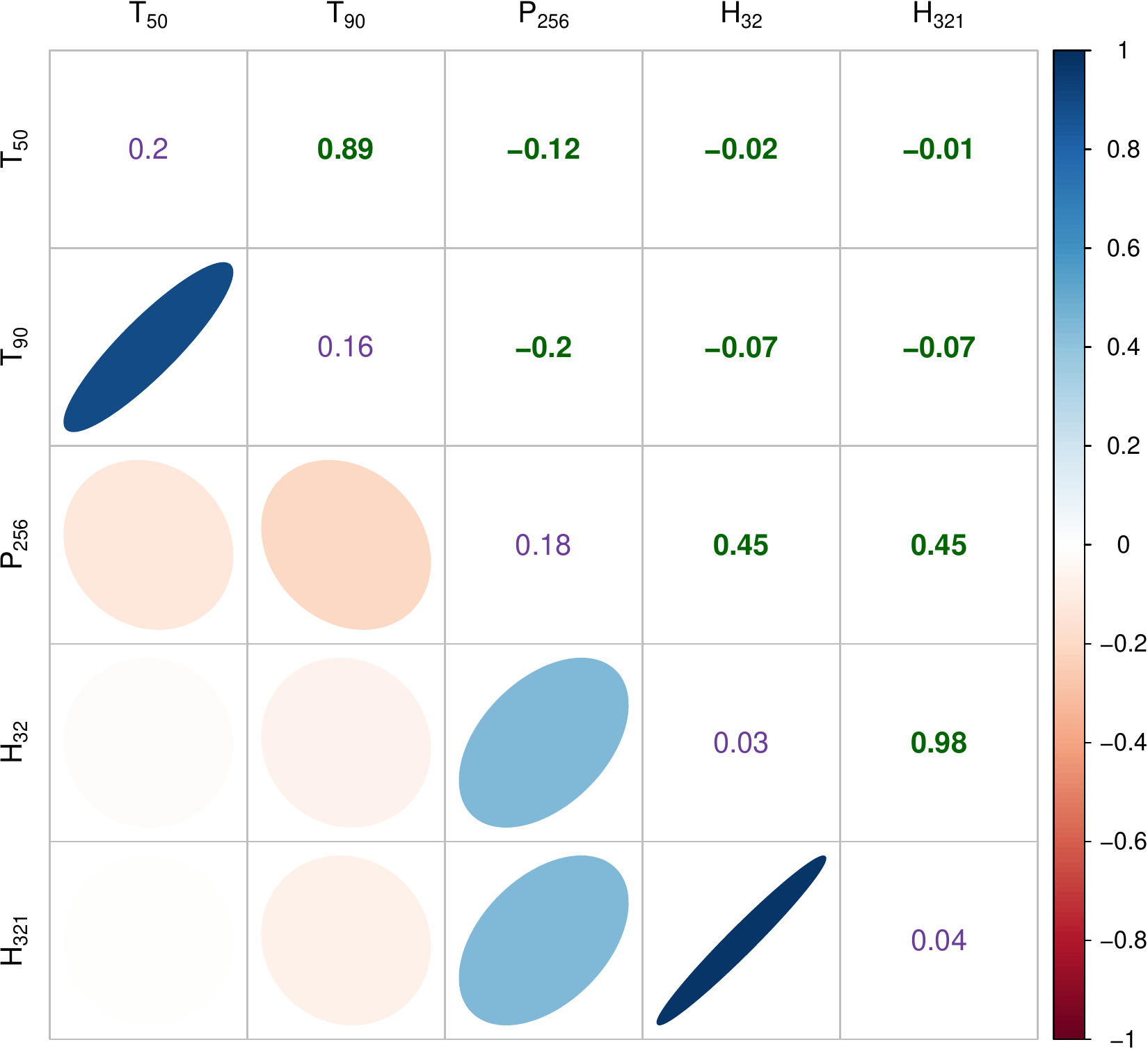}}}
  \mbox{
    \subfloat[Group 4]{\label{fig5:cor4}\includegraphics[width=0.33\textwidth]{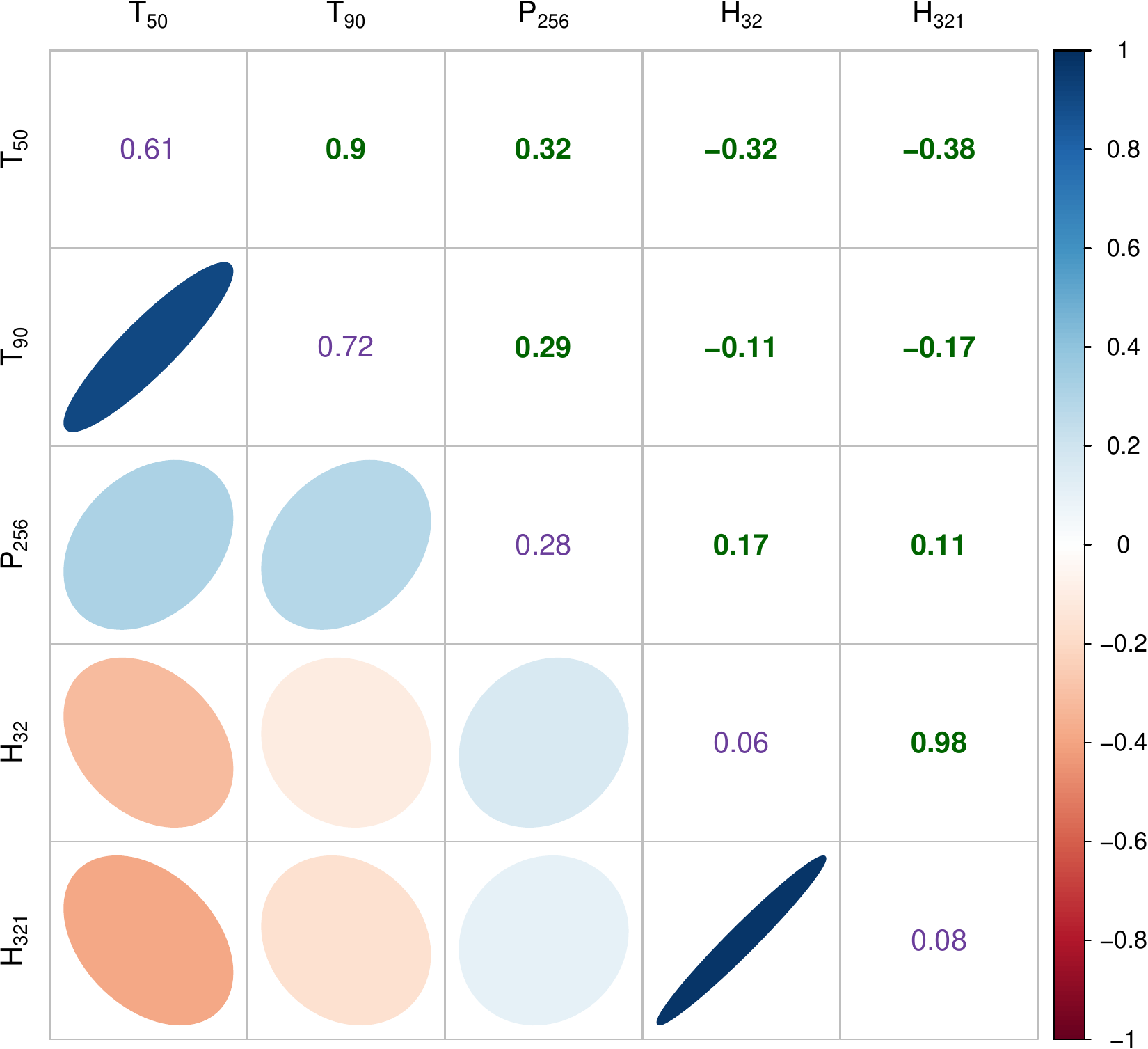}}
    \subfloat[Group 5]{\label{fig5:cor5}\includegraphics[width=0.33\textwidth]{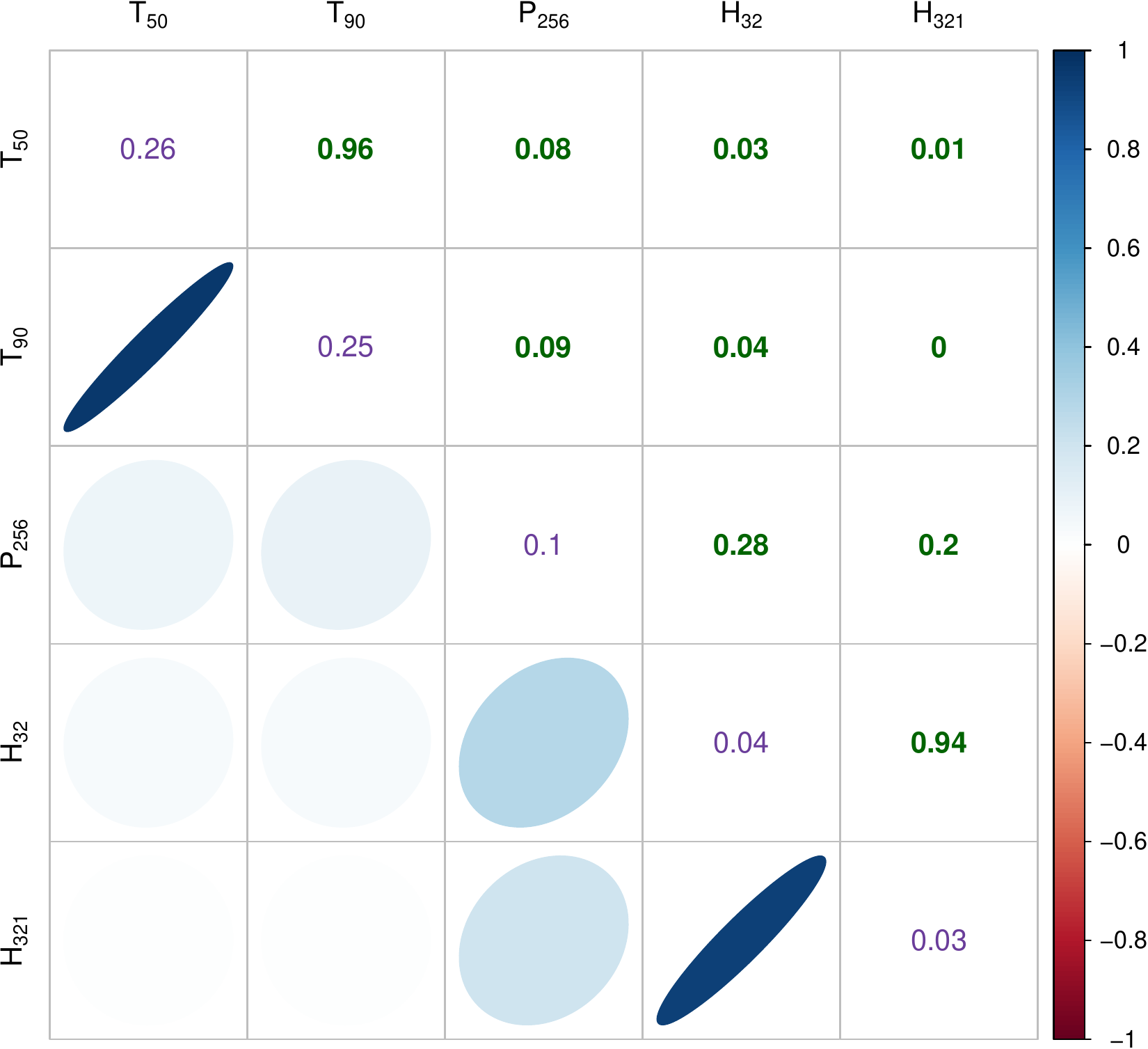}}}
  \caption{Variances and displays of the estimated correlations for each
    of the five   groups obtained by the 5-component GMMBC of the 1929
    five-parameter GRBs. For each group, the off-diagonal elements
    display correlation between the variables while the diagonals
    display the variances. Both correlations and variances are
    calculated for the variables in the base-10 logarithmic scale.}
  \label{corrplot.5var}
\end{figure*}

We also studied the associations between the five
parameters in the five groups. Figure~\ref{corrplot.5var}
displays the dispersion structures of the five GMMBC groups obtained
by GMMBC of the 1929 GRBs. The estimated group variances of the five
variables obtained by GMMBC is provided in the diagonals of each
display. A very strong positive association is exhibited between the
two duration variables ($\log_{10}T_{50}$ and $\log_{10}T_{90}$) and
the two hardness ratios ($\log_{10}H_{32}$ and $\log_{10}H_{321}$) for
all five groups while $\log_{10}T_{90}$ and $\log_{10}H_{321}$ exhibits
a moderate negative association in the first and the fourth  groups, a weak
negative association in the third group and a weak positive association in
the second group. Negligible association is also exhibited between
them in the fifth group.  On the other hand, $\log_{10}T_{90}$ and
$\log_{10}P_{256}$ exhibit a moderately  
negative association in the first and third groups. In the fourth group they
exhibit a moderate positive association and a weak positive
association in each of the other two groups. Finally, $\log_{10}P_{256}$ and
$\log_{10}H_{321}$ exhibit moderate positive association in the all
the groups barring the second one where they show negligible association. 

The results of our five-parameter GMMBC analysis on the 1929 GRBs also
indicate that 
there are five kinds of GRBs and reasonable agreement with the groups
obtained using the six-parameter analysis with the 1599
GRBs. However, the five-parameter analysis has resulted in less
distinct groups owing to the required dropping of  $\log_{10}F_t$
which was determined to be relevant for GMMBC as per
Table~\ref{tab:1}. It would therefore be important to develop methods
which could incorporate missing observations in the analysis. This
would permit the inclusion of $\log_{10}F_t$ for the 1599 cases for
which this information is available in the GMMBC analysis of all the
GRBs. 

\section{Conclusions}
\label{summary}
\citet{chattopadhyayetal07}  carried our $k$-means clustering and GMMBC on
the BATSE 4B catalog data and suggested that three classes were
adequate to describe the heterogeneity in the GRBs. Other researchers
have reported findings of between 2-3 groups of GRBs. These
conflicting accounts led us to
carry out a detailed review of nonhierarchical clustering methods
used in analyzing GRBs from the BATSE 4Br catalog.
We found $k$-means to be somewhat inconclusive for clustering GRBs
since in our own experiments using $k$-means, the jump 
statistic, the silhouette index and graphical SPCA projections  did
not show much  support for a few number of  homogeneous
spherically-dispersed groups. We feel that this may be due to
some of the factors involving $k$-means (and clustering algorithms in
general). Taking help from a simulated dataset we have reviewed and
demonstrated the limitations of $k$-means owing to inherent structural
assumptions of the clusters obtained.  The variables used by
\citet{mukherjeeetal98} for non 
parametric hierarchical clustering and model based clustering is
another point of study in this paper. Six variables were used by 
\citet{chattopadhyayetal07} for their evaluations, but only a subset
of these variables have been used by other 
researchers~\citep{mukherjeeetal98,horvath02,horvathetal08,zitounietal15,zhangetal16,veresetal10,horvathetal10}.
\citet{mukherjeeetal98} eliminated three of the six 
selected variables citing presence of redundancy among them. Using
model-based variable selection, we did not find much evidence of
redundancy among the six variables originally selected by
\citet{mukherjeeetal98}. We further perform GMMBC
using all the six variables, and used BIC to determine the optimum number
of groups and found five homogeneous groups in the BATSE 4Br GRB
data. To validate the clustering results obtained through GMMBC, we have
calculated the generalized overlap measures which all indicated five
distinct groups while modeling GRBs using GMMBC.
We thus provide evidence 
in favor of five groups in the BATSE 4Br dataset using model based
clustering. In terms of properties the five groups showed a good
degree of resemblance to the three groups obtained by
\citet{mukherjeeetal98} using model based clustering. Following the
procedure of \citet{mukherjeeetal98} we have classified the five
groups (using duration, fluence and hardness ratios) as
intermediate/faint/intermediate,
long/intermediate/soft, intermediate/intermediate/intermediate,
short/faint/hard and long/bright/intermediate.

Our primary analysis in this paper focused on 1599 GRBs from the
BATSE 4Br catalog for which complete information on all the six
parameters $\log_{10}T_{50}$,
$\log_{10}T_{90}$, $\log_{10}P_{256}$, $\log_{10}H_{32}$,
$\log_{10}H_{321}$ and $\log_{10}F_t$ were available. All these
parameters were determined to be relevant for GMMBC as per variable
selection methods. We next analyzed the 1929 GRBs for which complete
information is available only on five parameters ({\em i.e.}, the above
six parameters excluding $F_t$). Here also, we obtained five distinct
groups with good agreement in many of the classifications for the
common 1599 GRBs used in both groupings, but the obtained clusters
were less distinct in this case. We surmise that this may be on
account of the necessity to exclude $F_t$ in order to perform
GMMBC. 

There are a number of issues that could benefit from further
attention. For one, the identified groups could be analyzed further
in same manner as  in \citet{hakkilaetal00} or \citet{hakkilaetal03}. Further, 
It would be interesting to see if the groupings of GRBs
that we have found in this paper are also replicated for datasets from
catalogs such as {\em Swift} and {\em Fermi}. Further, in this paper,
we have used only six of the variables 
available in the BATSE catalog: it would be of interest to see if the
unused variables contain additional or more precise information
for clustering the GRBs. Further, in this paper, we have followed the
standard approach of analyzing the data in the logarithmic scale. It
would be of interest to see if more appropriate transformations
exist for cluster analysis. Finally, the analysis of the complete
BATSE 4Br catalog could benefit further if clustering methods can be
developed for use when some of the observations are missing.  We feel that GMMBC is particularly well-suited for adaptation
in this case because the underlying EM algorithm is originally
developed in the context of missing data problems. This would also
benefit analysis of other catalogs such as the Swift catalog.
Thus, we see that while we have some interesting findings in the
context of clustering GRBs, a number of issues meriting further consideration remain.

\section*{Acknowledgements}
%We thank Asis Kumar Chattopadhyay for introducing us to this
%problem. 
We sincerely thank Asis K. Chattopadhyay for introducing us to this problem and also for providing the BATSE 4B GRB dataset used in \citet{chattopadhyayetal07}. We are also very grateful to Charles A. Meegan and the BATSE GRB team for clarifying our query on the origin of the zero parameter values in the BATSE 4Br catalog. Our sincerest thanks are also extended to an anonymous reviewer and editor for their many comments during the review process of this article.
 
%%%%%%%%%%%%%%%%%%%%%%%%%%%%%%%%%%%%%%%%%%%%%%%%%%



\bibliographystyle{mnras}
\bibliography{references} % if your bibtex file is called example.bib

\begin{thebibliography}{}
\makeatletter
\relax
\def\mn@urlcharsother{\let\do\@makeother \do\$\do\&\do\#\do\^\do\_\do\%\do\~}
\def\mn@doi{\begingroup\mn@urlcharsother \@ifnextchar [ {\mn@doi@}
  {\mn@doi@[]}}
\def\mn@doi@[#1]#2{\def\@tempa{#1}\ifx\@tempa\@empty \href
  {http://dx.doi.org/#2} {doi:#2}\else \href {http://dx.doi.org/#2} {#1}\fi
  \endgroup}
\def\mn@eprint#1#2{\mn@eprint@#1:#2::\@nil}
\def\mn@eprint@arXiv#1{\href {http://arxiv.org/abs/#1} {{\tt arXiv:#1}}}
\def\mn@eprint@dblp#1{\href {http://dblp.uni-trier.de/rec/bibtex/#1.xml}
  {dblp:#1}}
\def\mn@eprint@#1:#2:#3:#4\@nil{\def\@tempa {#1}\def\@tempb {#2}\def\@tempc
  {#3}\ifx \@tempc \@empty \let \@tempc \@tempb \let \@tempb \@tempa \fi \ifx
  \@tempb \@empty \def\@tempb {arXiv}\fi \@ifundefined
  {mn@eprint@\@tempb}{\@tempb:\@tempc}{\expandafter \expandafter \csname
  mn@eprint@\@tempb\endcsname \expandafter{\@tempc}}}

\bibitem[\protect\citeauthoryear{Akaike}{Akaike}{1973}]{akaike73}
Akaike H.,  1973, in Second international symposium on information theory. pp
  267--281, \mn@doi{10.1007/978-1-4612-1694-0_15}

\bibitem[\protect\citeauthoryear{Akaike}{Akaike}{1974}]{akaike74}
Akaike H.,  1974, \mn@doi [IEEE Transactions on Automatic Control]
  {10.1109/TAC.1974.1100705}, 19, 716

\bibitem[\protect\citeauthoryear{Bagoly, M\'esz\'aros, Horv\'ath, Bal\'azs  \&
  M\'esz\'aros}{Bagoly et~al.}{1998}]{bagolyetal98}
Bagoly Z.,  M\'esz\'aros A.,  Horv\'ath I.,  Bal\'azs L.~G.,   M\'esz\'aros P.,
   1998, \mn@doi [\apj] {10.1086/305530}, 498, 342

\bibitem[\protect\citeauthoryear{Baudry, Raftery, Celeux, Lo  \&
  Gottardo}{Baudry et~al.}{2010}]{baudryetal10}
Baudry J.-P.,  Raftery A.~E.,  Celeux G.,  Lo K.,   Gottardo R.,  2010, \mn@doi
  [Journal of Computational and Graphical Statistics]
  {10.1198/jcgs.2010.08111}, 19, 332

\bibitem[\protect\citeauthoryear{Biernacki, Celeux  \& Govaert}{Biernacki
  et~al.}{2003}]{biernackietal03}
Biernacki C.,  Celeux G.,   Govaert G.,  2003, \mn@doi [Computational
  Statistics and Data Analysis] {10.1016/S0167-9473(02)00163-9}, 41, 561

\bibitem[\protect\citeauthoryear{Celeux \& Govaert}{Celeux \&
  Govaert}{1992}]{celeuxandgovaert92}
Celeux G.,  Govaert G.,  1992, \mn@doi [Computational Statistics and Data
  Analysis] {http://dx.doi.org10.1016/0167-9473(92)90042-E}, 14, 315

\bibitem[\protect\citeauthoryear{Chang}{Chang}{1983}]{chang83}
Chang W.~C.,  1983, Applied Statistics, 32, 267–275

\bibitem[\protect\citeauthoryear{Chattopadhyay, Misra, Chattopadhyay  \&
  Naskar}{Chattopadhyay et~al.}{2007}]{chattopadhyayetal07}
Chattopadhyay T.,  Misra R.,  Chattopadhyay A.~K.,   Naskar M.,  2007, \mn@doi
  [\apj] {https://doi.org/10.1086/520317}, 667, 1017

\bibitem[\protect\citeauthoryear{Chen \& Maitra}{Chen \&
  Maitra}{2011}]{chenandmaitra11}
Chen W.-C.,  Maitra R.,  2011, \mn@doi [Statistical Analysis and Data Mining]
  {10.1002/sam.10143}, 4, 567

\bibitem[\protect\citeauthoryear{Chen \& Maitra}{Chen \&
  Maitra}{2015a}]{Chen2015EMClusterpackage}
Chen W.-C.,  Maitra R.,  2015a, {EMCluster}: {EM} Algorithm for Model-Based
  Clustering of Finite Mixture Gaussian Distribution

\bibitem[\protect\citeauthoryear{Chen \& Maitra}{Chen \&
  Maitra}{2015b}]{Chen2015EMClustervignette}
Chen W.-C.,  Maitra R.,  2015b, A Quick Guide for the {EMCluster} Package (Ver.
  0.2-5)

\bibitem[\protect\citeauthoryear{Chen, Ostrouchov, Pugmire, Prabhat  \&
  Wehner}{Chen et~al.}{2014}]{chenetal13}
Chen W.-C.,  Ostrouchov G.,  Pugmire D.,  Prabhat  Wehner M.,  2014, \mn@doi
  [Technometrics] {10.1080/00401706.2013.826146}, 55, 513

\bibitem[\protect\citeauthoryear{Dempster, Laird  \& Rubin}{Dempster
  et~al.}{1977}]{dempsteretal77}
Dempster A.~P.,  Laird N.~M.,   Rubin D.~B.,  1977, \mn@doi [Jounal of the
  Royal Statistical Society, Series B] {10.2307/2984875}, 39, 1

\bibitem[\protect\citeauthoryear{{Dezalay}, {Barat}, {Talon}, {Syunyaev},
  {Terekhov}  \& {Kuznetsov}}{{Dezalay} et~al.}{1992}]{dezalayetal92}
{Dezalay} J.-P.,  {Barat} C.,  {Talon} R.,  {Syunyaev} R.,  {Terekhov} O.,
  {Kuznetsov} A.,  1992, in {Paciesas} W.~S.,  {Fishman} G.~J.,  eds,  American
  Institute of Physics Conference Series Vol. 265, American Institute of
  Physics Conference Series. pp 304--309

\bibitem[\protect\citeauthoryear{Everitt}{Everitt}{2011}]{everitt11}
Everitt B.,  2011, Cluster analysis.
Wiley, Chichester, West Sussex, U.K, \mn@doi{10.1002/9780470977811}

\bibitem[\protect\citeauthoryear{{Feigelson} \& {Babu}}{{Feigelson} \&
  {Babu}}{1998}]{feigelsonandbabu98}
{Feigelson} E.~D.,  {Babu} G.~J.,  1998, in {McLean} B.~J.,  {Golombek} D.~A.,
  {Hayes} J.~J.~E.,   {Payne} H.~E.,  eds,  IAU Symposium Vol. 179, New
  Horizons from Multi-Wavelength Sky Surveys. p.~363,
  \mn@doi{10.1007/978-94-009-1485-8_90}

\bibitem[\protect\citeauthoryear{Forgy}{Forgy}{1965}]{forgy65}
Forgy E.,  1965, Biometrics, 21, 768

\bibitem[\protect\citeauthoryear{Fraley \& Raftery}{Fraley \&
  Raftery}{2002}]{fraleyandraftery02}
Fraley C.,  Raftery A.~E.,  2002, \mn@doi [Journal of the American Statistical
  Association] {10.1198/016214502760047131}, 97, 611

\bibitem[\protect\citeauthoryear{Fraley, Raftery, Murphy  \& Scrucca}{Fraley
  et~al.}{2012}]{fraleyetal12}
Fraley C.,  Raftery A.~E.,  Murphy T.~B.,   Scrucca L.,  2012, Technical
  Report~597, {\tt mclust} Version 4 for R: Normal Mixture Modeling for
  Model-Based Clustering, Classification, and Density Estimation.
Department of Statistics, University of Washington

\bibitem[\protect\citeauthoryear{Garey \& Johnson}{Garey \&
  Johnson}{1979}]{gareyandjohnson79}
Garey M.~R.,  Johnson D.~S.,  1979, Computers and Intractability: A Guide to
  the Theory of {NP}-Completeness.
W. H. Freeman

\bibitem[\protect\citeauthoryear{Green \& Krieger}{Green \&
  Krieger}{1995}]{greenandkrieger95}
Green P.~E.,  Krieger A.~M.,  1995, Journal of the Market Research Society, 37,
  221–239

\bibitem[\protect\citeauthoryear{Hakkila, Haglin, Pendleton, Mallozzi, Meegan
  \& Roiger}{Hakkila et~al.}{2000}]{hakkilaetal00}
Hakkila J.,  Haglin D.~J.,  Pendleton G.~N.,  Mallozzi R.~S.,  Meegan C.~A.,
  Roiger R.~J.,  2000, \mn@doi [\apj] {10.1086/309107}, 538, 165

\bibitem[\protect\citeauthoryear{Hakkila, Giblin, Roiger, Haglin, Paciesas  \&
  Meegan}{Hakkila et~al.}{2003}]{hakkilaetal03}
Hakkila J.,  Giblin T.~W.,  Roiger R.~J.,  Haglin D.~J.,  Paciesas W.~S.,
  Meegan C.~A.,  2003, \mn@doi [\apj] {10.1086/344568}, 582, 320

\bibitem[\protect\citeauthoryear{Hartigan \& Wong}{Hartigan \&
  Wong}{1979}]{hartiganandwong79}
Hartigan J.~A.,  Wong M.~A.,  1979, \mn@doi [Applied statistics]
  {10.2307/2346830}, pp 100--108

\bibitem[\protect\citeauthoryear{Hintze \& Nelson}{Hintze \&
  Nelson}{1998}]{hintzeandnelson98}
Hintze J.~L.,  Nelson R.~D.,  1998, \mn@doi [The American Statistician]
  {10.1080/00031305.1998.10480559}, 52, 181

\bibitem[\protect\citeauthoryear{Horv\'ath}{Horv\'ath}{1998}]{horvathetal98}
Horv\'ath I.,  1998, \mn@doi [\apj] {10.1086/306416}, 508, 757

\bibitem[\protect\citeauthoryear{Horv\'ath}{Horv\'ath}{2002}]{horvath02}
Horv\'ath I.,  2002, \mn@doi [\aap] {10.1051/0004-6361:20020808}, 392, 791

\bibitem[\protect\citeauthoryear{Horv\'ath}{Horv\'ath}{2009}]{horvath09}
Horv\'ath I.,  2009, \mn@doi [\apss] {10.1007/s10509-009-0039-1}, 323

\bibitem[\protect\citeauthoryear{Horv{\'a}th \& T{\'o}th}{Horv{\'a}th \&
  T{\'o}th}{2016}]{horvathandtoth16}
Horv{\'a}th I.,  T{\'o}th B.~G.,  2016, \mn@doi [\apss]
  {10.1007/s10509-016-2748-6}, 361, 155

\bibitem[\protect\citeauthoryear{{Horv{\'a}th}, {M{\'e}sz{\'a}ros},
  {Bal{\'a}zs}  \& {Bagoly}}{{Horv{\'a}th} et~al.}{2004}]{horvathetal04}
{Horv{\'a}th} I.,  {M{\'e}sz{\'a}ros} A.,  {Bal{\'a}zs} L.~G.,   {Bagoly} Z.,
  2004, Baltic Astronomy, \href
  {http://adsabs.harvard.edu/abs/2004BaltA..13..217H} {13, 217}

\bibitem[\protect\citeauthoryear{{Horv{\'a}th}, {Bal{\'a}zs}, {Bagoly}, {Ryde}
  \& {M{\'e}sz{\'a}ros}}{{Horv{\'a}th} et~al.}{2006}]{horvathetal06}
{Horv{\'a}th} I.,  {Bal{\'a}zs} L.~G.,  {Bagoly} Z.,  {Ryde} F.,
  {M{\'e}sz{\'a}ros} A.,  2006, \mn@doi [\aap] {10.1051/0004-6361:20041129},
  \href {http://adsabs.harvard.edu/abs/2006A%26A...447...23H} {447, 23}

\bibitem[\protect\citeauthoryear{Horv\'ath, Bal\'azs, Bagoly  \&
  Veres}{Horv\'ath et~al.}{2008}]{horvathetal08}
Horv\'ath I.,  Bal\'azs L.~G.,  Bagoly Z.,   Veres P.,  2008, \mn@doi [\aap]
  {10.1051/0004-6361:200810269}, 489, L1

\bibitem[\protect\citeauthoryear{Horv\'ath, Bagoly, Bal\'azs, de
  Ugarte~Postigo, Veres  \& M\'asz\'aros}{Horv\'ath
  et~al.}{2010}]{horvathetal10}
Horv\'ath I.,  Bagoly Z.,  Bal\'azs L.~G.,  de Ugarte~Postigo A.,  Veres P.,
  M\'asz\'aros A.,  2010, \mn@doi [\apj] {10.1088/0004-637X/713/1/552}, 713,
  552

\bibitem[\protect\citeauthoryear{Hotelling}{Hotelling}{1933a}]{hotelling33a}
Hotelling H.,  1933a, \mn@doi [Journal of Educational Psychology]
  {10.1037/h0071325}, 24, 417

\bibitem[\protect\citeauthoryear{Hotelling}{Hotelling}{1933b}]{hotelling33b}
Hotelling H.,  1933b, \mn@doi [Journal of Educational Psychology]
  {10.1037/h0070888}, 24, 498

\bibitem[\protect\citeauthoryear{Inselberg}{Inselberg}{1985}]{inselberg85}
Inselberg A.,  1985, \mn@doi [The Visual Computer] {10.1007/BF01898350}, 1, 69

\bibitem[\protect\citeauthoryear{Kass \& Raftery}{Kass \&
  Raftery}{1995}]{kassandraftery95}
Kass R.~E.,  Raftery A.~E.,  1995, \mn@doi [Journal of the American Statistical
  Association] {10.1080/01621459.1995.10476572}, 90, 773

\bibitem[\protect\citeauthoryear{Keribin}{Keribin}{2000}]{keribin00}
Keribin C.,  2000, \mn@doi [Sankhy\={a}] {10.2307/25051289}, 62, 49

\bibitem[\protect\citeauthoryear{Kettenring}{Kettenring}{2006}]{kettenring06}
Kettenring J.~R.,  2006, \mn@doi [Journal of {C}lassification]
  {10.1111/j.1541-0420.2008.01160.x}, 23, 3

\bibitem[\protect\citeauthoryear{Koren \& Carmel}{Koren \&
  Carmel}{2004}]{korenandcarmel04}
Koren Y.,  Carmel L.,  2004, \mn@doi [IEEE {T}ransactions on {V}isualization
  and {C}omputer {G}raphics] {10.1109/TVCG.2004.17}, 10, 459

\bibitem[\protect\citeauthoryear{{Kouveliotou}, {Meegan}, {Fishman}, {Bhat},
  {Briggs}, {Koshut}, {Paciesas}  \& {Pendleton}}{{Kouveliotou}
  et~al.}{1993}]{kouveliotouetal93}
{Kouveliotou} C.,  {Meegan} C.~A.,  {Fishman} G.~J.,  {Bhat} N.~P.,  {Briggs}
  M.~S.,  {Koshut} T.~M.,  {Paciesas} W.~S.,   {Pendleton} G.~N.,  1993,
  \mn@doi [\apjl] {10.1086/186969}, \href
  {http://adsabs.harvard.edu/abs/1993ApJ...413L.101K} {413, L101}

\bibitem[\protect\citeauthoryear{Kulkarni \& Desai}{Kulkarni \&
  Desai}{2017}]{kulkarnianddesai17}
Kulkarni S.,  Desai S.,  2017, \mn@doi [\apss] {10.1007/s10509-017-3047-6},
  362, 70

\bibitem[\protect\citeauthoryear{Lloyd}{Lloyd}{1982}]{lloyd82}
Lloyd S.,  1982, \mn@doi [IEEE Transactions on Information Theory]
  {10.1109/TIT.1982.1056489}, 28, 129

\bibitem[\protect\citeauthoryear{MacQueen}{MacQueen}{1967}]{macqueen67}
MacQueen J.,  1967, in Proceedings of the Fifth Berkeley Symposium on
  Mathematical Statistics and Probability, Volume 1: Statistics. University of
  California Press, Berkeley, California, pp 281--297, \url
  {http://projecteuclid.org/euclid.bsmsp/1200512992}

\bibitem[\protect\citeauthoryear{Maitra}{Maitra}{2001}]{maitra01}
Maitra R.,  2001, \mn@doi [Technometrics] {10.1198/004017001316975925}, 43, 336

\bibitem[\protect\citeauthoryear{Maitra}{Maitra}{2009}]{maitra09}
Maitra R.,  2009, \mn@doi [IEEE/ACM Transactions on Computational Biology and
  Bioinformatics] {10.1109/TCBB.2007.70244}, 6, 144

\bibitem[\protect\citeauthoryear{Maitra}{Maitra}{2010}]{maitra10}
Maitra R.,  2010, \mn@doi [NeuroImage]
  {http://dx.doi.org/10.1016/j.neuroimage.2009.11.070}, 50, 124

\bibitem[\protect\citeauthoryear{Maitra \& Melnykov}{Maitra \&
  Melnykov}{2010}]{maitraandmelnykov10}
Maitra R.,  Melnykov V.,  2010, \mn@doi [Journal of Computational and Graphical
  Statistics] {10.1198/jcgs.2009.08054}, 19, 354

\bibitem[\protect\citeauthoryear{Maitra \& Ramler}{Maitra \&
  Ramler}{2009}]{maitraandramler09}
Maitra R.,  Ramler I.~P.,  2009, \mn@doi [Biometrics]
  {10.1111/j.1541-0420.2008.01064.x}, 65, 341

\bibitem[\protect\citeauthoryear{Maitra, Melnykov  \& Lahiri}{Maitra
  et~al.}{2012}]{maitraetal12}
Maitra R.,  Melnykov V.,   Lahiri S.,  2012, \mn@doi [Journal of the American
  Statistical Association] {http://dx.doi.org/10.1080/01621459.2011.646935},
  107, 378

\bibitem[\protect\citeauthoryear{Maugis, Celeux  \& Martin-Magniette}{Maugis
  et~al.}{2009}]{maugisetal09}
Maugis C.,  Celeux G.,   Martin-Magniette M.-L.,  2009, \mn@doi [Biometrics]
  {10.1111/j.1541-0420.2008.01160.x}

\bibitem[\protect\citeauthoryear{{Mazets} et~al.,}{{Mazets}
  et~al.}{1981}]{mazetsetal81}
{Mazets} E.~P.,  et~al., 1981, \mn@doi [\apss] {10.1007/BF00649140}, \href
  {http://adsabs.harvard.edu/abs/1981Ap%26SS..80....3M} {80, 3}

\bibitem[\protect\citeauthoryear{McLachlan \& Krishnan}{McLachlan \&
  Krishnan}{2008}]{mclachlanandkrishnan08}
McLachlan G.,  Krishnan T.,  2008, The {EM} Algorithm and Extensions, second
  edn.
Wiley, New York, \mn@doi{10.2307/2534032}

\bibitem[\protect\citeauthoryear{McLachlan \& Peel}{McLachlan \&
  Peel}{2000}]{mclachlanandpeel00}
McLachlan G.,  Peel D.,  2000, Finite Mixture Models.
John Wiley and Sons, Inc., New York, \mn@doi{10.1002/0471721182}

\bibitem[\protect\citeauthoryear{Melnykov \& Maitra}{Melnykov \&
  Maitra}{2010}]{melnykovandmaitra10}
Melnykov V.,  Maitra R.,  2010, \mn@doi [Statist. Surv.] {10.1214/09-SS053}, 4,
  80

\bibitem[\protect\citeauthoryear{Melnykov \& Maitra}{Melnykov \&
  Maitra}{2011}]{melnykovandmaitra11}
Melnykov V.,  Maitra R.,  2011, Journal of Machine Learning Research, 12, 69

\bibitem[\protect\citeauthoryear{Melnykov, Chen  \& Maitra}{Melnykov
  et~al.}{2012}]{melnykovetal13}
Melnykov V.,  Chen W.-C.,   Maitra R.,  2012, \mn@doi [Journal of Statistical
  Software] {10.18637/jss.v051.i12}, 51, 1

\bibitem[\protect\citeauthoryear{{Mukherjee}, {Feigelson}, {Jogesh Babu},
  {Murtagh}, {Fraley}  \& {Raftery}}{{Mukherjee}
  et~al.}{1998}]{mukherjeeetal98}
{Mukherjee} S.,  {Feigelson} E.~D.,  {Jogesh Babu} G.,  {Murtagh} F.,  {Fraley}
  C.,   {Raftery} A.,  1998, \mn@doi [\apj] {10.1086/306386}, \href
  {http://adsabs.harvard.edu/abs/1998ApJ...508..314M} {508, 314}

\bibitem[\protect\citeauthoryear{Nakar}{Nakar}{2007}]{nakar07}
Nakar E.,  2007, \mn@doi [Physics Reports]
  {http://dx.doi.org/10.1016/j.physrep.2007.02.005}, 442, 166

\bibitem[\protect\citeauthoryear{Neath \& Cavanaugh}{Neath \&
  Cavanaugh}{2012}]{neathandcavanaugh12}
Neath A.~A.,  Cavanaugh J.~E.,  2012, \mn@doi [Wiley Interdisciplinary Reviews:
  Computational Statistics] {10.1002/wics.199}, 4, 199

\bibitem[\protect\citeauthoryear{{Norris}, {Cline}, {Desai}  \&
  {Teegarden}}{{Norris} et~al.}{1984}]{norrisetal84}
{Norris} J.~P.,  {Cline} T.~L.,  {Desai} U.~D.,   {Teegarden} B.~J.,  1984,
  \mn@doi [\nat] {10.1038/308434a0}, \href
  {http://adsabs.harvard.edu/abs/1984Natur.308..434N} {308, 434}

\bibitem[\protect\citeauthoryear{Paciesas et~al.,}{Paciesas
  et~al.}{1999}]{paciesasetal99}
Paciesas W.~S.,  et~al., 1999, \mn@doi [\apjs] {10.1086/313224}, 122, 465

\bibitem[\protect\citeauthoryear{{Paczy{\'n}ski}}{{Paczy{\'n}ski}}{1998}]{paczynski98}
{Paczy{\'n}ski} B.,  1998, \mn@doi [\apjl] {10.1086/311148}, \href
  {http://adsabs.harvard.edu/abs/1998ApJ...494L..45P} {494, L45}

\bibitem[\protect\citeauthoryear{Pearson}{Pearson}{1901}]{pearson1901}
Pearson K.,  1901, \mn@doi [Philosophical Magazine Series 6]
  {10.1080/14786440109462720}, 2, 559

\bibitem[\protect\citeauthoryear{{Pendleton} et~al.,}{{Pendleton}
  et~al.}{1994}]{pendletonetal94}
{Pendleton} G.~N.,  et~al., 1994, \mn@doi [\apj] {10.1086/174495}, \href
  {http://adsabs.harvard.edu/abs/1994ApJ...431..416P} {431, 416}

\bibitem[\protect\citeauthoryear{Pendleton et~al.,}{Pendleton
  et~al.}{1997}]{pendletonetal97}
Pendleton G.~N.,  et~al., 1997, \mn@doi [The Astrophysical Journal]
  {10.1086/304763}, 489, 175

\bibitem[\protect\citeauthoryear{Piran}{Piran}{2005}]{piran05}
Piran T.,  2005, \mn@doi [Rev. Mod. Phys.] {10.1103/RevModPhys.76.1143}, 76,
  1143

\bibitem[\protect\citeauthoryear{{R Core Team}}{{R Core Team}}{2016}]{R}
{R Core Team} 2016, R: A Language and Environment for Statistical Computing.
R Foundation for Statistical Computing, Vienna, Austria, \url
  {https://www.R-project.org/}

\bibitem[\protect\citeauthoryear{Raftery \& Dean}{Raftery \&
  Dean}{2006}]{rafteryanddean06}
Raftery A.~E.,  Dean N.,  2006, \mn@doi [Journal of the American Statistical
  Association] {10.1198/016214506000000113}, 101, 168

\bibitem[\protect\citeauthoryear{Rousseeuw}{Rousseeuw}{1987}]{rousseeuw87}
Rousseeuw P.~J.,  1987, \mn@doi [Journal of Computational and Applied
  Mathematics] {10.1016/0377-0427(87)90125-7}, 20, 53

\bibitem[\protect\citeauthoryear{Schwarz}{Schwarz}{1978}]{schwarz78}
Schwarz G.,  1978, \mn@doi [Ann. Statist.] {10.1214/aos/1176344136}, 6, 461

\bibitem[\protect\citeauthoryear{Scrucca \& Raftery}{Scrucca \&
  Raftery}{2015}]{clustvarselR}
Scrucca L.,  Raftery A.~E.,  2015, (submitted to) Journal of Statistical
  Software, ??, ??

\bibitem[\protect\citeauthoryear{{Shahmoradi} \& {Nemiroff}}{{Shahmoradi} \&
  {Nemiroff}}{2015}]{shahmoradiandnemiroff15}
{Shahmoradi} A.,  {Nemiroff} R.~J.,  2015, \mn@doi [\mnras]
  {10.1093/mnras/stv714}, \href
  {http://adsabs.harvard.edu/abs/2015MNRAS.451..126S} {451, 126}

\bibitem[\protect\citeauthoryear{Sugar \& James}{Sugar \&
  James}{2003}]{sugarandjames03}
Sugar C.~A.,  James G.~M.,  2003, \mn@doi [Journal of the American Statistical
  Association] {10.1198/016214503000000666}, 98, 750

\bibitem[\protect\citeauthoryear{Tarnopolski}{Tarnopolski}{2015}]{tarnopolski15}
Tarnopolski M.,  2015, \mn@doi [\aap] {10.1051/0004-6361/201526415}, 581, A29

\bibitem[\protect\citeauthoryear{Veres, Bagoly, Horv\'ath, M\'esz\'aros  \&
  Bal\'azs}{Veres et~al.}{2010}]{veresetal10}
Veres P.,  Bagoly Z.,  Horv\'ath I.,  M\'esz\'aros A.,   Bal\'azs L.~G.,  2010,
  \mn@doi [\apj] {https://doi.org/10.1088/0004-637X/725/2/1955}, 725, 1955

\bibitem[\protect\citeauthoryear{Ward}{Ward}{1963}]{ward63}
Ward J.,  1963, \mn@doi [Journal of the American Statistical Association]
  {10.2307/2282967}, 58, 236

\bibitem[\protect\citeauthoryear{Wegman}{Wegman}{1990}]{wegman90}
Wegman E.,  1990, \mn@doi [Journal of the American Statistical Association]
  {10.1080/01621459.1990.10474926}, 85, 664

\bibitem[\protect\citeauthoryear{Witten \& Tibshirani}{Witten \&
  Tibshirani}{2010}]{wittenandtibshirani10}
Witten D.~M.,  Tibshirani R.~J.,  2010, \mn@doi [Journal of the American
  Statistical Association] {10.1198/jasa.2010.tm09415}, 105, 294

\bibitem[\protect\citeauthoryear{Woosley \& Bloom}{Woosley \&
  Bloom}{2006}]{woosleyandbloom06}
Woosley S.,  Bloom J.,  2006, \mn@doi [\araa]
  {10.1146/annurev.astro.43.072103.150558}, 44, 507

\bibitem[\protect\citeauthoryear{Xu \& Wunsch}{Xu \&
  Wunsch}{2009}]{xuandwunsch09}
Xu R.,  Wunsch D.~C.,  2009, Clustering.
John Wiley and Sons, Inc, NJ, Hoboken, \mn@doi{10.1002/9780470382776}

\bibitem[\protect\citeauthoryear{Yeung \& Ruzzo}{Yeung \&
  Ruzzo}{2001}]{yeungandruzzo01}
Yeung K.~Y.,  Ruzzo W.~L.,  2001, Bioinformatics, 17, 763–774

\bibitem[\protect\citeauthoryear{Zhang, Yang, Choi  \& Chang}{Zhang
  et~al.}{2016}]{zhangetal16}
Zhang Z.-B.,  Yang E.-B.,  Choi C.-S.,   Chang H.-Y.,  2016, \mn@doi [\mnras]
  {10.1093/mnras/stw1835}, 462, 3243

\bibitem[\protect\citeauthoryear{Zitouni, Guessoum, Azzam  \&
  Mochkovitch}{Zitouni et~al.}{2015}]{zitounietal15}
Zitouni H.,  Guessoum N.,  Azzam W.~J.,   Mochkovitch R.,  2015, \mn@doi
  [\apss] {10.1007/s10509-015-2311-x}, 357, 7

\bibitem[\protect\citeauthoryear{{de Ugarte Postigo} et~al.,}{{de Ugarte
  Postigo} et~al.}{2011}]{deugartepostigoetal11}
{de Ugarte Postigo} A.,  et~al., 2011, \mn@doi [\aap]
  {10.1051/0004-6361/201015261}, \href
  {http://adsabs.harvard.edu/abs/2011A%26A...525A.109D} {525, A109}

\makeatother
\end{thebibliography}

%%%%%%%%%%%%%%%%%%%%%%%%%%%%%%%%%%%%%%%%%%%%%%%%%%

%%%%%%%%%%%%%%%%% APPENDICES %%%%%%%%%%%%%%%%%%%%%

\begin{comment}
\appendix

\section{Some extra material}

If you want to present additional material which would interrupt the flow of the main paper,
it can be placed in an Appendix which appears after the list of references.

%%%%%%%%%%%%%%%%%%%%%%%%%%%%%%%%%%%%%%%%%%%%%%%%%%

\end{comment}
% Don't change these lines
\bsp	% typesetting comment
\label{lastpage}
\end{document}